\documentclass[11pt,twoside]{article} 
\usepackage{epsfig,amsfonts,color}
\usepackage{amsmath}
\usepackage{amssymb, geometry,url}
\usepackage{soul}
\usepackage{ulem}
\usepackage[margin = 1cm]{caption}

\usepackage{amsmath}
\usepackage{blkarray}
\usepackage{bm}
\usepackage{algorithm}
\usepackage{algorithmic}
\usepackage{multirow}
\usepackage{graphicx}
\usepackage{caption}
\usepackage{subcaption}
\usepackage{placeins}
\usepackage{marvosym}
\usepackage{amsthm}

\newcommand{\bd}{\boldsymbol{d}}

\newcommand{\bX}{\boldsymbol{X}}

\newcommand{\dn}{d^{\mathcal{N}}}
\newcommand{\de}{d^{\mathcal{E}}}
\newcommand{\dg}{d^{\mathcal{G}}}
\newcommand{\mn}{\mathcal{N}}
\newcommand{\me}{\mathcal{E}}
\newcommand{\mg}{\mathcal{G}}
\newcommand{\ma}{\mathcal{A}}
\newcommand{\ml}{\mathcal{L}}
\newcommand{\mdg}{\mathcal{DG}}

\theoremstyle{definition}
\newtheorem{definition}{Definition}[section]

\usepackage{accents}
\usepackage{listings}
\DeclareCaptionFont{white}{\color{white}}
\DeclareCaptionFormat{listing}{{\parbox{\dimexpr\textwidth-2\fboxsep\relax}{#1#2#3}}}
\usepackage[dvipsnames]{xcolor}

\lstdefinelanguage{julia}
{
keywordsprefix=\@,
morekeywords={
exit,whos,edit,load,is,isa,isequal,typeof,tuple,ntuple,uid,hash,finalizer,convert,promote,
subtype,typemin,typemax,realmin,realmax,sizeof,eps,promote_type,method_exists,applicable,
invoke,dlopen,dlsym,system,error,throw,assert,new,Inf,Nan,pi,im,begin,while,for,in,return,
break,continue,macro,quote,let,if,elseif,else,try,catch,end,bitstype,ccall,do,using,module,
import,export,importall,baremodule,immutable,local,global,const,Bool,Int,Int8,Int16,Int32,
Int64,Uint,Uint8,Uint16,Uint32,Uint64,Float32,Float64,Complex64,Complex128,Any,Nothing,None,
function,type,typealias,abstract,get_node,create_estimation_model,set_solution, Matrix, Dict,
solve, get_solution, solve_ss_problem, create_estimation_problem, addnode, nothing
},
morekeywords = [2]{},
sensitive=true,
morecomment=[l]{\#},
morestring=[b]',
morestring=[b]"
}
\usepackage{multirow}
\usepackage{authblk}

\usepackage[colorlinks=true,linkcolor=blue,citecolor=blue,urlcolor=blue]{hyperref}

\usepackage{fancyhdr}
\begin{document}
\title{{\tt PlasmoData.jl} -- A Julia Framework for \\ Modeling and Analyzing Complex Data as Graphs}

\author{David L. Cole${}^{\dag}$and Victor M. Zavala${}^{\dag\ddag}$\thanks{Corresponding Author: victor.zavala@wisc.edu}
}

\date{\small
  ${}^\dag$Department of Chemical and Biological Engineering, \\[0in]
  University of Wisconsin-Madison, Madison, WI 53706 United States of America\\[.05in]
  ${}^\ddag$Mathematics and Computer Science Division, \\
  Argonne National Laboratory,  Lemont, IL 60439 United States of America\\[-2in]
}

\maketitle

\begin{abstract}
Datasets encountered in scientific and engineering applications appear in complex formats (e.g., images, multivariate time series, molecules, video, text strings, networks). Graph theory provides a unifying framework to model such datasets and enables the use of powerful tools that can help analyze, visualize, and extract value from data.  In this work, we present {\tt PlasmoData.jl}, an open-source, {\tt Julia} framework that uses concepts of graph theory to facilitate the modeling and analysis of complex datasets. The core of our framework is a general data modeling abstraction, which we call a {\tt DataGraph}. We show how the abstraction and  software implementation can be used to represent diverse data objects as graphs and to enable the use of tools from topology, graph theory, and machine learning (e.g., graph neural networks) to conduct a variety of tasks.  We illustrate the versatility of the framework by using real datasets: i) an image classification problem using topological data analysis to extract features from the graph model to train machine learning models; ii) a disease outbreak problem where we model multivariate time series as graphs to detect abnormal events; and iii) a technology pathway analysis problem where we highlight how we can use graphs to navigate connectivity. Our discussion also highlights how {\tt PlasmoData.jl} leverages native {\tt Julia} capabilities to enable compact syntax, scalable computations, and interfaces with diverse packages. Overall, we show that the {\tt DataGraph} abstraction and {\tt PlasmoData.jl} {\tt Julia} package are able to {\it model} data within graphs and enable useful analysis.
\end{abstract}


\section{Introduction}

Data appears in complex formats that require the use of advanced tools for its representation and processing. In this context, it is important to highlight that any data object (e.g., image, video, audio signal, a text string, time series) needs to be represented as a mathematical model (a mathematical abstraction) to enable processing and analysis. Examples of mathematical models that are available to do this include those from linear algebra (e.g., data is modeled as vectors, matrices, higher-order tensors), graph/network theory (e.g., data is modeled as graphs or hypergraphs), statistics (e.g., data is modeled as random variables), and topology/geometry (e.g., data is modeled as graphs, manifolds, and simplicial complexes).  Understanding how to model data and to recognize alternative representations of data is key in facilitating its analysis and in extracting information/value \cite{willett2016embedded, robertson1991methodology, torres2021and}. For example, an image can be represented as a matrix (each entry is a pixel), as a graph (each node is a pixel and edges represent connectivity of neighboring pixels), or as a function/manifold (the additional dimension represents the light intensity). These various representations reveal different (and often complementary) aspects/features of the data object; for instance, a matrix representation can help reveal correlation structures, while a graph representation can reveal connected structures. Importantly, how to represent a dataset is a {\it modeling decision} that can impact the information extracted from the data (can influence the value of the dataset); as such, it is necessary to have proper tools that can help experiment with different data models and associated analysis tools. 
\\

A modeling abstraction that has gained increased popularity in data science is {\em graphs} (networks). Graphs provide a unifying framework to capture diverse data objects because they have intuitive interpretation and they enable the use of powerful processing techniques. Simply stated, a graph is a mathematical model that comprises a set of nodes that are connected via edges; the nodes and edges are abstract objects that can encode data that may appear in different forms. In a typical graph, data attached to nodes and edges are scalar values or vectors, but more advanced graph abstractions can attach abstract objects such as algebraic models, text, or other graphs \cite{jalving2022graph,hajij2023topological}. Graph representations have been used in a wide variety of applications such as chemical processes \cite{shao2020modularity}, biological systems \cite{pavlopoulos2011using}, brain networks \cite{bullmore2009complex, rubinov2010complex}, hydrology \cite{cole2023hydrographs, heckmann2015graph, king2021lake}, disease transmission \cite{de2022euler,taylor2015topological}, and social networks \cite{wasserman_faust_1994}. An application that has recently gained substantial attention is the representation of molecules as graphs \cite{qin2021predicting,qin2023capturing, wu2018moleculenet}; here, atoms are represented by nodes, bonds are represented by edges, and nodes/edges can embed diverse attributes such as atom type and bond length (represented as vectors). These representations have been used to predict diverse properties of molecules and to predict potential reactions that can occur between them. 
\\

Recently, the field of topological data analysis (TDA) has opened interesting and new perspectives on data modeling and processing For example, a matrix (e.g., a grayscale image) can be represented as a node-weighted graph. Here, nodes are matrix entries, node weights are numerical values (data) attached to the matrix entries, and edges capture adjacency of the matrix entries \cite{smith2021euler,sanchez-lengeling2021a}. This representation can be applied to high-dimensional data (tensors), where entries of the higher dimensional object are connected to adjacent entries. Symmetric matrices (e.g., correlation or covariance matrices) can also be represented as edge-weighted graphs, where the nodes represent the variable associated with a column/row and edge weights capture the degree of correlation/covariance. The graph representation of correlation matrices is widely used in neuroscience to analyze brain connectivity  \cite{smith2021euler,bullmore2009complex,rubinov2010complex} and in disease transmission to identify extreme events \cite{de2022euler}. This is done via the use of tools of graph theory and topology, which enable the quantification of the graph structure or shape. For example, the Euler Characteristic (EC) is a topological descriptor that quantifies the structure of an unweighted graph by computing the number of connected components and the number of cycles in the graph \cite{smith2021euler}. To quantify the structure of weighted graphs (graphs with attached values to nodes and edges) one can use filtration/percolation operations; as the name suggests, these operations filter out nodes/edges with weight values below a certain threshold values. Filtration operations keep track of how topological descriptors appear or disappear at different weight threshold values, and this information is summarized in the form of topological summaries, such as the EC curve. Topological descriptors can reveal aspects of the data that might not be accessed by tools from linear algebra or statistics. For instance, eigenvalues and eigenvectors are linear algebra descriptors that extract information from matrices in the form of energy/variance, not shape. 
\\

Graph representations of data enables access to a wide variety of software tools for analysis and visualization. These tools range from broad to specific application scope. Some of the most general graph modeling tools include {\tt NetworkX} \cite{hagberg2008networkx} (in Python), {\tt Graphs.jl} \cite{Graphs2021} (in Julia), {\tt igraph} \cite{igraph2006} (in Python and R), {\tt graph-tool} \cite{peixoto2014graphtool} (in Python), and the {\tt graph} and {\tt digraph} functions of Matlab \cite{matlab_graphs}. These packages are general and allow the user to define their own graph data structures. They also include functions for analyzing graph connectivity and other metrics/descriptors. These tools also allow the user to define abstract properties or attributes on nodes and edges (in the case of {\tt Graphs.jl}, this ability is implemented in the package extension {\tt MetaGraphs.jl}). These packages use different methods for storing the data; depending on the amount or nature of the data, this can have significant implications on scalability and application scope. These packages are limited in the methods that they use for processing and analyzing data; for instance, these tools do not implement topological analysis operations (e.g., filtration operations and computation of topological descriptors). Furthermore, these packages do not provide capabilities to transform data models to graph models (e.g., obtain a graph from a tensor). 
\\

In addition to general graph modeling packages, several other packages exist that have more specific application scope. For instance, there are several graph tools specifically designed for studying brain function, including {\tt Graphvar} \cite{kruschwitz2015graphvar}, {\tt brain-connectivity-toolbox} \cite{rubinov2010complex}, {\tt GRETNA} \cite{wang2015gretna}, and {\tt BRAPH} \cite{mijalkov2017braph}. Because of this broad spectrum, a complete review of available tools is not practical here, but we will discuss the most pertinent and prominent options. Some application-specific packages are targeted towards creating models for machine learning, such as graph neural networks (GNNs). For instance, {\tt MoleculeNet} \cite{wu2018moleculenet} is designed for representing molecules as graphs, which can then be used for GNNs to enable predictions. The deep graph library \cite{wang2019dgl}, {\tt scikit-network} \cite{bonald2020}, and {\tt GraphNeuralNetworks.jl} \cite{lucibello2021GNN} all provide an interface for defining graphs with data that can then be used for training and testing GNNs. These packages are effective at conducting machine learning tasks, but they are not designed for performing transformations of data objects and topological analysis of the graph structure (e.g., finding the EC or connected components). In summary, existing graph modeling tools are often limited in their ability to represent and process wide spans of data. 
\\

The goal of this work is to provide a general framework for modeling data as graphs. Recently, we have developed a framework for modeling optimization problems as graphs\cite{jalving2022graph}; in this abstraction, nodes and edges can embed algebraic optimization problems. This abstraction has been shown to unify a wide range of structures found in optimization (e.g., optimal control, stochastic optimization, and network optimization) and this unification has enabled a number of advances in theory and algorithms. Moreover, the abstraction has facilitated access to a broad range of graph analysis and software tools that facilitate visualization and processing (e.g., graph partitioning and aggregation).  Our work aims to expand this graph abstraction to model general data and with this unify data objects found across domains and leverage the use of tools from such domains. Our abstraction is implemented as an open-source {\tt Julia} package that we call {\tt PlasmoData.jl}. This package has been designed to readily represent diverse data objects as graphs (e.g., images, matrices, and tensors as node-weighted graphs and symmetric matrices as edge-weighted graphs) and to store data within a user-defined graph structure. Our modeling framework interfaces to diverse software packages that enables processing of graph data objects (e.g., via filtration, partition, and aggregation), facilitates the computation of descriptors using tools from graph theory and topology, and enables the use of machine learning tools (e.g., graph neural networks or GNNs). Our software design principle is analogous to those of algebraic modeling languages (e.g., {\tt JuMP} or {\tt Plasmo.jl}), which provide provide interfaces to packages that process/solve such models. 
\\

\begin{figure}[!htp]
    \centering
    \includegraphics[width = \textwidth]{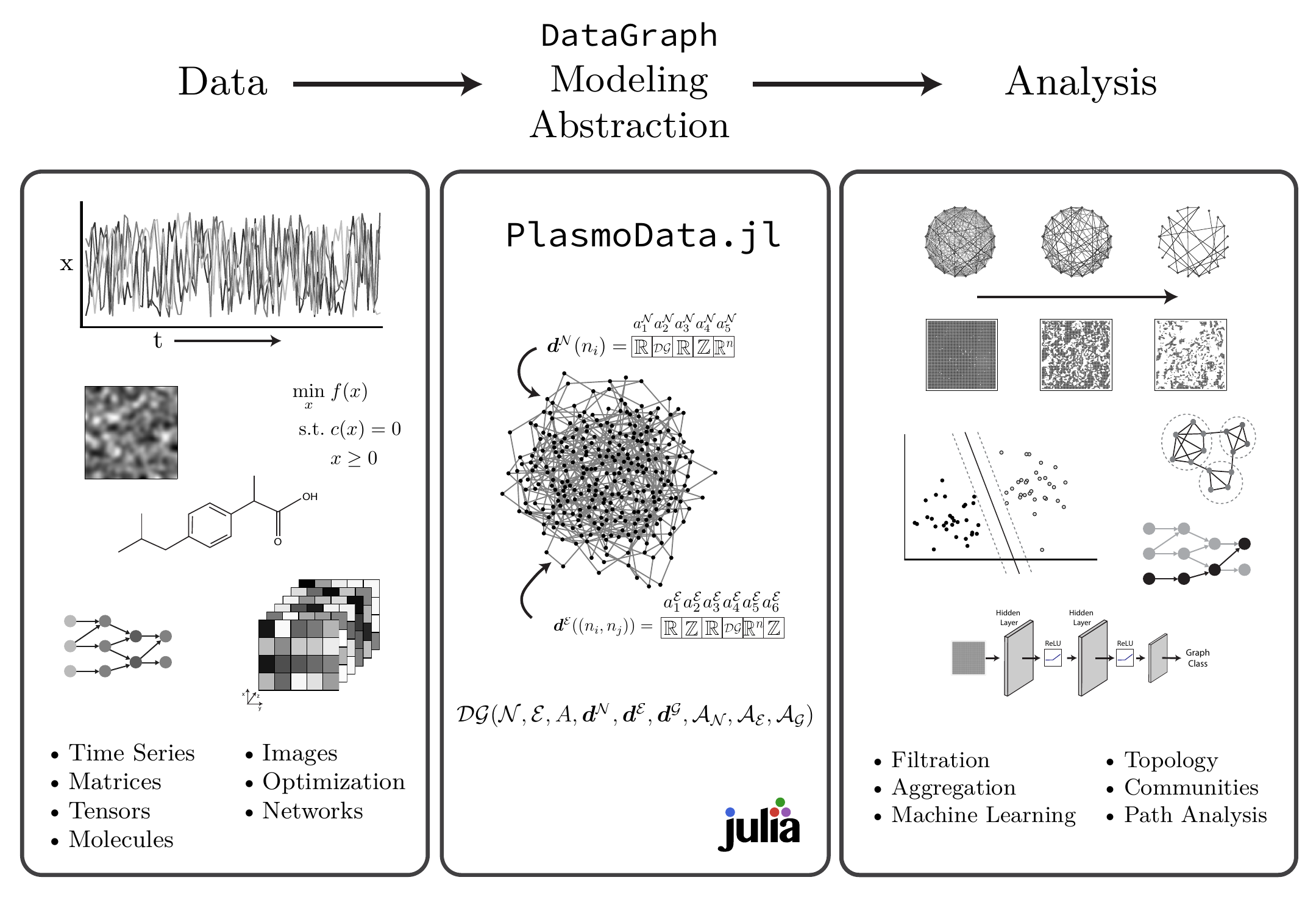}
    \captionof{figure}{Overview of the {\tt PlasmoData.jl} modeling framework. Diverse data objects can be modeled under the {\tt DataGraph} abstraction; this provides access to a diverse analysis tools and techniques for analyzing the data}
    \label{fig:GA}
\end{figure}

{\it The goal behind the implementation of {\tt PlasmoData.jl} is thus to change the way the user thinks about data (as a modeling task rather than just an analysis task).} We believe that the focus on modeling can bring significant benefits in the way we explore alternative data representations that are suitable for applications, in selecting suitable tools to enable data processing, and in interpreting analysis results. We demonstrate the versatility of {\tt PlasmoData.jl} and of its unifying modeling abstraction by using applications that appear in quite distinct application domains. A graphical representation of this work is shown in Figure \ref{fig:GA} and highlights how a variety of data can be modeled within {\tt PlasmoData.jl}, which in turn provides access to diverse analysis tools and techniques.
\\

The paper is structured as follows. Section \ref{sec:overview} provides an overview of the {\tt DataGraph} abstraction for modeling data as graphs, discusses how it is implemented in {\tt PlasmoData.jl}, and illustrates how some common data structures can be represented under the proposed abstraction. Section \ref{sec:data_analysis} provides an overview of data analysis that can be performed for data modeled as graphs. Section \ref{sec:examples} provides applications that illustrate the versatility of the modeling framework. Section \ref{sec:conclusion} provides concluding remarks and future directions. 

\section{Graph Abstraction and Software Implementation}\label{sec:overview}

This section introduces the {\tt DataGraph} modeling abstraction along with its implementation in {\tt PlasmoData.jl}, available at \url{https://github.com/zavalab/PlasmoData.jl}. We introduce a mathematical formulation for the {\tt DataGraph} and discuss how it can be applied to a variety of common data structures. We also provide brief code snippets showing how these representations are implemented in {\tt PlasmoData.jl}. Versions of these code snippets are also available in the repository \url{https://github.com/zavalab/JuliaBox/tree/master/PlasmoData_examples}, along with the package and {\tt Julia} versions used.

\subsection{Mathematical Formulation}

A graph is a mathematical model that contains a set of nodes $\mathcal{N}$ and a set of edges $\mathcal{E}$ (which connect nodes). A node $n\in \mathcal{N}$ and an edge $e\in \mathcal{E}$ can encode diverse abstract objects (e.g., values, text, equations). To indicate data embedded in the graph, we define the set of node (data) attributes as $\mathcal{A}_\mathcal{N}$, the set of edge (data) attributes as $\mathcal{A}_\mathcal{E}$, and a set of (global) graph attributes as $\mathcal{A}_\mathcal{G}$.  To associate data to specific nodes, edges, and to the graph we make the following definitions:

\begin{equation}\label{eq:node_data}
    d^\mathcal{N}_a(n) \in \mathcal{D}_a, \quad n \in \mathcal{N}, a \in \mathcal{A}_\mathcal{N}
\end{equation}

\begin{equation}\label{eq:edge_data}
    \de_a(e) \in \mathcal{D}_a, \quad e \in \mathcal{E}, a \in \ma_\me
\end{equation}

\begin{equation}\label{eq:graph_data}
    \dg_a \in \mathcal{D}_a, \quad  a \in \ma_\mg
\end{equation}

Here, $\mathcal{D}_a$ is the set of possible values for a data attribute $a$, $\dn_a(n)$ is the data stored on the node $n$ for attribute $a$, $\de_a(e)$ is the data stored on the edge $e$ for attribute $a$, and $\dg_a$ is the data stored on the graph for attribute $a$. Importantly, our definitions are abstract and we make no distinction on the form that any of that data must take. The data could be scalars, vectors, matrices, graphs, an optimization model, text strings, other graphs, or any other kind of information (see \cite{andrews2013beyond, huang2019text, jalving2022graph} for application examples for data that can be embedded in graphs). We use the notation $\bd^\mn$, $\bd^\me$, and $\bd^\mg$ to define the set of all node, edge, and graph data corresponding to $\ma_\mn$, $\ma_\me$, and $\ma_\mg$, respectively. We will use the notation $\bd^\mn(n)$ for the set of node data on node $n$ for the attribute set $\ma_\mn$ and $\bd^\me(e)$ for the set of node data on edge $e$ for the attribute set $\ma_\me$. Note that, under these definitions, the {\it attribute} sets are used for indicating subsets of data (and do not represent the data itself). 
\\

We use the previous definitions to define our modeling object, which we call a {\tt DataGraph}. This modeling abstraction has a graph structure and it stores data on the nodes, edges, and the graph itself; the object has the following form: 
\begin{equation}\label{eq:datagraph}
    \mathcal{DG} \big( \mn, \me, A, \bd^\mn, \bd^\me, \bd^\mg, \ma_\mn, \ma_\me, \ma_\mg \big).
\end{equation}
A {\tt DataGraph} object is fully defined by its nodes, edges, adjacency matrix, ($A$ which encodes the graph connectivity), node data, edge data, graph data, node attributes, edge attributes, and graph attributes. Importantly, the adjacency matrix also encodes the type of edges in the graph (directed or undirected). 

\subsection{Software Implementation}

\begin{figure}[!htp]
    \centering
    \includegraphics[width = \textwidth]{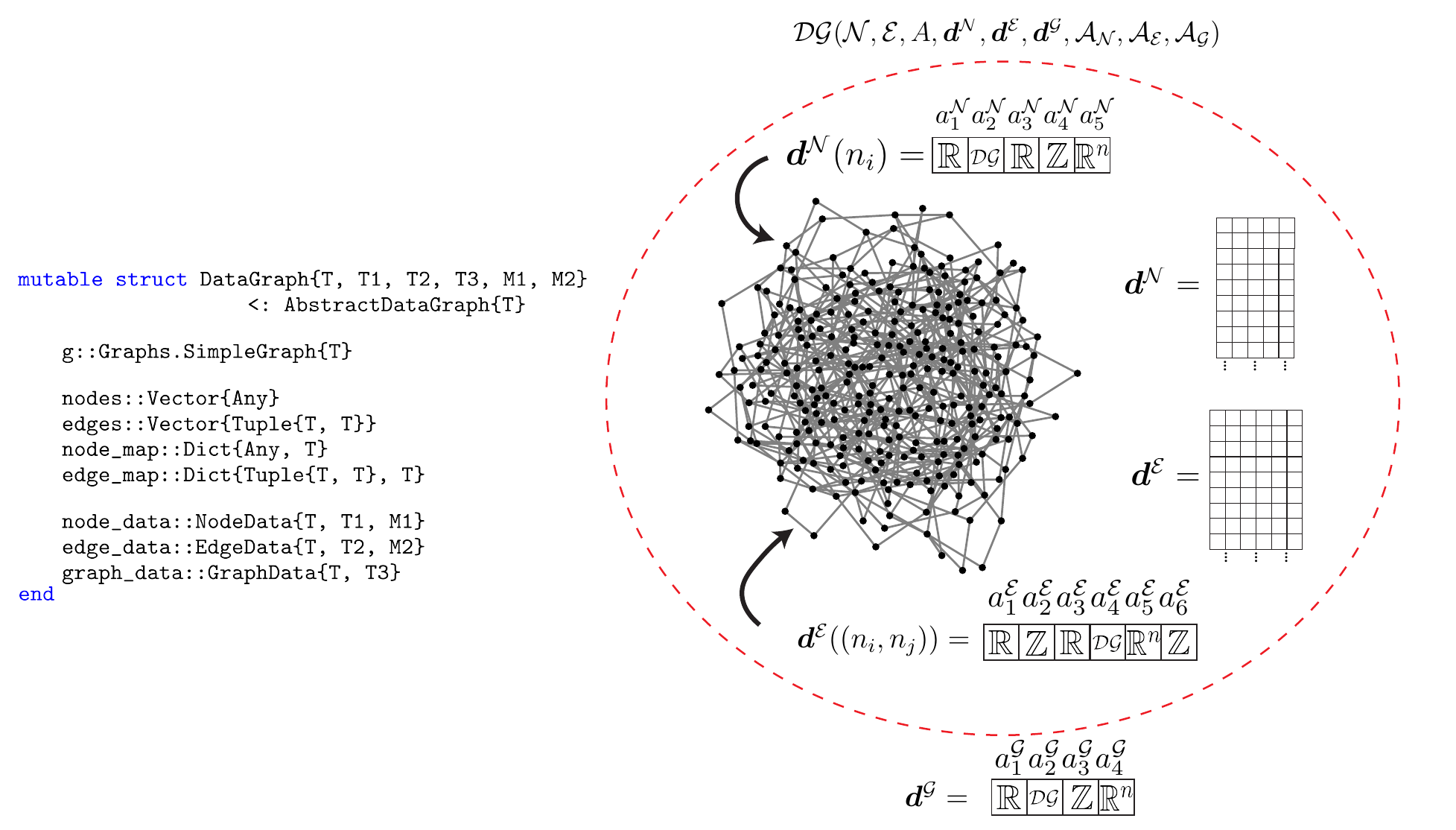}
    \captionof{figure}{Visualization of the {\tt DataGraph} object. The {\tt PlasmoData.DataGraph} (left) contains fundamental attributes ({\tt g}, {\tt nodes}, {\tt edges}, {\tt node\_map}, {\tt edge\_map}, {\tt node\_data}, {\tt edge\_data}, {\tt graph\_data}) which correspond to attributes of the mathematical {\tt DataGraph} object (right) containing nodes ($\mn$), edges ($\me$), directionality ($A$), node data ($\bd^\mn$), edge data ($\bd^\me$), graph data ($\bd^\mg$), and node, edge, and graph attributes ($\ma_\mn$, $\ma_\me$, and $\ma_\mg$, respectively). Each node and edge can embed data of a variety of forms, such as scalar values, vectors, or other graphs}
    \label{fig:datagraph_object}
\end{figure}

The {\tt DataGraph} abstraction is implemented in {\tt Julia} in the package {\tt PlasmoData.jl} (see Figure \ref{fig:datagraph_object}). {\tt PlasmoData.jl} allows for any data structure to be encoded on the nodes, edges, or on the graph itself. To differentiate references to the mathematical formulation from  the computer implementation, we will refer to the software counterpart as {\tt PlasmoData.DataGraph}. 
\\

The {\tt PlasmoData.DataGraph} object implemented contains the following fundamental attributes (defined in the software data structure): 

\begin{itemize}
    \item {\tt g}: A {\tt Graphs.SimpleGraph} (or a {\tt Graphs.SimpleDiGraph} for directed graphs) object. The {\tt Graphs.SimpleGraph} object is defined within the {\tt Graphs.jl} package, and it stores graph structure efficiently by storing only the number of edges and a vector of neighbors for each node. {\tt PlasmoData.jl} has {\tt Graphs.jl} as a dependency and stores the {\tt SimpleGraph} object to enable access to a variety of graph algorithms and to provide a simpler interface with other tools. 
    
    \item {\tt nodes}: A vector of node names for each node in the graph. Within {\tt PlasmoData.jl}, nodes can be ``named" or defined using any data type. This is different from some other graph modeling tools. For example, {\tt Graphs.jl} does not directly name any nodes, but instead only records an adjacency list for each node, and thus implicitly names nodes only by an integer index. {\tt nodes}, corresponds to $\mathcal{N}$ in \eqref{eq:datagraph}.
    
    \item {\tt edges}: A vector of pairs (tuples) of integers corresponding to each edge of the graph. The integer pairs correspond to the index of nodes in {\tt nodes} (e.g., pair $(1, 2)$ corresponds to an edge between the first and second nodes defined in {\tt nodes}). Restricting {\tt edges} to only contain integer pairs (rather than the node pairs themselves) is done to reduce the amount of memory required to store the graph. {\tt edges}, corresponds to $\mathcal{E}$ in \eqref{eq:datagraph}.
    
    \item {\tt node\_map}: A dictionary mapping the node names to their index in the {\tt nodes} vector. This index also corresponds to the adjacency list saved in the {\tt Graphs.jl} {\tt g} object.
    
    \item {\tt edge\_map}: A dictionary mapping the edges (integer pairs) to their index in the {\tt edges} vector.
    
    \item {\tt node\_data}: A mutable structure that contains a matrix of all node data where each row of the matrix corresponds to a node in the graph and each column corresponds to a node attribute. The order of the rows is identical to the order in {\tt nodes}. This data structure also contains a list of all node attributes and their mapping to the data matrix. Node attributes are restricted to be strings. {\tt node\_data} corresponds to $\bd^\mathcal{N}$ in \eqref{eq:datagraph}. 
    
    \item {\tt edge\_data}: A mutable structure that contains a matrix of all edge data where each row of the matrix corresponds to an edge in the graph and each column corresponds to an edge attribute. The order of the rows is identical to the order in {\tt edges}. This data structure also contains a list of all edge attributes and their mapping to the data matrix. Edge attributes are restricted to be strings. {\tt edge\_data} corresponds to $\bd^\me$ in \eqref{eq:datagraph}.
    
    \item {\tt graph\_data}: A mutable structure that contains a vector of graph data. This data structure also contains a list of all graph attributes and their mapping to the data vector. Graph attributes are restricted to be strings. {\tt graph\_data} corresponds to $\bd^\mg$ in \eqref{eq:datagraph}.
\end{itemize}

While the mathematical abstraction encodes directionality (or lack thereof) via the adjacency matrix, $A$, {\tt PlasmoData.jl} includes {\tt PlasmoData.DataGraph} and a {\tt PlasmoData.DataDiGraph} modeling objects for undirected and directed graphs, respectively. This terminology is consistent with other graph modeling tools (e.g., {\tt Graphs.jl} and {\tt NetworkX}). For our software design, how we chose to store node, edge, and graph data was an important consideration that deserves some explanation as {\tt PlasmoData.jl} stores this data differently than some alternative tools. A discussion of this decision, as well as a comparison of memory allocation for storing data in {\tt PlasmoData.jl} and other graph modeling tools, is provided in the supporting information.

\subsubsection{{\tt PlasmoData.jl} Tutorial Example}

{\tt PlasmoData.jl} implements the {\tt DataGraph} abstraction as shown above, and it provides a user-friendly interface and suite of functions; basic funcationality for constructing the graph is illustrated in Code Snippet \ref{code:overview}. An empty {\tt PlasmoData.DataGraph} is instantiated on Line \ref{line:dg_init}. The data types within the {\tt \{\}} define the types of data that can be defined in the node, edge, and graph data (see Figure \ref{fig:datagraph_object}). In this example, we use {\tt Any} for the data type in the node, edge, and graph data which will be important when adding data to this graph. Alternatively, a user can call {\tt DataGraph()} without the information in {\tt \{\}}, and the default data types in the node, edge, and graph data will be {\tt Float64}. Once the {\tt PlasmoData.DataGraph} object is instantiated, nodes, edges, and data can be added to the graph. Nodes can be added to the {\tt PlasmoData.DataGraph} by calling the {\tt add\_node!} function (Lines \ref{line:dg_add_node_start} - \ref{line:dg_add_node_end}). {\tt PlasmoData.jl} allows for any {\tt Julia} object to be defined as a node. For example, as shown in the code snippet, the node names can be integers, strings, or symbols. This can be important, as will be shown later when representing matrices as {\tt PlasmoData.DataGraph}s. Edges can be added to the {\tt PlasmoData.DataGraph} by calling the {\tt add\_edge!} function (Lines \ref{line:dg_add_edge_start1} - \ref{line:dg_add_edge_end2}). This function takes pairs of node names as arguments which represent the edge between those two nodes. These pairs of nodes can be passed as separate arguments (Lines \ref{line:dg_add_edge_start1} - \ref{line:dg_add_edge_end1}) or as a tuple of two node names (Lines \ref{line:dg_add_edge_start2} - \ref{line:dg_add_edge_end2}). This progressively builds the adjacency matrix of the graph object. 
\\

\begin{figure}[!htp]
\begin{minipage}[t]{0.9\linewidth}
    \begin{scriptsize}
    \lstset{language=Julia, breaklines = true}
    \begin{lstlisting}[caption = {Example showing the basic functionality for building a {\tt PlasmoData.DataGraph} in {\tt PlasmoData.jl}, including adding nodes, edges, and data and basic plotting functionality}, label = code:overview] 
using PlasmoData, PlasmoDataPlots

dg = DataGraph{Int, Any, Any, Any, Matrix{Any}, Matrix{Any}}()|\label{line:dg_init}|

# add_node!(datagraph, node)
add_node!(dg, 1) |\label{line:dg_add_node_start}|
add_node!(dg, 2)
add_node!(dg, 3)
add_node!(dg, "node4")
add_node!(dg, :node5) |\label{line:dg_add_node_end}|

# add_edge!(datagraph, node1, node2)
# add_edge!(datagraph, (node1, node2))
add_edge!(dg, 1, 2)|\label{line:dg_add_edge_start1}|
add_edge!(dg, 2, 3)
add_edge!(dg, "node4", 1)|\label{line:dg_add_edge_end1}|
add_edge!(dg, (:node5, 2))|\label{line:dg_add_edge_start2}|
add_edge!(dg, (3, "node4")) |\label{line:dg_add_edge_end2}|

# add_node_data!(datagraph, node, node_data, data_attribute)
add_node_data!(dg, 1, [6, 3, 4], "node_data1") |\label{line:dg_add_node_data_start}|
add_node_data!(dg, 2, 3.4, "node_data1")
add_node_data!(dg, 3, "this is on node 3", "node_data1")
add_node_data!(dg, "node4", [1 2; 3 4], "node_data1")
add_node_data!(dg, :node5, DataGraph(), "node_data1")|\label{line:dg_add_node_data_end}|

# add_edge_data!(datagraph, node1, node2, edge_data, edge_attribute)
# add_edge_data!(datagraph, (node1, node2), edge_data, edge_attribute)
add_edge_data!(dg, 1, 2, DataGraph(), "edge_data1")|\label{line:dg_add_edge_data_start}|
add_edge_data!(dg, 2, 3, [1 2 ; 5 7], "edge_data1")
add_edge_data!(dg, "node4", 1, 1.0, "edge_data1")
add_edge_data!(dg, (:node5, 2), -0.00001, "edge_data1")
add_edge_data!(dg, (3, "node4"), Dict(), "edge_data1") |\label{line:dg_add_edge_data_end}|

# add_graph_data!(dg, graph_data, graph_attribute)
add_graph_data!(dg, 1.0, "graph_data1")|\label{line:dg_add_graph_data}|

PlasmoDataPlots.plot_graph(dg; xdim = 400, ydim = 400) |\label{line:dg_plot_graph}|
    \end{lstlisting}
    \end{scriptsize}
\end{minipage}
\end{figure}

\begin{figure}[!htp]
    \centering
    \includegraphics[scale = .4]{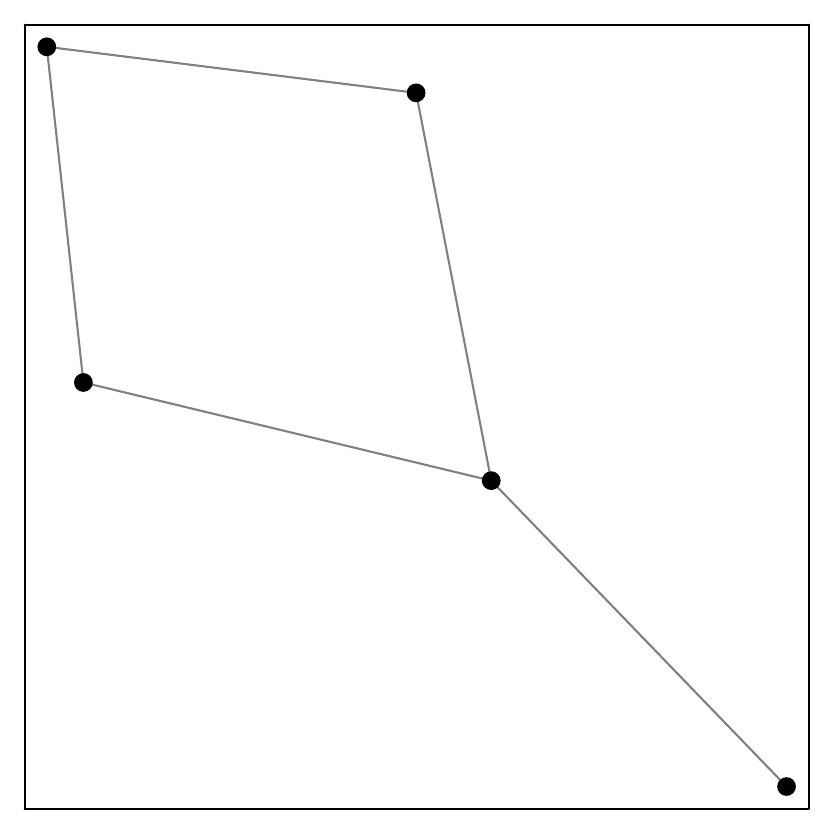}
    \captionof{figure}{Visualization of the {\tt PlasmoData.DataGraph} associated with Code Snippet \ref{code:overview}}
    \label{fig:dg_overview}
\end{figure}

Data can also be embedded within the defined graph structure through API functions. Data is added to nodes through the {\tt add\_node\_data!} function (Lines \ref{line:dg_add_node_data_start} - \ref{line:dg_add_node_data_end}). This function takes the following arguments: i) the {\tt PlasmoData.DataGraph}, ii) the node name, iii) the node data, and iv) the attribute name for the data (must be a string). The node data is restricted to be the type defined on the {\tt PlasmoData.DataGraph} (in this case, the {\tt PlasmoData.DataGraph} was instantiated with type {\tt Any}, so any data can be passed). As seen in the code snippet, data have different formats, including floats, strings, vectors, matrices, or another {\tt PlasmoData.DataGraph}. Data is added to edges through the {\tt add\_edge\_data!} function (Lines \ref{line:dg_add_edge_data_start} - \ref{line:dg_add_edge_data_end}). This function behaves similarly to the {\tt add\_node\_data!} function, but in place of the node name argument, a pair of nodes (two separate arguments) or a tuple of two node names (one argument) is supplied. Data is added to the graph by the {\tt add\_graph\_data!} function (Line \ref{line:dg_add_graph_data}). This function only takes the graph data and the graph attribute, as shown in the snippet. Additional node, edge, or graph data can be defined on the {\tt DataGraph} by passing the data with a different attribute name. There is no limit to the number of attributes that can be added to the nodes, edges, or graph. 
\\

One of the benefits of graph representations is that they facilitate visualization of structure. For basic plotting capabilities, we have developed an accompanying  package called {\tt PlasmoDataPlots.jl} (available at \url{https://github.com/zavalab/PlasmoDataPlots.jl}) that interfaces to {\tt PlasmoData.jl} and that enables visualizations of {\tt PlasmoData.DataGraph} objects. This package provides basic functionalities for plotting both {\tt PlasmoData.DataGraph}s and {\tt PlasmoData.DataDiGraph}s primarily through the Julia packages {\tt Plots.jl} and {\tt GraphMakie.jl}. The visualization in Figure \ref{fig:dg_overview} is created by calling the {\tt PlasmoDataPlots.jl} function {\tt plot\_graph} (Line \ref{line:dg_plot_graph}). 
\\

The methods presented above for building {\tt PlasmoData.DataGraph}s, adding nodes and edges, and adding data also work for {\tt DataDiGraph}s. The only difference is that the instantiation function is called {\tt DataDiGraph} and the edge order is maintained in the internal structure of the {\tt DataDiGraph}. 

\subsection{Modeling Common Data Objects as {\tt DataGraph}s}

With the general mathematical formulation for the {\tt DataGraph} defined, we next outline how specific data objects (matrices, 3D tensors, symmetric matrices) can be represented using the graph abstraction and how these can be implemented in {\tt PlasmoData.jl}.Importantly, a benefit of representing matrices and tensors as graphs is that graphs are modeling objects that do not live in a Euclidean space. As such, all that matters from a graph perspective is connectivity (not location of nodes and edges in space). As such, a graph object is not affected by rotations (matrices are). This property is quite useful when processing images/video and also when processing data objects from molecular simulations (in which nodes move randomly in space) \cite{zavala2023outlook}. Thus the approaches in this section could be applied to a variety of problems in different fields.

\subsubsection{Matrices as a Node-Weighted Graphs}\label{sec:mat_to_graph}

We can represent a matrix $M \in \mathbb{R}^{p \times q}$ (where $p$ and $q$ are positive integers) as a node-weighted graph with a mesh structure \cite{smith2021euler,parihar2018image}, and we can express the resulting graph using the {\tt DataGraph} abstraction. We use the notation $m_{i,j}\in \mathbb{R}$ for the matrix entry at row $i$ and column $j$. For convenience, we define the integer set $\mathbb{N}_k := \{1, 2, ..., k\}$. To represent the matrix as a graph, we define a node, $n_*$, for every matrix entry and place edges between all adjacent entries of the matrix. On each node, $m_{i, j}$ is stored as the node data. As there is a single weight in each node, we define the attribute set as $\mathcal{A}_\mathcal{N} = \{ a \}$ (a singleton). Mathematically, this is represented with the {\tt DataGraph} abstraction as:
\begin{equation}
    \begin{split}
    &\mathcal{DG}(\mathcal{N}, \mathcal{E}, A, \bd^\mn, \emptyset, \emptyset, \ma_\mn, \emptyset, \emptyset)\\
    \textrm{where} &\; \mathcal{N} = \{n_{i, j}: i \in \mathbb{N}_p, j \in \mathbb{N}_q \} \\
    &\; \mathcal{E} = \{(n_{i, j}, n_{i + 1, j}): i \in \mathbb{N}_{p - 1}, j \in \mathbb{N}_{q} \}\cup\{(n_{i, j}, n_{i, j+1}): i \in \mathbb{N}_{p},  j \in \mathbb{N}_{q -1} \} \\
    &\; \mathcal{A}_\mathcal{N} = \{ a \} \\
    &\; d^\mn_{a}(n_{i, j}) = m_{i, j}, i \in \mathbb{N}_p, j \in \mathbb{N}_q
    \end{split}
\end{equation}
where $A$ is a symmetric matrix ($\mdg$ is undirected) matching the connectivity defined by $\me$. An alternative structure can also be constructed by also adding edges between diagonal elements of the matrix \cite{janakiraman2010image}. In this case, the set of edges becomes $\mathcal{E} = \{(n_{i, j}, n_{i + 1, j}): i \in \mathbb{N}_{p - 1}, j \in \mathbb{N}_{q} \}\cup\{(n_{i, j}, n_{i, j+1}): i \in \mathbb{N}_{p}, j \in \mathbb{N}_{q -1} \} \cup \{ (n_{i, j}, n_{i + 1, j + 1}): i \in \mathbb{N}_{p - 1}, j \in \mathbb{N}_{q - 1} \} \cup \{ (n_{i, j + 1}, n_{i + 1, j}): i \in \mathbb{N}_{p -1}, j \in \mathbb{N}_{q -1} \}$. 
\\

\begin{figure}[!htp]
    \centering
    \includegraphics[scale = .6]{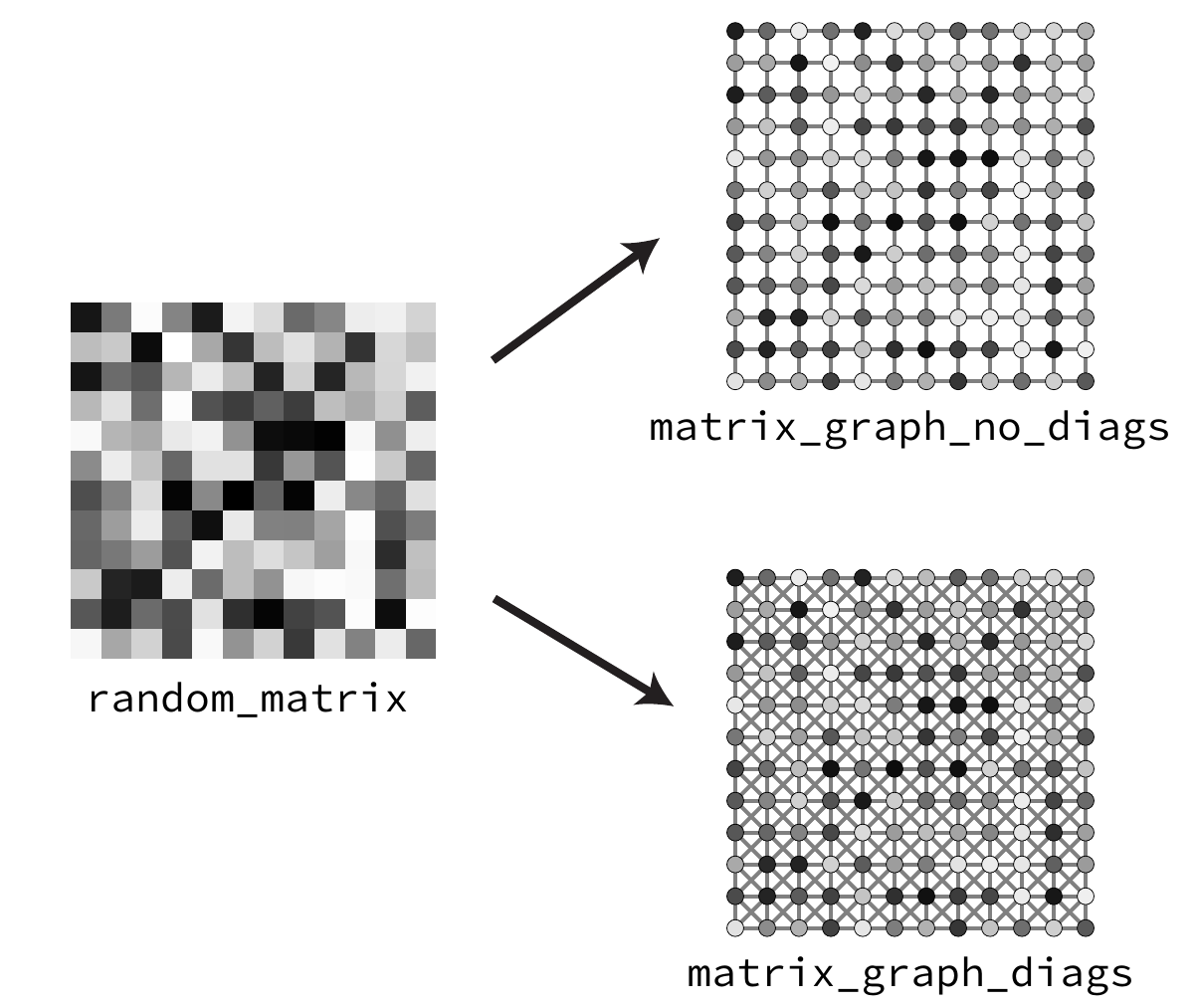}
    \captionof{figure}{Visualization of graph representations of a matrix. The top graph corresponds to the {\tt matrix\_graph\_no\_diags} {\tt PlasmoData.DataGraph} object from Code Snippet \ref{code:matrix_to_graph}, and the lower graph corresponds to the {\tt matrix\_graph\_diags} {\tt PlasmoData.DataGraph} object from Code Snippet \ref{code:matrix_to_graph} and is the output of the code snippet. Matrix entries are represented by nodes with the matrix entry values stored as data on each node}
    \label{fig:matrix_to_graph}
\end{figure}

 {\tt PlasmoData.jl} facilitates the representation of matrices as graphs. Code Snippet \ref{code:matrix_to_graph} shows how a {\tt PlasmoData.DataGraph} object can be constructed automatically by calling the {\tt matrix\_to\_graph} function. Any matrix containing real numbers can be passed to this function. The boolean keyword argument {\tt diagonal} can be passed to identify whether the diagonal edges should be included in the graph. The resulting graphs can also be visualized using {\tt PlasmoDataPlots.jl}. In Line \ref{line:set_node_pos}, nodes are given a fixed position (stored in the node data of the {\tt PlasmoData.DataGraph}) for visualization. In Lines \ref{line:plot_mat_graph_start} - \ref{line:plot_mat_graph_end}, the graph containing diagonals is plotted, where the weight data from the original matrix is used for coloring the nodes (see Line \ref{line:node_z_mat_graph}). Figure \ref{fig:matrix_to_graph} visualizes the alternative representations. Because matrices themselves are a general abstraction for representing data, {\tt PlasmoData.DataGraph} can be applied to many application. For instance, multivariate time series and grayscale images can be represented as matrices (and thus as graphs) \cite{morris1986graph, perret2019higra, salembier2000binary}.

\begin{figure}[!htp]
\begin{minipage}[t]{0.9\linewidth}
    \begin{scriptsize}
    \lstset{language=Julia, breaklines = true}
    \begin{lstlisting}[label = code:matrix_to_graph, caption = Example of representing a matrix as a {\tt PlasmoData.DataGraph}] 
using PlasmoData, Random, PlasmoDataPlots
    
Random.seed!(15)
random_matrix = rand(12, 12)

# Convert matrix to node-weighted graph
matrix_graph_diags = matrix_to_graph(random_matrix; diagonal = true)
matrix_graph_no_diags = matrix_to_graph(random_matrix; diagonal = false)

# Fix the node positions
set_matrix_node_positions!(matrix_graph_diags, random_matrix) |\label{line:set_node_pos}|

# Plot the graph
plot_graph( |\label{line:plot_mat_graph_start}|
    matrix_graph_diags,
    nodesize = 12,
    linewidth = 5,
    nodecolor = :grays,
    node_z = get_node_data(matrix_graph_diags, "weight"), |\label{line:node_z_mat_graph}|
    rev = true,
) |\label{line:plot_mat_graph_end}|
    \end{lstlisting}
    \end{scriptsize}
\end{minipage}
\end{figure}

\subsubsection{3D Tensors as Node-Weighted Graphs}\label{sec:tensor_to_graph}

Below we outline how a 3D tensor can be represented as a graph. While tensors are not restricted to be 3D, when we refer to tensors in this work, we will be referring to 3D tensors unless otherwise noted. Any tensor $T \in \mathbb{R}^{p \times q \times r}$ (where $p$, $q$, and $r$ are positive integers) can be represented as a node-weighted graph. There are at least a couple of methods for how to represent $T$. The first approach is to construct a node-weighted mesh following the description outlined in Section \ref{sec:mat_to_graph} based on the first couple of dimensions of $T$ (i.e., the mesh structure will be $p \times q$). The third dimension is then represented as a vector of weights attached to every element of the mesh (with $r = |\ma_\mn|$). This can be used, for instance, to represent color images (which have different color channels). In addition, we use the notation $t_{i, j, k}$ for the tensor entry, where $i, j,$ and $k$ are the entries in the first, second, and third dimensions, respectively. Under the {\tt DataGraph} abstraction, this has the form:

\begin{equation}\label{eq:ten_to_graph1}
    \begin{split}
    &\mathcal{DG}(\mathcal{N}, \mathcal{E}, A, \bd^\mn, \emptyset, \emptyset, \ma_\mn, \emptyset, \emptyset)\\
    \textrm{where} &\; \mathcal{N} = \{ n_{i, j}: i \in \mathbb{N}_p, j \in \mathbb{N}_q \}  \\
    &\; \mathcal{E} = \{(n_{i, j}, n_{i + 1, j}): i \in \mathbb{N}_{p - 1}, j \in \mathbb{N}_{q} \}\cup\{(n_{i, j}, n_{i, j+1}): i \in \mathbb{N}_{p}, j \in \mathbb{N}_{q -1} \} \\
    &\; \mathcal{A}_\mathcal{N} = \{ a_1, a_2, ..., a_r\} \\
    &\; d^\mn_{a_k}(n_{i, j}) = t_{i, j, k}, i \in \mathbb{N}_p, j \in \mathbb{N}_q, k \in \mathbb{N}_r
    \end{split}
\end{equation}
where $A$ is a symmetric matrix ($\mdg$ is undirected) matching the connectivity defined by $\me$. As with the matrix form in Section \ref{sec:mat_to_graph}, edges can also be placed between the entries that are diagonal to one another in the mesh structure. 
\\

Alternatively, each tensor entry can be represented by a node with each node/entry connected to the adjacent nodes/entries (similar to the mesh structure, but now across a third dimension). The value of each  tensor entry is then embedded in the corresponding node, so each node only contains one node weight. Under the {\tt DataGraph} abstraction, this has the form: 

\begin{equation}\label{eq:ten_to_graph2}
    \begin{split}
    &\mathcal{DG}(\mathcal{N}, \mathcal{E}, A, \bd^\mn, \emptyset, \emptyset, \ma_\mn, \emptyset, \emptyset)\\
    \textrm{where} &\; \mathcal{N} = \{ n_{i, j, k}: i \in \mathbb{N}_p, j \in \mathbb{N}_q, k \in \mathbb{N}_r \} \\
    &\; \mathcal{E} = \{(n_{i, j, k}, n_{i + 1, j, k}): i \in \mathbb{N}_{p - 1}, j \in \mathbb{N}_{q}, k \in \mathbb{N}_r \}\cup\{(n_{i, j, k}, n_{i, j+1, k}):\\
    & \quad \quad \quad i \in \mathbb{N}_{p}, j \in \mathbb{N}_{q -1}, k \in \mathbb{N}_r \} \cup \{(n_{i, j, k}, n_{i, j, k + 1}): i \in \mathbb{N}_p, j \in \mathbb{N}_q, k \in \mathbb{N}_{r - 1} \}\\
    &\; \mathcal{A}_\mathcal{N} = \{ a \} \\
    &\; d^\mn_a(n_{i, j, k}) = t_{i, j, k}, i \in \mathbb{N}_p, j \in \mathbb{N}_q, k \in \mathbb{N}_r
    \end{split}
\end{equation}
where $A$ is a symmetric matrix ($\mdg$ is undirected) matching the connectivity defined by $\me$. While \eqref{eq:ten_to_graph2} is defined for 3D tensors, it could easily be extended to higher-order tensors.
\\

\begin{figure}[!htp]
    \centering
    \includegraphics[scale = .4]{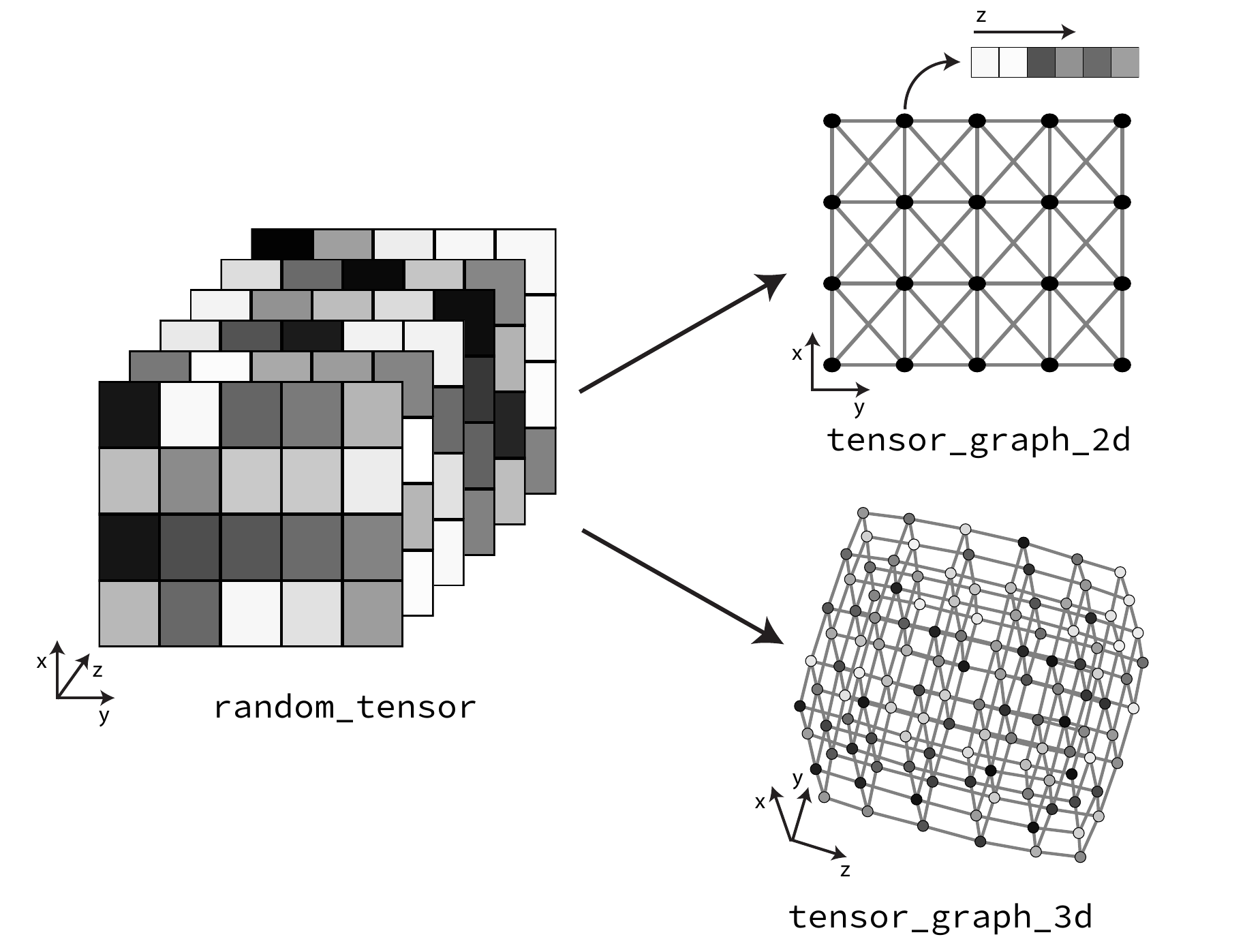}
    \captionof{figure}{Visualization of graph representations of a tensor. The upper graph corresponds to the {\tt tensor\_graph\_2d} {\tt PlasmoData.DataGraph} object from Code Snippet \ref{code:tensor_to_graph}, and the lower graph corresponds to the {\tt tensor\_graph\_3d} {\tt PlasmoData.DataGraph} object from Code Snippet \ref{code:tensor_to_graph}}
    \label{fig:tensor_to_graph}
\end{figure}

{\tt PlasmoData.jl} facilitates both of the representations discussed. In Code Snippet \ref{code:tensor_to_graph}, the random tensor can be formed into the mesh structure discussed in Section \ref{sec:mat_to_graph} by calling the {\tt matrix\_to\_graph} function (Line \ref{line:matrix_to_graph_tensor}). This function recognizes that this is a 3D array and creates the mesh structure based on the first two dimensions. The third dimension are the weights on each node. The default name for these weights is the string {\tt "weight"} with a number (e.g., {\tt "weight1"}, {\tt "weight2"}), but a user can define their own names for the attributes in $\ma_\mn$. Alternatively, the user can call the function {\tt tensor\_to\_graph} (Line \ref{line:tensor_to_graph}) which performs the second method discussed above. These graphs can be visualized using {\tt PlasmoDataPlots.jl} as before (Lines \ref{line:tensor_to_graph_plot_start} - \ref{line:tensor_to_graph_plot_end}). In this case, the node locations are determined automatically within {\tt PlasmoDataPlots.jl} through the {\tt NetworkLayout.jl} package. A visualization of the methods in the code snippet is presented in Figure \ref{fig:tensor_to_graph}. Note that at this time, {\tt PlasmoData.jl} only automates representing 3D tensors as graphs and not higher-order tensors.

\begin{figure}[!htp]
    \begin{minipage}[t]{0.9\linewidth}
        \begin{scriptsize}
        \lstset{language=Julia, breaklines = true}
        \begin{lstlisting}[label = code:tensor_to_graph, caption = Example of representing a tensor as a {\tt PlasmoData.DataGraph} using two different approaches] 
using PlasmoData, Random, PlasmoDataPlots

Random.seed!(15)
random_tensor = rand(4, 5, 6)

# Convert the tensor to a node weighted graph with 4 x 5 nodes
tensor_graph_2d = matrix_to_graph(random_tensor; diagonal = true)|\label{line:matrix_to_graph_tensor}|

# Convert the tensor to a node weighted graph 4 x 5 x 6 nodes
tensor_graph_3d = tensor_to_graph(random_tensor) |\label{line:tensor_to_graph}|

# Plot the graph
plot_graph( |\label{line:tensor_to_graph_plot_start}|
    tensor_graph_3d,
    nodesize = 8,
    linewidth = 4,
    nodecolor = :grays,
    node_z = get_node_data(tensor_graph_3d, "weight"),
    rev = true,
) |\label{line:tensor_to_graph_plot_end}|
        \end{lstlisting}
        \end{scriptsize}
    \end{minipage}
\end{figure}
    
The above graph representations are flexible and can be applied to many types of datasets. For example, videos or space-time data can be represented as a tensor (i.e., a 2D field that changes over a third dimension of time) and data embedded in a 3D space (e.g., temperature in a room) can be stored as a tensor. Color images (e.g., RGB or hyperspectral images) can also be represented as a tensor, where the third dimension corresponds to the light intensity of different color channels. 

\subsubsection{Symmetric Matrices as Edge-Weighted Graphs}\label{sec:sym_mat_to_graph}

While any matrix can be represented as a node-weighted graph, a symmetric matrix $S \in \mathbb{R}^{p \times p}$ (e.g., a correlation matrix) can also be represented as an edge-weighted graph. We will use the notation $s_{i,j}$ for the entry of the $i$th row and $j$th column of $S$. In this case, $s_{i, j}$ is the edge weight between node $i$ and node $j$. Under the {\tt DataGraph} abstraction, this has the form:
\begin{equation}\label{eq:sym_mat_to_graph}
    \begin{split}
    &\mathcal{DG}(\mathcal{N}, \mathcal{E}, A, \emptyset, \bd^\me, \emptyset, \emptyset, \ma_\me, \emptyset)\\
    \textrm{where} &\; \{ n_{i}: i \in \mathbb{N}_p\} \\
    &\; \mathcal{E} = \{(n_{i}, n_{j}): i \in \mathbb{N}_{p - 1}, j \in \mathbb{N}_p \setminus \mathbb{N}_i\}\\
    &\; \mathcal{A}_\mathcal{E} = \{ a \} \\
    &\; d^\me_a((n_i, n_j)) = s_{i, j}, i \in \mathbb{N}_{p - 1}, j \in \mathbb{N}_p \setminus \mathbb{N}_i
    \end{split}
\end{equation}
where $A$ is a symmetric matrix ($\mdg$ is undirected) matching the connectivity defined by $\me$. 

\begin{figure}[!htp]
    \centering
    \includegraphics[scale = .5]{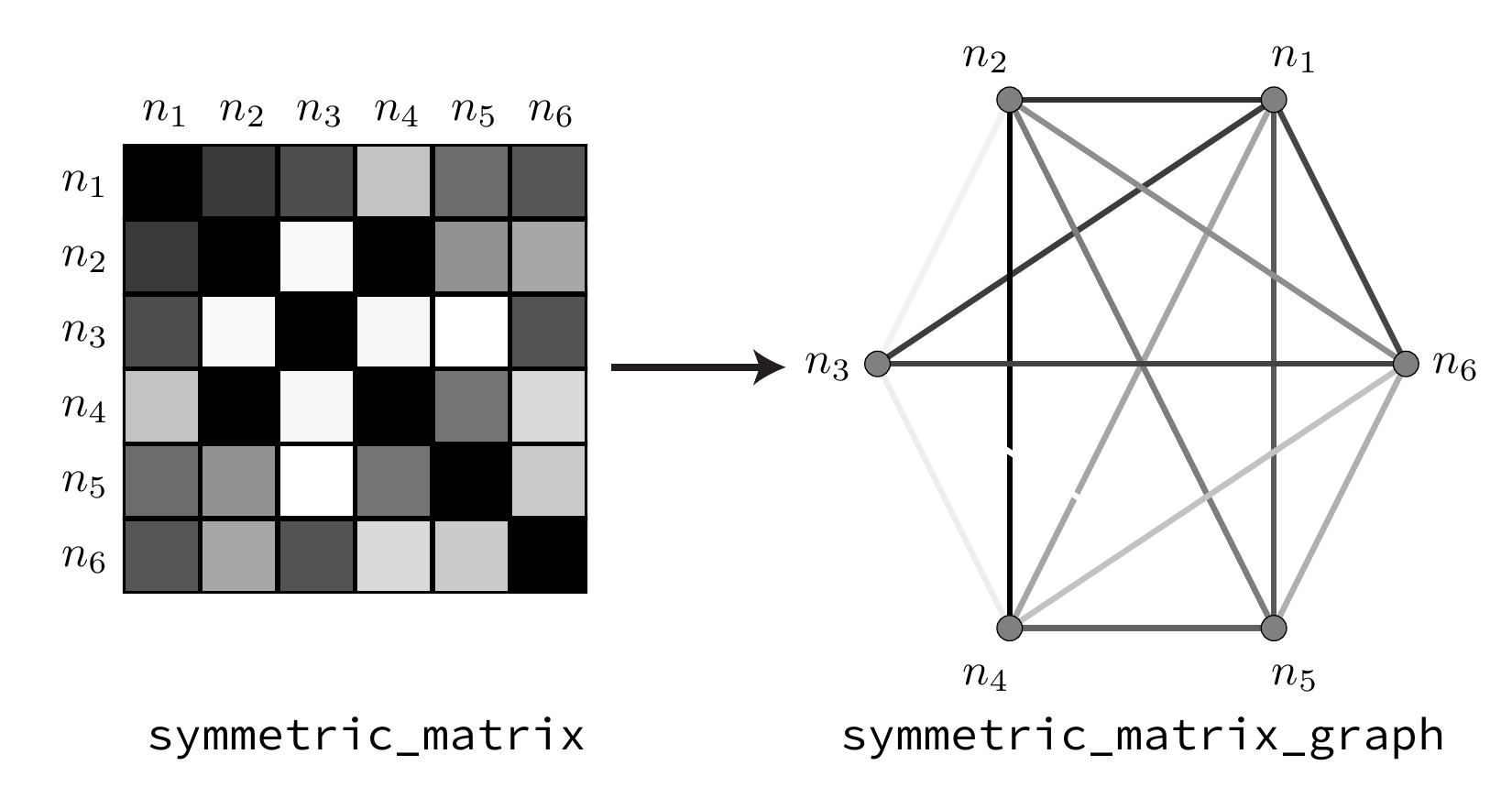}
    \captionof{figure}{Visualization of a graph representation of a symmetric matrix where the graph corresponds to the {\tt symmetric\_matrix\_graph} {\tt PlasmoData.DataGraph} object from Code Snippet \ref{code:sym_mat_to_graph}. Values of matrix entries are stored as data on the edges of the graph}
    \label{fig:sym_mat_to_graph}
\end{figure}

{\tt PlasmoData.jl} also facilitates this symmetric matrix graph representation. The implementation for this representation is shown in Code Snippet \ref{code:sym_mat_to_graph}. The symmetric matrix is passed to the function {\tt symmetric\_matrix\_to\_graph} (Line \ref{line:sym_mat_to_graph}) which forms the edge weighted {\tt PlasmoData.DataGraph}. The visualization of this is shown in Figure \ref{fig:sym_mat_to_graph}. The {\tt PlasmoData.DataGraph} can also be visualized using {\tt PlasmoDataPlots.jl}, where we first set the node positions using {\tt set\_circle\_node\_positions!}, which provides the circular structure shown in Figure \ref{fig:sym_mat_to_graph}. The graph is visualized with the {\tt plot\_graph} function (Lines \ref{line:sym_mat_plot_start} - \ref{line:sym_mat_plot_end}), where we color edges based on the edge weight through the {\tt line\_z} argument and the {\tt get\_edge\_data} API from {\tt PlasmoData.jl} (Line \ref{line:get_edge_data}).

\begin{figure}[!htp]
    \begin{minipage}[t]{0.9\linewidth}
        \begin{scriptsize}
        \lstset{language=Julia, breaklines = true}
        \begin{lstlisting}[label = code:sym_mat_to_graph, caption = Example of representing a symmetric matrix as a {\tt PlasmoData.DataGraph}] 
using PlasmoData, Random, PlasmoDataPlots, LinearAlgebra

# Create symmetric matrix
Random.seed!(5)
random_matrix = rand(6, 6)
symmetric_matrix = (random_matrix .+ random_matrix|'|) / 2
symmetric_matrix[diagind(symmetric_matrix)] .= 1

# Convert symmetric matrix to edge weighted graph
symmetric_matrix_graph = symmetric_matrix_to_graph(symmetric_matrix) |\label{line:sym_mat_to_graph}|

# Set node positions
set_circle_node_positions!(symmetric_matrix_graph) |\label{line:circle_node_pos}|

# Plot the graph
plot_graph( |\label{line:sym_mat_plot_start}|
    symmetric_matrix_graph,
    nodesize = 12,
    nodecolor = "gray",
    linewidth = 5,
    linecolor = :binary,
    line_z = get_edge_data(symmetric_matrix_graph, "weight"), |\label{line:get_edge_data}|
) |\label{line:sym_mat_plot_end}|
    \end{lstlisting}
    \end{scriptsize}
\end{minipage}
\end{figure}
    
Data is also commonly found directly as symmetric matrices and the above approach has been used for different analyses in literature \cite{bullmore2009complex,rubinov2010complex,de2022euler,smith2021euler}. For instance, symmetric matrices include correlation/covariance matrices (e.g., from a time series), Hessian matrices, or a matrix of distances. The representation of symmetric matrices as graphs can be useful in extracting connectivity information from covariance/correlation matrices, which cannot be extracted using traditional tools such as eigenvalue decompositions(e.g., principal component analysis-PCA).


\section{Data Analysis using the {\tt DataGraph} Abstraction}\label{sec:data_analysis}

Modeling data as graphs enable the use of tools for processing, manipulating, and analyzing the data. In this section, we highlight how the graph structure can be manipulated to reveal hidden aspects and reduce dimensionality. In addition, we provide an introduction to tools of topological data analysis (TDA) that can be used to extract information from the data model. Finally, we discuss some of the limitations of graph representations.

\subsection{Structure Manipulation}

Graph structures can be manipulated by filtering out (eliminating) nodes/edges based on data encoded on them or by aggregating sets of nodes. These processes can help reduce the data and can provide valuable insight into the data. 

\subsubsection{Node and Edge Removal}

Perhaps the simplest form of structure manipulation is removal of a single node or edge. To formalize this process for the {\tt DataGraph} model, we define the following operations:

\begin{definition}[Node-Removal Function $f^\mathcal{N}_{rem}$]
    For a {\tt DataGraph}, $\mathcal{DG}\left(\mathcal{N}, \mathcal{E}, A, \bd^\mn, \bd^\me, \bd^\mg, \ma_\mn, \ma_\me, \ma_\mg \right)$ where $\mathbb{G}$ is the set of all possible {\tt DataGraph}s and $\mathbb{V}$ is the set of all possible nodes, the node-removal function $f_\mathcal{N}: (\mathbb{G}, \mathbb{V}) \rightarrow \mathbb{G}$ is defined
    \begin{equation}\label{eq:node_removal}
        \begin{split}
            f^\mn_{rem}\big(&\mathcal{DG}, n_{rem} \big) = \mathcal{DG}_r\big(\mathcal{N}_r, \mathcal{E}_r, A_r, \bd^\mn_r, \bd^\me_r, \bd^\mg, \ma_\mn, \ma_\me, \ma_\mg \big) \\
            \textrm{where} &\; \mn_r = \mn \setminus \{n_{rem}\}\\
            &\; \me_r = \{(n_i, n_j) : n_i \in \mn_r, n_j \in \mn_r, (n_i, n_j) \in \me \} \\
            &\; \bd^\mn_r(n) = \bd^\mn(n), \quad n \in \mn_r\\
            &\; \bd^\me_r(e) = \bd^\me(e), \quad e \in \me_r
        \end{split}
    \end{equation}
    where $n_{rem} \in \mathcal{N}$ is the node removed and $A_r$ is the adjacency matrix for the new set of edges $\me_r$.
\end{definition}

\begin{definition}[Edge-Removal Function $f^\mathcal{E}_{rem}$]
    For a {\tt DataGraph}, $\mathcal{DG}\left(\mathcal{N}, \mathcal{E}, A, \bd^\mn, \bd^\me, \bd^\mg, \ma_\mn, \ma_\me, \ma_\mg \right)$ where $\mathbb{E}$ is the set of possible edges, the edge-removal function $f^\mathcal{E}_{rem}: (\mathbb{G}, \mathbb{E}) \rightarrow \mathbb{G}$ is defined such that
    \begin{equation}\label{eq:edge_removal}
        \begin{split}
            f^\me_{rem}\big(&\mathcal{DG}, e_{rem} \big) = \mathcal{DG}_r\big(\mathcal{N}, \mathcal{E}_r, A_r, \bd^\mn, \bd^\me_r, \bd^\mg, \ma_\mn, \ma_\me, \ma_\mg \big) \\
            \textrm{where} &\; \me_r = \me_r \setminus \{ e_{rem}\} \\
            &\; \bd^\me_r(e) = \bd^\me(e), \quad e \in \me_r
        \end{split}
    \end{equation}
    where $e_{rem} \in \mathcal{E}$ is the edge to be removed, and $A_r$ is the adjacency matrix for the new set of edges $\me_r$.
\end{definition}
\noindent The node- and edge-removal functions are simple operations but form the basis of more complex operations that manipulate the graph structure. They also provide a basis for different analysis, such as analyzing the topology of a graph after the removal of a node or edge. The node- and edge-removal functions are implemented in {\tt PlasmoData} as functions {\tt remove\_node!} and {\tt remove\_edge!}.

\subsubsection{Graph Filtration}

Filtering out nodes or edges of a graph involves removing nodes or edges whose data does not meet specified (logical) criteria. We formalize the filtration process for {\tt DataGraph}s by defining the following:

\begin{definition}[Node-Filtration Function $f_\mathcal{N}$]
    For a {\tt DataGraph}, $\mathcal{DG}\left(\mathcal{N}, \mathcal{E}, A, \bd^\mn, \bd^\me, \bd^\mg, \ma_\mn, \ma_\me, \ma_\mg \right)$, containing a set of node attributes, $\mathcal{A}_\mathcal{N}$, such that $\bd^\mn \ne \emptyset$, and where $\mathbb{G}$ is the set of all possible {\tt DataGraph}s and $\mathbb{L}$ is the set of all logic sets, the node-filtration function $f_\mathcal{N}: (\mathbb{G}, \mathbb{L}) \rightarrow \mathbb{G}$ is defined
    \begin{equation}\label{eq:node_filtration}
        \begin{split}
            f_\mathcal{N}\big(&\mathcal{DG}, \mathcal{L} \big) = \mathcal{DG}_f\big(\mathcal{N}_f, \mathcal{E}_f, A_f, \bd^\mn_f, \bd^\me_f, \bd^\mg, \ma_\mn, \ma_\me, \ma_\mg \big) \\
            \textrm{where} &\; \mathcal{N}_f = \{ n : \bd^\mathcal{N}(n) \in \mathcal{L}, n \in \mathcal{N}\}\\
            &\; \mathcal{E}_f = \{(n_i, n_j) : (n_i, n_j) \in \mathcal{E}, n_i \in \mathcal{N}_f, n_j \in \mathcal{N}_f \}\\
            &\; \bd^\mn_f(n) = \bd^\mn(n), \quad n \in \mathcal{N}_f\\  
            &\; \bd^\me_f(e) = \bd^\me(e), \quad e\in \me_f
        \end{split}
    \end{equation}
    for the logic set $\mathcal{L}$, where $A_f$ is the adjacency matrix for the edge set $\me_f$.
\end{definition}

\begin{definition}[Edge-Filtration Function $f_\mathcal{E}$]
    For a {\tt DataGraph}, $\mathcal{DG}\left(\mathcal{N}, \mathcal{E}, A, \bd^\mn, \bd^\me, \bd^\mg, \ma_\mn, \ma_\me, \ma_\mg \right)$, containing a set of edge attributes, $\mathcal{A}_\mathcal{E}$, such that $\bd^\me \ne \emptyset$, the edge-filtration function $f_\mathcal{E}: (\mathbb{G}, \mathbb{L}) \rightarrow \mathbb{G}$ is defined such that
    \begin{equation}\label{eq:edge_filtration}
        \begin{split}
            f_\mathcal{E}(&\mathcal{DG}, \mathcal{L}) = \mathcal{DG}_f\big(\mathcal{N}, \mathcal{E}_f, A_f \bd^\mn, \bd^\me_f, \bd^\mg, \ma_\mn, \ma_\me, \ma_\mg \big) \\
            \textrm{where} &\; \mathcal{E}_f = \{e : \bd^\me(e) \in \mathcal{L}, e \in \mathcal{E}\}\\
            &\; \bd^\me_f(e) = \bd^\me(e), \quad e \in \me
        \end{split}
    \end{equation}
    for the logic set $\mathcal{L}$, where $A_f$ is the adjacency matrix for the edge set $\me_f$.
\end{definition}

\noindent Here, we refer to a ``logic set", $\mathcal{L}$, as a set of data corresponding to the data and attributes on nodes or edges of the graph. For a set of attributes $\{ a_1, a_2, ..., a_r\}$, we use the notation
$$ \ml = \begin{cases} a_1 &: \mathcal{Z}_1 \\a_2 &: \mathcal{Z}_2 \\ 
\; \vdots &\; \; \;  \vdots\\ a_r &: \mathcal{Z}_r \end{cases}$$
where $a_1 : A_1$ indicates that the set $\mathcal{Z}_1 \subseteq \mathcal{D}_{a_1}$ is used for comparing the data of attribute $a_1$. As the data on the nodes, edges, or graph can be of many forms, $\mathcal{L}$ can span multiple attributes and take many different shapes. Filtration is often applied to scalar weights (e.g., filtering out all nodes whose weight is less than some threshold value); however, filtration can be much more general than the scalar weight case. For example, the data attached to nodes could be text, matrices, or even an optimization model; as such, the filtration can be performed by using logic defined to attributes of the data (e.g., filtering out any optimization problems that are ``nonlinear" or ``unconstrained"). In addition, we note that edge-filtrations do not alter the nodes or the node data while the node-filtrations can result in edges being removed when nodes are filtered out. 
\\

\begin{figure}[!htp]
    \centering
    \includegraphics[width = .9\textwidth]{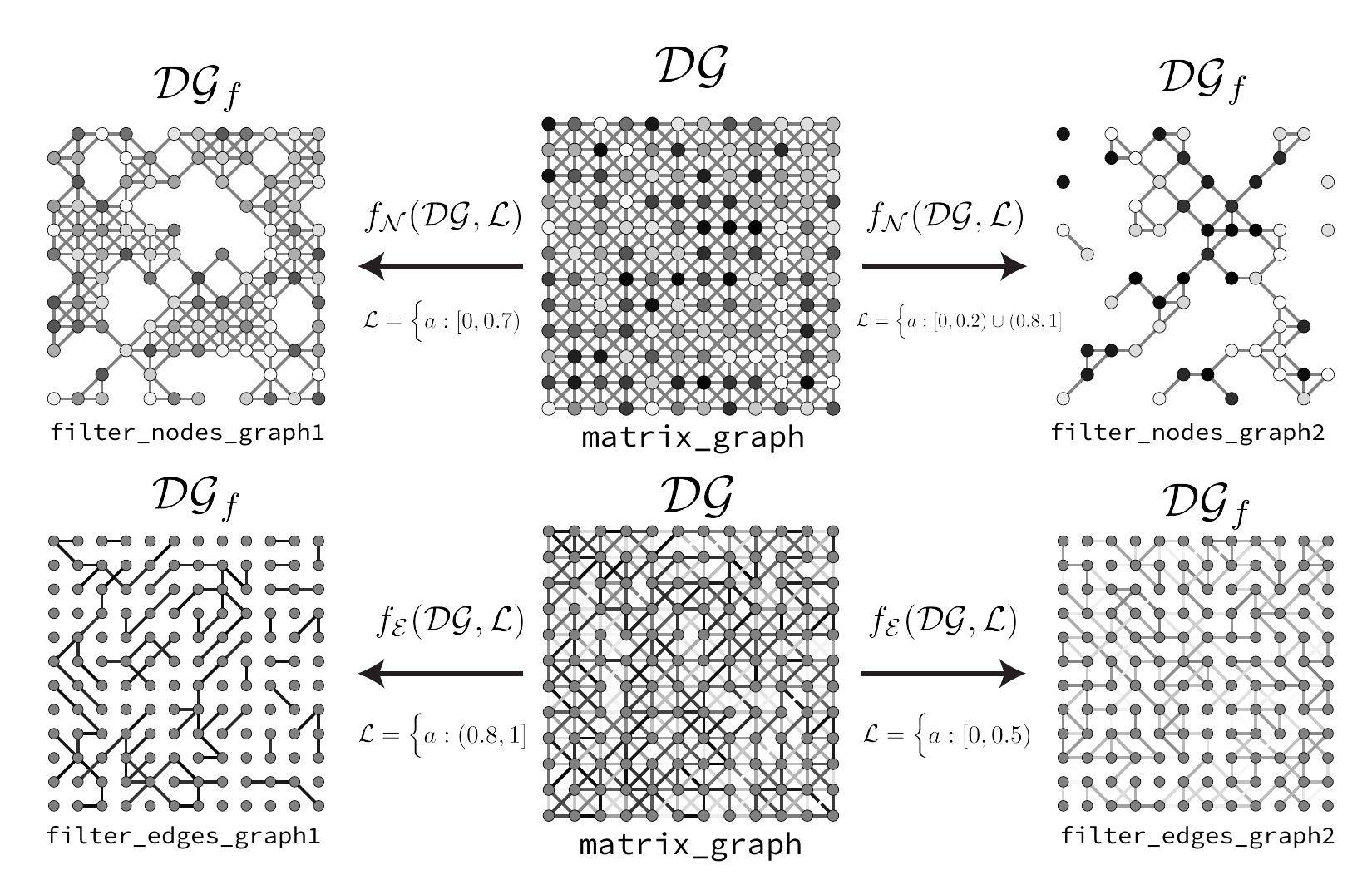}
    \captionof{figure}{Visualization of different examples of node or edge filtration associated with Code Snippet \ref{code:filtration}, where different logic sets are used for removing nodes or edges of the graph}
    \label{fig:filtration}
\end{figure}

The above node- and edge-filtrations are implemented in {\tt PlasmoData.jl} as shown in Code Snippet \ref{code:filtration} and visualized in Figure \ref{fig:filtration}. The graph defined on Line \ref{line:filter_graph} can be filtered by calling the function {\tt filter\_nodes} (the software implementation of $f_\mn$). This function takes arguments of the {\tt PlasmoData.DataGraph}, a threshold value (often used in the filter function), the attribute name, and an optional keyword argument {\tt fn}, which is the filter function used for the filtration. Note that all nodes where the filter function, {\tt fn}, does not return true are filtered out. On Line \ref{line:filter_nodes1}, we filter out all nodes whose ``weight" attribute is greater than $0.7$ through the {\tt Base.isless} function. On Lines \ref{line:filter_function_start} - \ref{line:filter_function_end}, we define our own function for filtration called {\tt extreme\_vals}, which we pass to {\tt filter\_nodes} on Line \ref{line:filter_nodes2} and which filters out all values that are not less than $0.2$ or greater than $0.8$. This highlights how user-defined functions can be used to filter out a graph. On Line \ref{line:add_edge_dataset}, we add random weights to all the edges in the graph (note that {\tt add\_edge\_dataset!} has a similar function as {\tt add\_edge\_data}, but instead adds data of a single attribute to multiple edges, whereas the latter function only adds data to a single edge). Once data is defined on the edges, we can likewise perform filtrations based on the edge data, as shown on lines \ref{line:filter_edges1} and \ref{line:filter_edges2} where we now call {\tt filter\_edges} (the software implementation of $f_\me$). This function behaves similarly to {\tt filter\_nodes}, but now operates on the edge data rather than the node data. All arguments are the same for these functions, except that {\tt filter\_edges} requires an attribute defined on the edges rather than on the nodes. While these functions are shown for {\tt PlasmoData.DataGraph}s, they also apply for {\tt PlasmoData.DataDiGraph}s.

\begin{figure}[!htp]
    \begin{minipage}[t]{0.9\linewidth}
        \begin{scriptsize}
        \lstset{language=Julia, breaklines = true}
        \begin{lstlisting}[label = code:filtration, caption = Example of filtering a {\tt PlasmoData.DataGraph} by node or edge data, escapechar = ~] 
using PlasmoData, Random

Random.seed!(15)
random_matrix = rand(12, 12)

matrix_graph = matrix_to_graph(random_matrix; diagonal = true) ~\label{line:filter_graph}~

# Filter out/remove nodes with weight >= 0.7
filter_nodes_graph1 = filter_nodes(matrix_graph, 0.7, "weight", fn = Base.isless) ~\label{line:filter_nodes1}~

function extreme_vals(a, b) ~\label{line:filter_function_start}~
    return ((a <= .2) ~||~ a >= .8)
end ~\label{line:filter_function_end}~

# Filter out/remove nodes with weight >= 0.2 and <= 0.8
filter_nodes_graph2 = filter_nodes(matrix_graph, nothing, "weight", fn = extreme_vals) ~\label{line:filter_nodes2}~

n_edges = length(matrix_graph.edges)
add_edge_dataset!(matrix_graph, rand(n_edges), "weight") ~\label{line:add_edge_dataset}~

# Filter out/remove edges with weight <= 0.8
filter_edges_graph1 = filter_edges(matrix_graph, 0.8, "weight", fn = Base.isgreater)~\label{line:filter_edges1}~

# Filter out/remove edges with weight >= 0.5
filter_edges_graph2 = filter_edges(matrix_graph, 0.5, "weight", fn = Base.isless)~\label{line:filter_edges2}~
    \end{lstlisting}
    \end{scriptsize}
\end{minipage}
\end{figure}

\subsubsection{Graph Aggregation}

Aggregation is a useful structural manipulation that can help reduce the dimensionality of the graph and enable scalable analysis. Graph aggregation is also often referred to as coarsening. For a {\tt DataGraph} $\mdg(\mathcal{N}, \mathcal{E}, A, \bd^\mn, \bd^\me, \bd^\mg, \ma_\mn, \ma_\me, \ma_\mg)$, aggregation combines a subset of nodes $\bar{\mathcal{N}} \subseteq \mathcal{N}$ into a new node, $n_{agg}$. This aggregation can also require aggregation of node and/or edge data. We formalize the aggregation process for {\tt DataGraph}s by defining the following:

\begin{definition}[Aggregation Function $f_{+}$]
    For an undirected {\tt DataGraph} \\$\mdg \left(\mn, \me, A, \bd^\mn, \bd^\me, \bd^\mg, \ma_\mn, \ma_\me, \ma_\mg\right)$ and a subset of nodes to be aggregated, $\bar{\mn} \in \mn$, the aggregation function collapses the node set into a single new node and aggregates node and edge data where applicable. The aggregation function $f_{+}: (\mathbb{G}, \mathbb{V}, \mathbb{F}, \mathbb{F}) \rightarrow \mathbb{G}$ (where $\mathbb{V}$ is the set of all possible nodes and $\mathbb{F}$ is the set of all possible functions for aggregating sets of node or edge data) is defined as
    \begin{equation}\label{eq:aggregation}
        \begin{split}
            f_{+}(& \mdg, \mn', f^\mn_{+}, f^\me_{+}) = \mdg_{agg}\big( \mn_{agg}, \me_{agg}, A_{agg}, \bd^\mn_{agg}, \bd^\me_{agg}, \bd^\mg, \ma_\mn, \ma_\me, \ma_\mg  \big)\\
            \textrm{where} &\; \mathcal{N}_{agg} = \{ n_{agg}\} \cup \mathcal{N} \setminus \bar{\mn}\\
            &\; \bar{\me}_{agg} = \{(n_i, n_{agg}): n_i \in \mn \setminus \bar{\mn}, n_j \in \bar{\mn}, (n_i, n_j) \in \mathcal{E} \vee (n_j, n_i) \in \mathcal{E}\} \\
            &\; \mathcal{E}_{agg} = \{ (n_i, n_j): n_i \not\in \bar{\mn}, n_j \not\in \bar{\mn}, (n_i, n_j) \in \mathcal{E} \} \cup \bar{\me}_{agg}\\
            &\; \bd^\mn_{agg}(n) = \bd^\mn(n), \quad n \in \mathcal{N}\setminus \bar{\mn}\\
            &\; \bd^\mn_{agg}(n_{agg}) = f^\mn_{+}(\bd^\mn, \bar{\mn})\\
            &\; \bd^\me_{agg}(e) = \bd^\me(e), \quad e \in \{ (n_i, n_j): (n_i, n_j) \in \mathcal{E}, n_i \not\in \bar{\mn}, n_j \not\in \bar{\mn}\}\\
            &\; \bd^\me_{agg}((n_i, n_{agg})) = f^\me_{+}\big(\bd^\me, \{(n_i, n_j): n_j \in \bar{\mn}, (n_i, n_j) \in \me \} \cup \\
            &\;\qquad \qquad \qquad \qquad \qquad \qquad \qquad \{(n_j, n_i): n_j \in \bar{\mn}, (n_j, n_i) \in \me \}\big) \quad (n_i, n_{agg}), \in \bar{\me}_{agg}
        \end{split}
    \end{equation}
    where $f^\mn_{+}$ is a function that aggregates node data in $\bd^\mn$ for nodes $\bar{\mn}$ into node data for a single node ($n_{agg}$) and $f^\me_{+}$ is a function that aggregates edge data in $\bd^\me$ for the set of edges being passed to the function, and $A_{agg}$ is the adjacency matrix for edge set $\me_{agg}$. 
\end{definition}

In the above definition, we introduce two new functions, $f^\mn_{+}$ and $f^\me_{+}$. These functions can take many forms because of the variety of data that can be stored on the nodes and edges. In addition, the function for aggregating edge data, $f^\me_{+}$, is only needed when there are multiple nodes in $\bar{\mn}$ connected to some node in $\mn\setminus \bar{\mn}$; in this case, there are multiple edges that will be replaced by a single edge for undirected graphs. We also note that the aggregation function for directed graphs is different because the order of the edges will matter; consequently, rather than the function $f^\me_{+}$ acting on the combined set of edges $\{(n_i, n_j): n_j \in \bar{\mn}, (n_i, n_j) \in \me \}$ and $\{(n_j, n_i): n_j \in \bar{\mn}, (n_j, n_i) \in \me \}$, the function could need to be applied to multiple, independent sets. In addition, the set $\bar{\me}$ would need to be expanded to consider the edges where $n_{agg}$ could be the source node (not just the destination) for an aggregated edge. For both directed and undirected graphs, $f^\mn_{+}$ and $f^\me_{+}$ must be permutation invariant. \\
 
\begin{figure}[!htp]
    \centering
    \includegraphics[width = .7\textwidth]{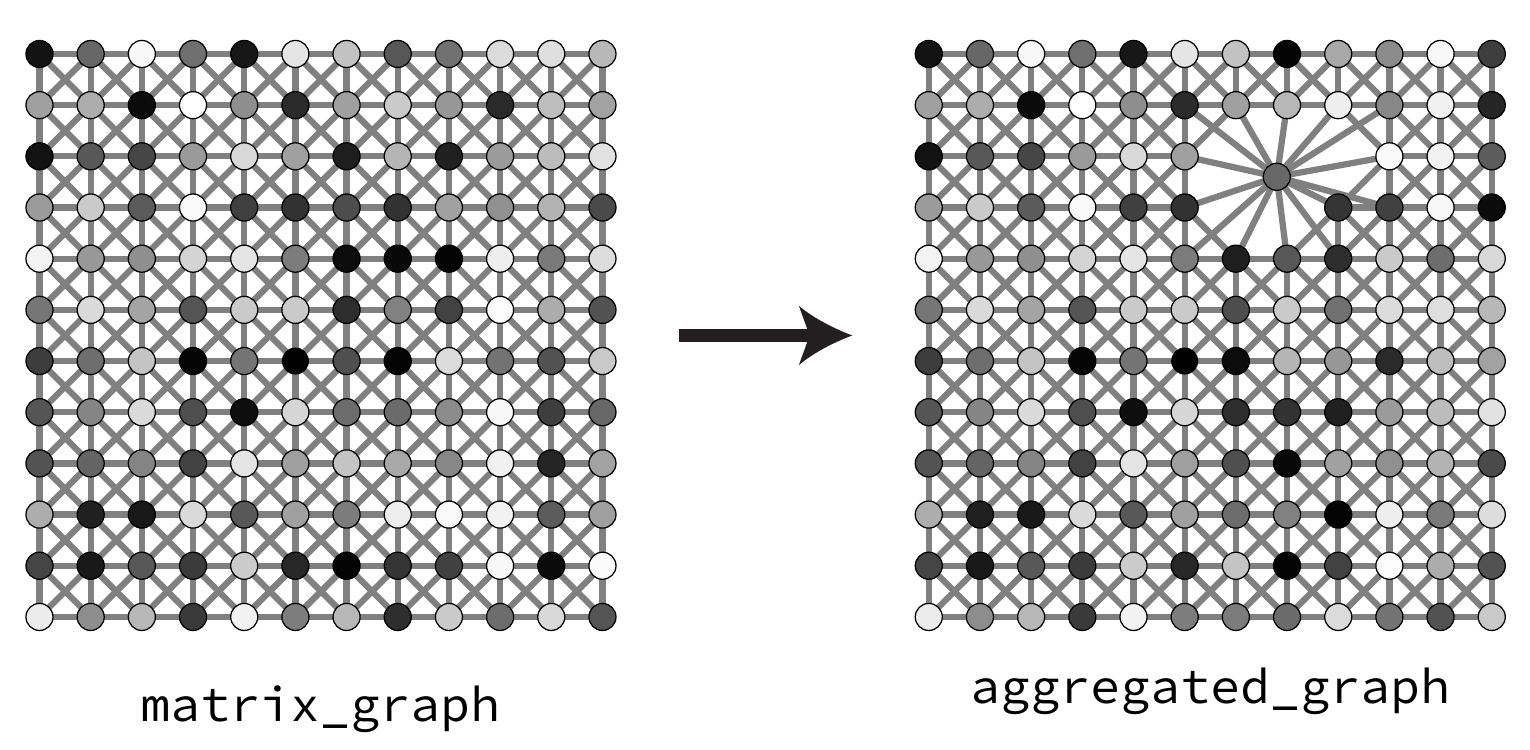}
    \captionof{figure}{Visualization associated with Code Snippet \ref{code:aggregation} where a subset of nodes is aggregated into a single new node with the node weights on the aggregated node averaged to get a new node weight for the new node}
    \label{fig:aggregation}
\end{figure}

{\tt PlasmoData.jl} implements aggregation for graph objects, as shown in Code Snippet \ref{code:aggregation} and Figure \ref{fig:aggregation}. Here, a node set is defined in Line \ref{line:aggregate_set}, and the node set is aggregated in line \ref{line:aggregate_call} using the function {\tt aggregate}. The default for aggregating node data or edge data is to average the values of the data for each attribute, but one can also use, for instance, a $\max$ operator (as done in graph neural networks). The user can use these different functions    by passing the function to the key word arguments {\tt node\_fn} and {\tt edge\_fn}. \\

\begin{figure}[!htp]
    \begin{minipage}[t]{0.9\linewidth}
        \begin{scriptsize}
        \lstset{language=Julia, breaklines = true}
        \begin{lstlisting}[label = code:aggregation, caption = Example of aggregating a set of nodes in a {\tt PlasmoData.DataGraph}, escapechar = ~] 
using PlasmoData, Random

Random.seed!(15)
random_matrix = rand(12, 12)

matrix_graph = matrix_to_graph(random_matrix; diagonal = true)

# Define nodes to be aggregated
nodes_for_aggregation = [(3, 7), (3, 8), (3, 9), (4, 7), (4, 8)] ~\label{line:aggregate_set}~

# Aggregate nodes in the graph
aggregated_graph = aggregate(matrix_graph, nodes_for_aggregation, "agg_node") ~\label{line:aggregate_call}~

        \end{lstlisting}
        \end{scriptsize}
    \end{minipage}
\end{figure}

There is an extension of \eqref{eq:aggregation} that could be considered. Under \eqref{eq:aggregation}, any data on the edges between nodes in $\bar{\mn}$ is lost, but such data could be important in different applications. We could extend \eqref{eq:aggregation} by defining a function $f_+^{\me \rightarrow \mn}$ that reduces $\bd^\me(e), e \in \{(n_i, n_j):(n_i, n_j) \in \me, n_i \in \bar{\mn}, n_j \in \bar{\mn} \}$ into new data on $n_{agg}$ for an additional set of attributes, $\ma_{\me \rightarrow \mn}$. Such attributes would not be restricted to be in either $\ma_\mn$ or $\ma_\me$. This would result in a slightly different {\tt DataGraph} such that $\mdg_{agg}\big( \mn_{agg}, \me_{agg}, A_{agg}, \bd^\mn_{agg}, \bd^\me_{agg}, \bd^\mg, \ma_\mn \cup \ma_{\me \rightarrow \mn}, \ma_\me, \ma_\mg  \big)$, where $\bd^\mn_{agg}$ now contains additional data for $\ma_{\me \rightarrow \mn}$. {\tt PlasmoData.jl} does not (yet) support this extension, so we have restricted \eqref{eq:aggregation} to follow the implementation of {\tt PlasmoData.jl}, but this could be an area of future work.

note that aggregation here refers to reducing the number of nodes in the graph and requires a defined set of nodes to be aggregated. The definition for fN+ and fE+ must be permutation invariant; in addition, 

\subsection{Graph Structural Analysis}

We now discuss different tools for analyzing and quantifying the structure of the {\tt DataGraph} model, such as Topological Data Analysis (TDA) and community detection. The implementation of these procedures (or interfacing to software that conducts these procedures) is greatly facilitated by the proposed graph abstraction. These procedures can also be combined with machine learning tools to enable supervised and unsupervised learning tasks (e.g., for classifying and clustering graphs or for predicting emerging properties from graphs). In addition, these graph analysis tools are often enhanced by the ability to manipulate the graph structure as discussed above. 
\\

Community detection and clustering in graphs focuses on how to identify organization within a graph, typically by identifying communities/clusters of nodes that have several connections within that grouping but relatively few connections to nodes outside of the grouping \cite{fortunato2010community, javed2018community}. Community detection has been applied to a variety of problems including optimization programs represented as graphs \cite{allman2019decode, mitrai2020decomposition}, biological networks \cite{jonsson2006cluster}, and fraud detection \cite{gangopadhyay2016health}. Identifying communities can give insight into the data. In addition, there are numerous algorithms for identifying communities \cite{javed2018community}, and different algorithms will identify different communities. Some of these algorithms (such as clique percolation \cite{palla2005uncovering}, label propagation \cite{raghavan2007near}, or Newman's modularity \cite{newman2004finding}) are already implemented within the {\tt Graphs.jl} package and are thus are easily accessible by {\tt PlasmoData.jl}.
\\

TDA is a growing field in data science that develops tools for analyzing and quantifying the shape/structure of data objects \cite{carlsson2009topology, chazal2021introduction, munch2017user, wasserman2018topological}. Graphs are topologically invariant objects that can be approached through the lens of TDA. Many data objects may not have an inherent topology in their mathematical definition (e.g., matrices); modeling these types of data as graphs enables TDA applications that otherwise might not be possible. Many TDA tools can also be combined with the structural manipulation based on data (e.g., graph filtration) to elucidate further insights into the data.  We now outline several TDA concepts and tools that can be used to analyze and quantify the shape/structure of graphs. 

\subsubsection{Graph Connectivity}

There are several general metrics/descriptors describing graph connectivity which could be considered within TDA. These descriptors could be used to help describe a graph, or they could be used in combination with the structure manipulation (e.g., analyzing these metrics after performing filtration). A few of these descriptors include:

\begin{itemize}

    \item {\bf Paths} - In graphs, a walk is a way for moving from one node to another node (via edges), and a path is a walk in which no node appears more than once \cite{wilson1979introduction}. There are existing algorithms for finding the shortest path between a couple of nodes or for finding whether nodes are connected. The paths in a graph could yield information about the graph and its data. For example, paths are frequently used for analyzing biological systems \cite{aittokallio2006graph, dost2008qnet, naderi2020revisiting}. As a further example, in a supply chain problem, we may be interested in finding paths (or the number of paths) between raw materials and products (see for example \cite{kim2006supply, wilson2022exploring}), as well as how those paths change if nodes or edges are filtered out. In addition, we may generally be interested in how the length of the shortest path between two nodes changes as the graph topology changes (e.g., via filtration). {\tt PlasmoData.jl} provides functions for analyzing paths by extending functions within {\tt Graphs.jl}, such as the {\tt has\_path} or {\tt get\_path} functions. 
    
    \item {\bf Cycles} - A cycle is a walk which starts and ends at the same node using entirely distinct edges. Cycles are used in many graph analyses, including construction \cite{perera2014cycle} and electrical networks \cite{berger2004minimum, dorfler2018electrical, kavitha2009cycle}. The number of cycles or length of cycles could be used in analyzing data represented as graphs. The function {\tt cycle\_basis} is extended from {\tt Graphs.jl} for {\tt PlasmoData.DataGraph}s. 
   
    \item {\bf Connected Components} - Connected components are sets of nodes that are all connected via paths. The number of connected components can describe how ``connected" a graph may be, and this metric could give insight into how connected or separated the data may be. The function {\tt connected\_components} is extended from {\tt Graphs.jl} for {\tt PlasmoData.DataGraph}s. 
    
    \item {\bf Node Degree} - The node degree is the number of nodes a node is connected to (i.e., the number of neighbors of a given node). The average node degree can also be computed for an entire graph. This metric has been used in node ranking \cite{liu2021survey} and sensor analysis \cite{jain2013node, vieira2016link}.
    
    \item {\bf Diameter} - The diameter of a graph is the length of the shortest path connecting the most distanced nodes (i.e., the maximum shortest path length for any two nodes in a graph). This metric also provides an idea of the ``connectedness" of the data, and it has been used in analyzing the degree of separation for different entities \cite{borassi2015fast}. The function {\tt diameter} is extended from {\tt Graphs.jl} for {\tt PlasmoData.DataGraph}s. 
\end{itemize}

The above metrics can yield important insights into data. These metrics may also change after filtration/aggregation of nodes and edges, which could likewise elucidate information about the dataset (e.g., aggregating a graph can help uncover features that are not apparent when dealing with the full-resolution graph). The {\tt DataGraph} abstraction therefore provides a framework for obtaining to topological descriptors.

\subsubsection{Euler Characteristic}\label{sec:EC}

The EC is a valuable descriptor for topological objects, including graphs. The EC is a scalar integer value that summarizes the shape of a topological space \cite{smith2021euler,munkres2018elements}. For a graph, this can be defined as \cite{smith2021euler}: 
\begin{equation}
    \mathcal{X} = \textrm{\# Connected Components} - \textrm{\# Holes} = |\mathcal{N}| - |\mathcal{E}| = \beta_0 - \beta_1
\end{equation}
\noindent where $\beta_i$ are the $i$th betti numbers of the graph, the "holes" are equivalent to cycles in the graph, and $|\cdot|$ is the cardinality of the set. We will use the notation $EC(\mdg)$ for the EC of a {\tt DataGraph} $\mdg(\mn, \me, A, \bd^\mn, \bd^\me, \bd^\mg, \ma_\mn, \ma_\me, \ma_\mg)$. \\

The EC has also been combined with filtration of node- or edge-weighted graphs to create an EC curve \cite{smith2021euler,smith2022topological}. The EC curve is formed by filtering out nodes and/or edges at different threshold values (e.g., removing all nodes with a weight value less than some threshold value) and computing the EC for the filtered graph. The shape of the EC curve can give insight into how the shape of the data changes with the filtration (e.g., where connected components dominate or where holes dominate). For instance, the EC curve has been used for analysis of liquid crystal sensors and MD data \cite{smith2021euler,smith2022topological, jiang2023scalable}. In Section \ref{sec:LCA}, we will show how the EC curve can be used in a classification model with real data. 

For a {\tt DataGraph}, we can generalize the EC curve for a filtration process for node filtration as follows (edge filtration has a similar definition but for $f_\me$ instead of $f_\mn$). For a {\tt DataGraph}, $\mdg(\mn, \me, A, \bd^\mn, \bd^\me, \bd^\mg, \ma_\mn, \ma_\me, \ma_\mg)$, with $\ma_\mn \ne \emptyset$ and $\bd^\mn \ne \emptyset$, there exist a sequence of logic sets $\mathcal{L}_1, \mathcal{L}_2, ..., \mathcal{L}_r$ such that the filtration forms a sequence of nested subgraphs given in \eqref{eq:nested_graphs} associated with the logic sets, where $\mdg_{\ml_i} = f_\mn(\mdg, \ml_i)$. For a graph to be considered a ``nested subgraph" of another graph, the nodes and edges of the ``nested subgraph" must be subsets of the nodes and edges, respectively, of the other graph. 
\begin{equation}\label{eq:nested_graphs}
    \mdg_{\ml_1} \subseteq \mdg_{\ml_2} \subseteq ... \subseteq \mdg_{\ml_r}
\end{equation}
The EC curve can then be given by the vector $(EC(\mdg_{\ml_1}), EC(\mdg_{\ml_2}), ..., EC(\mdg_{\ml_r}))$. This approach for generating an EC curve is more general than the idea of using threshold values because the {\tt DataGraph} could include a variety of data types. In the case of single scalar node or edge weights where all weights are in $[0, 1]$, the logic sets could, for example, be a series of ranges such as $[0, 0.1], [0, 0.2], ..., [0, 1]$ (thus effectively filtering out all nodes with weights greater than a single value). However, an EC curve of sorts could be generated for more complex data, such as considering data with multiple weights. For {\tt PlasmoData.DataGraph}s with scalar data for each attribute, computing the EC curve can be done with the functions {\tt run\_EC\_on\_nodes} and {\tt run\_EC\_on\_edges}.

\section{Case Studies}\label{sec:examples}

We now provide case studies that highlight different applications and benefits of the {\tt DataGraph} abstraction. All scripts and data necessary to replicate the results in this section are available at \url{https://github.com/zavalab/JuliaBox/tree/master/PlasmoData_examples}. These examples include image analysis (representing matrices and tensors as graphs), multivariate time series analysis (representing symmetric matrices as graphs), and connectivity analysis (analyzing pathways within graphs).

\subsection{Image Analysis}\label{sec:LCA}

We consider a case study arising in image classification and show how the {\tt DataGraph} abstraction and {\tt PlasmoData.jl} help integrate with machine learning tools and facilitate feature extraction. We will show how these images of surfaces can be analyzed by representing these images as 3D tensors the representing these tensors using the {\tt DataGraph} abstraction. We will show how we use {\tt PlasmoData.jl} for extracting features from this data and use those features to train a machine learning model. We will also highlight how {\tt PlasmoData}'s data structure can be employed for GNNs.

\begin{figure}[!htp]
    \centering
    \includegraphics[scale = .35]{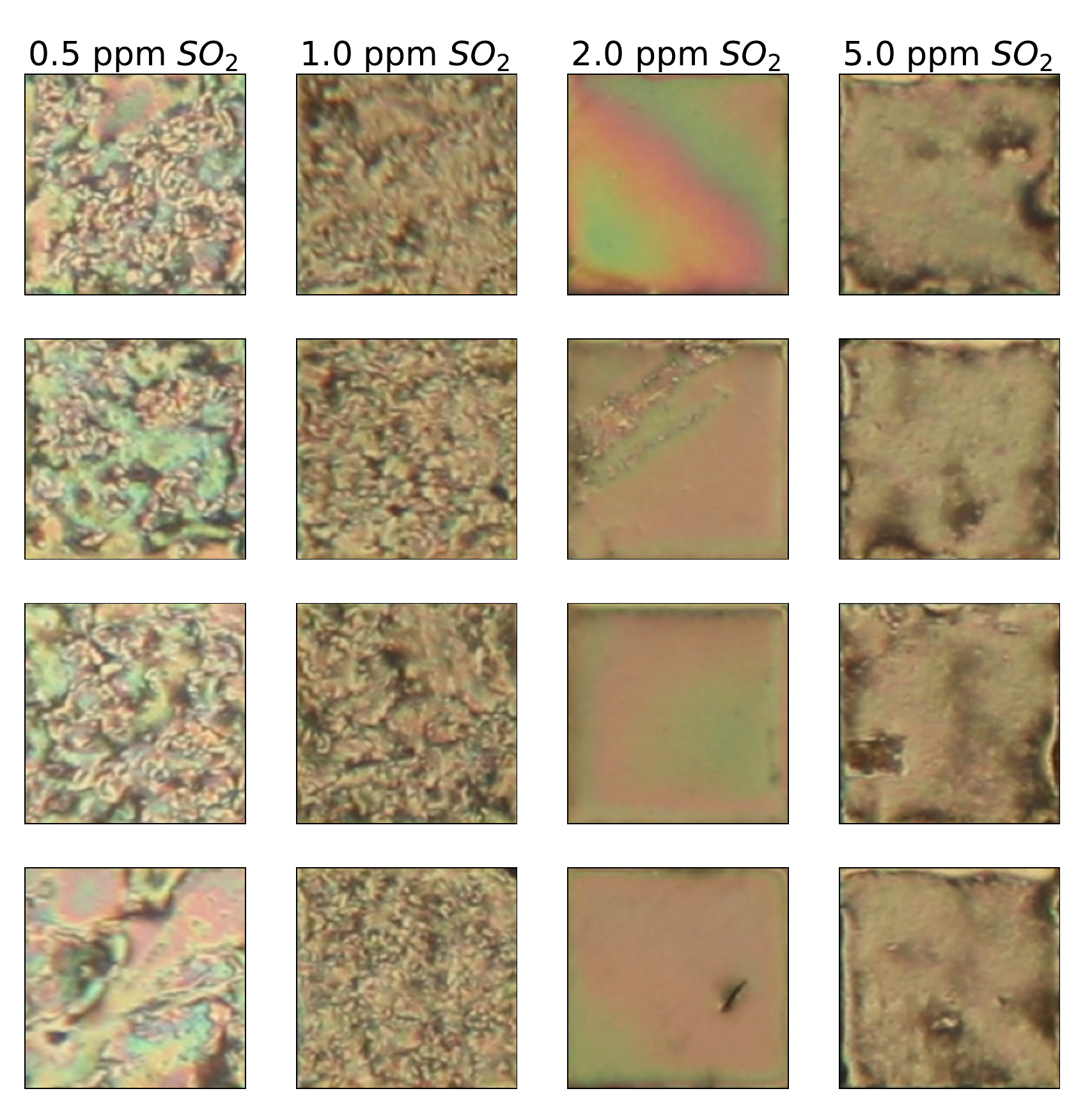}
    \captionof{figure}{Images of LCs exposed to gaseous environments with different concentrations of SO\textsubscript{2} \cite{jiang2023scalable, bao2022ordering}}
    \label{fig:4class_images}
\end{figure}

The images we consider are of chemoresponsive liquid crystals (LCs) which have garnered increased interest because of their potential as sensors for gas contaminants \cite{shah2001principles, carlton2013chemical, esteves2020seeing}. LCs have been shown to elicit different, visible chemical responses to varying chemical environments as the LCs change their orientation ordering \cite{bao2022ordering, szilvasi2018redox}. One of the challenges that naturally arises from these sensors is identifying the chemical environment surrounding a LC based on the appearance of the LC's surface. We are therefore interested in being able to classify images of LCs based on the chemical environment into which the LC has been placed. The dataset we consider are images of LCs exposed to four different concentrations  of sulfur dioxide (SO\textsubscript{2}): $0.5$, $1.0$, $2.0$, and $5.0$ ppm (i.e., four classes of images corresponding to each concentration). Each image is $134 \times 134$ pixels, and there are 72 images in each class (288 total). Examples of images from each class can be seen in Figure \ref{fig:4class_images} \cite{jiang2023scalable, bao2022ordering}. \\

\subsubsection{Classification using TDA and SVMs}

To classify these images, we will represent these images as tensors and then each tensor as a graph. Each of the 288 images can be treated as a tensor, $T \in [0, 1]^{134 \times 134 \times 3}$. Under the {\tt DataGraph} abstraction, we can represent the $i$th tensor as a graph with \eqref{eq:image_to_graph} using the first method shown in Section \ref{sec:tensor_to_graph}, where the third dimension of the tensor represents the red, green, and blue channels of the pixel intensity and where $t_{j, k, l}$ is the $j, k, l$ entry of $T$.

\begin{equation}\label{eq:image_to_graph}
    \begin{split}
    &\mathcal{DG}_i(\mathcal{N}, \mathcal{E}, A, \bd^\mn, \emptyset, \emptyset, \ma_\mn, \emptyset, \emptyset)\\
    \textrm{where} &\; \mathcal{N} = \{ n_{j, k}: j \in \mathbb{N}_{134}, k \in \mathbb{N}_{134} \}  \\
    &\; \mathcal{E} = \{(n_{j, k}, n_{j + 1, k}): j \in \mathbb{N}_{133}, k \in \mathbb{N}_{134} \}\cup\{(n_{j, k}, n_{j, k+1}): j \in \mathbb{N}_{134}, k \in \mathbb{N}_{133} \} \cup \\
    &\; \quad \quad \{ (n_{j, k}, n_{j + 1, k + 1}): j \in \mathbb{N}_{133}, k \in \mathbb{N}_{133} \} \cup \{ (n_{j, k + 1}, n_{j + 1, k}): j \in \mathbb{N}_{133}, k \in \mathbb{N}_{133} \} \\
    &\; \mathcal{A}_\mathcal{N} = \{ a_{red}, a_{green}, a_{blue}\} \\
    &\; d^\mn_{a_{red}}(n_{j, k}) = t_{j, k, 1}, j \in \mathbb{N}_{134}, k \in \mathbb{N}_{134}\\
    &\; d^\mn_{a_{green}}(n_{j, k}) = t_{j, k, 2}, j \in \mathbb{N}_{134}, k \in \mathbb{N}_{134}\\
    &\; d^\mn_{a_{blue}}(n_{j, k}) = t_{j, k, 3}, j \in \mathbb{N}_{134}, k \in \mathbb{N}_{134}
    \end{split}
\end{equation}
where $A$ is a symmetric matrix ($\mdg$ is undirected) matching the connectivity defined by $\me$.

With the {\tt DataGraph}, $\mdg_i$, defined for each image, we can generate an EC curve (or a set of EC curves) for the graph. We will generate an EC curve for the data for each attribute in $\ma_\mn$. Because we have scalar weights, we will filter the nodes by removing nodes that are greater than a threshold value, and we will do this for a series of threshold values. Under the formulation given in Section \ref{sec:EC}, we can define a sequence of logic sets for the data corresponding to each attribute in $\ma_\mn$ such that
\begin{equation*}
    \ml_{j, a_{red}} = \begin{cases} a_{red} &: [0, (j - 1) \times 0.005]\\
        a_{green} &: [0, 1]\\
        a_{blue} &: [0, 1]
    \end{cases}, \quad \quad j \in \mathbb{N}_{201}
\end{equation*}

\begin{equation*}
    \ml_{j, a_{green}} = \begin{cases} a_{red} &: [0, 1]\\
        a_{green} &: [0, (j - 1) \times 0.005]\\
        a_{blue} &: [0, 1]
    \end{cases}, \quad \quad j \in \mathbb{N}_{201}
\end{equation*}

\begin{equation*}
    \ml_{j, a_{blue}} = \begin{cases} a_{red} &: [0, 1]\\
        a_{green} &: [0, 1]\\
        a_{blue} &: [0, (j - 1) \times 0.005]
    \end{cases}, \quad \quad j \in \mathbb{N}_{201}
\end{equation*} 

One approach to classify these images is to feed a topological descriptor of the weighted graph (e.g., the EC curve) to a machine learning model. We generate an EC curve for the $i$th image and for each color channel, and we denote these curves for attribute (color channel) $a$ as 
$$\boldsymbol{\mathcal{X}}_{i, a} = (EC(f_\mn(\mdg_i, \ml_{1, a})), EC(f_\mn(\mdg_i, \ml_{2, a})), ..., EC(f_\mn(\mdg_i, \ml_{201,a}))). $$
To perform the classification for the $i$th image, we concatenate the EC curves $\boldsymbol{\mathcal{X}}_{i, a_{red}}, \boldsymbol{\mathcal{X}}_{i, a_{green}}$, and $\boldsymbol{\mathcal{X}}_{i, a_{blue}}$ into a vector $\boldsymbol{\mathcal{X}}_i$ and train a linear support vector machine (SVM) on sets of $\boldsymbol{\mathcal{X}}_i$ for $i \in \mathbb{N}_{288}$. \\  

The code for performing the tasks of generating the graph representations, building the EC curves, and training an SVM can be seen in Code Snippet \ref{code:EC_SVMs}. The data is in the format of a 4D array of size $(288, 134, 134, 3)$. The graph is formed by treating each pixel of the image (dimensions 2 and 3 of the array) as a node using {\tt PlasmoData.jl}'s {\tt matrix\_to\_graph} function (Line \ref{line:mattograph}). The intensity values for each RGB channel (dimension 4 of the array) are saved as weights on each node of the graph. After each graph is formed, we then find the EC curve for each weight (channel) on the graph and store this in the array {\tt so2\_ECs} (Lines \ref{line:runECstart} - \ref{line:runECend}). The three EC curves (one for each channel) are then concatenated into a vector and used to train a linear SVM using 5-fold cross validation (CV) (Lines \ref{line:SVMstart} - \ref{line:SVMend}).\\

\begin{figure}[!htp]
    \begin{minipage}[t]{\textwidth}
    \begin{scriptsize}
    \lstset{language=Julia, breaklines = true}
    \begin{lstlisting}[label = {code:EC_SVMs}, caption = Classifying images from the EC using SVMs] 
using PlasmoData, JLD, MLUtils, LIBSVM

### Construct matrices as graphs and get EC curves ###

# Load in the SO2 data; so2_data is size (288, 134, 134, 3)
so2_data = load("so2_data.jld")["data"]
so2_classes = load("so2_classes.jld")["classes"]

# Define threshold range for each EC curve
thresh = 0:.005:1 |\label{line:process_start}|

# Define a matrix for the EC curves; EC curves will be concatenated
so2_ECs = Array{Float64, 2}(undef, (length(thresh)*3, 288))
        
for i in 1:288
    # Build a graph from a 3-D array (134 x 134 x 3)
    mat_graph = matrix_to_graph(so2_data[i, :, :, :]) |\label{line:mattograph}|
        
    for j in 1:3 |\label{line:runECstart}|
        # Iterate through each channel and concatenate the EC curve
        range_bounds = (1 + (j - 1) * length(thresh)):(j * length(thresh))
        so2_ECs[range_bounds, i] = run_EC_on_nodes(mat_graph, thresh, "weight$j", false) 
    end |\label{line:runECend}|
end |\label{line:process_end}|

### Perform 5 fold CV with SVMs on EC data ###

# shuffle data
Xs, ys = shuffleobs((so2_ECs, so2_classes)) |\label{line:SVMstart}|

# define a function for calculating accuracy
function get_accuracy(yhat, ytest)
    num_errors = 0
    for i in 1:length(yhat)
        if yhat[i] != ytest[i]
            num_errors += 1
        end
    end
    return 1 - num_errors/length(yhat)
end

# Perform 5-fold CV 
accuracy_values = []
for (train_data, val_data) in kfolds((Xs, ys); k = 5)
    model = svmtrain(train_data[1], train_data[2], kernel = Kernel.Linear)
    yhat, decision_values = svmpredict(model, val_data[1]
    accuracy = get_accuracy(yhat, val_data[2])
    push!(accuracy_values, accuracy)
end |\label{line:SVMend}|
    \end{lstlisting}
    \end{scriptsize}
    \end{minipage}
\end{figure}

The method can effectively classify these images in a reasonable time. The classification accuracy, based on 5-fold CV, was $94.8\%$ with a standard deviation of $2.1\%$. Running this script on an Intel(R) Core i9-10885H (2.40GHz) processor on a single thread with Julia 1.7.3 resulted in a data processing time (Lines \ref{line:process_start} - \ref{line:process_end}) of $36.2$ s and a 5-fold CV (Lines \ref{line:SVMstart} - \ref{line:SVMend}) time of $0.09$ s. This is comparable to what Jiang and co-workers \cite{jiang2023scalable} found, where they had $87.5\%$ accuracy from other topological methods that took less time to process and train ($\sim 14$ s). For comparison, Jiang and co-workers \cite{jiang2023scalable} also trained a convolutional neural network (CNN) for classifying these images, and the CNN was able to classify with $95.1\%$ accuracy while taking $154$ seconds to train. Here, we do not aim to show superiority of one method over another or provide a rigorous comparison in processing time but rather to provide a general validation as to whether our methods yield reasonable results. In particular, our results suggest that it is possible to obtain high accuracy while performing these operations in a competitive time.\\

We also note---as Jiang and co-workers \cite{jiang2023scalable} did in their study---that there are benefits to topological methods such as those we followed above. First, the linear SVM above had relatively few parameters---each EC curve contained $201$ points before concatenation, so there were only $603$ parameters in the SVM. In contrast, CNNs can use high numbers of parameters (thousands to millions). They also can be sensitive to rotation, whereas graphs are rotationally invariant (e.g., the EC curve does not depend on the orientation of the graph). In addition, Jiang and co-workers \cite{jiang2023scalable} note that the CNN in their study was trained on a GPU, whereas our above methods (and Jiang and co-workers \cite{jiang2023scalable} TDA methods) functioned efficiently on a CPU.\\

\begin{figure}[!htp]
    \centering
    \includegraphics[width = 0.95\textwidth]{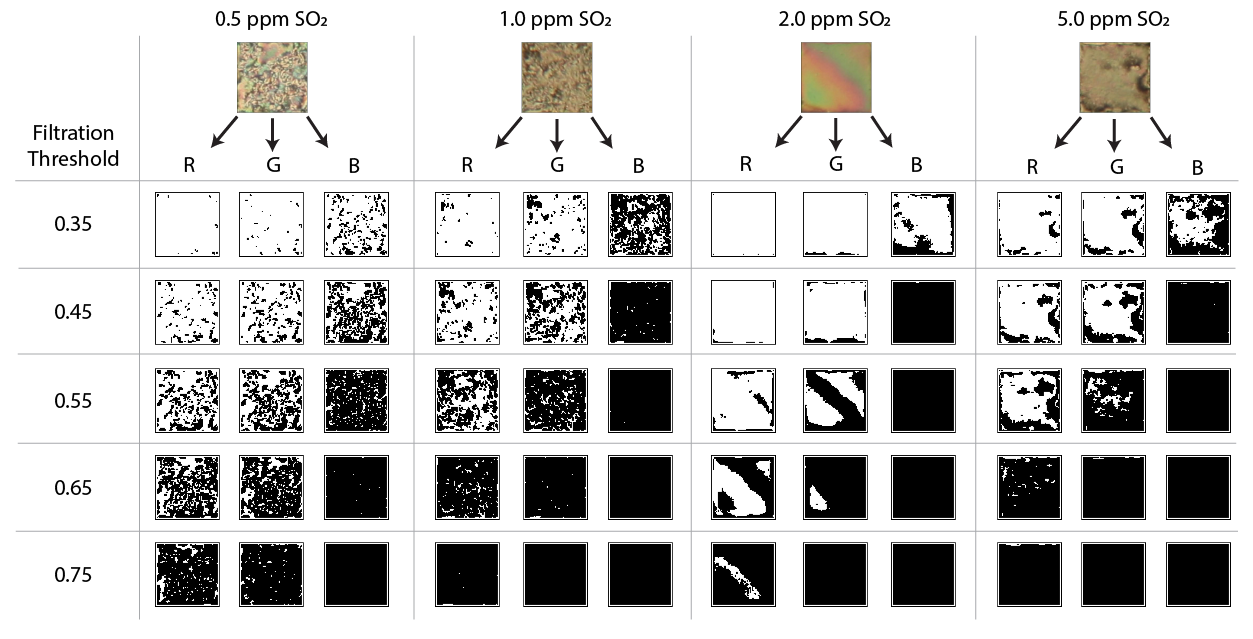}
    \captionof{figure}{Examples by class of graph filtrations at different threshold values. Each image has a value corresponding to the RGB channels, and these channels result in different topologies at the filtration thresholds. The EC curve captures these different topologies across channels and across classes. Data comes from the work of Bao and co-workers \cite{bao2022ordering} and Jiang and co-workers \cite{jiang2023scalable}}
    \label{fig:EC_filtrations}
\end{figure}

\begin{figure}[!htp]
    \centering
    \includegraphics[scale = .7]{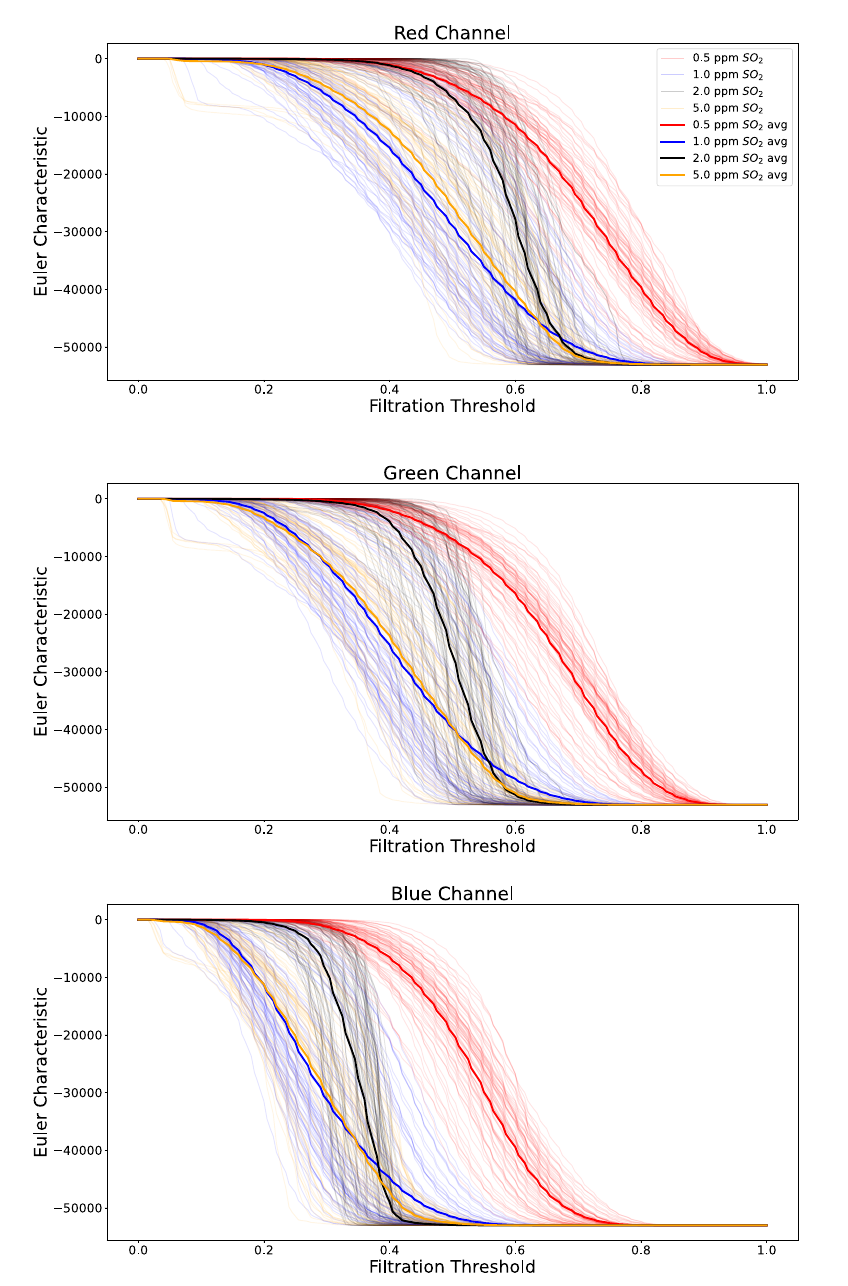}
    \captionof{figure}{EC Curves for the RGB channels of four classes of images for LC data from Bao and co-workers \cite{bao2022ordering} and Jiang and co-workers \cite{jiang2023scalable}}
    \label{fig:EC_curves}
\end{figure}

A further benefit of TDA is that the results can be interpretable and potentially provide insight into the data. For example, the EC curve can reveal how the topology of a graph is changing throughout a filtration. For example, we show in Figure \ref{fig:EC_filtrations} an example of what the filtration looks like for an example image from each class. In addition, the EC curves for the three different channels are shown in Figure \ref{fig:EC_curves}. In Figure \ref{fig:EC_filtrations}, different topologies are evident at different filtration thresholds. For example, the red channel of the $0.5$ ppm SO\textsubscript{2} class has several small, scattered holes throughout, and its corresponding EC curve (Figure \ref{fig:EC_curves}) is less steep than the $2.0$ ppm SO\textsubscript{2} class which exhibits larger and fewer holes in the topology. The channels also exhibit slightly different topologies at different filtration levels and can thus provide different information in differentiating the images. For example, the average EC curves for the $0.5$ and $5.0$ ppm SO\textsubscript{2} classes (Figure \ref{fig:EC_curves}) are almost on top of each other for most of the blue channel, whereas the two curves are more offset within the red channel.\\

The EC curve for each graph (and each channel) could provide insight into the image. Because the EC of a graph is only the number of nodes minus the number of edges, the EC curve begins at the origin but can end far from the origin. In the case of images, the full graph (i.e., when nothing is filtered out) has a large negative EC since the number of edges is much more than the number of nodes. Thus, the EC curve of a graph tends to show when and how quickly edges and cycles begin to appear. For example, in the case of images, if the filtration results mostly in nodes that are far apart, there are few if any edges between them and the EC is more positive. In contrast, if the filtration results in groups of clustered nodes, there are more edges connecting nodes and the EC is more negative. This helps explain why the EC curves for the $0.5$ ppm SO\textsubscript{2} class tend to be more positive at a given threshold value than the other classes (Figure \ref{fig:EC_curves}) since, as can be seen in Figure \ref{fig:EC_filtrations}, the filtration results in more spread out, small clusters at the lower threshold values for the $0.5$ ppm SO\textsubscript{2} class than the other classes.\\

The EC curves also reveal some minor differences across color channels, and this helps highlight how the multiple node weights within a graph can be used for more effectively analyzing the data. If only gray-scale images are used, the classification accuracy of these images is at least $10\%$ lower (see the Supporting Information). We can thus analyze our graph structure using different weights to gain different information. One area that could particularly benefit from this approach is hyperspectral imaging \cite{li2019deep,shenming2022new}. In hyperspectral images, rather than recording only 3 channels of colors, each image can contain several ($10$s - $100$s) weights at each pixel corresponding to different wavelengths of light. Thus, it is possible for each node of the graph to contain dozens or hundreds of weights that could be analyzed. {\tt PlasmoData.jl} is able to readily handle this type of data. 
\\

In this example, we have described how {\tt PlasmoData.jl} can be used to combine TDA with machine learning. There are more details that could be valuable on these topics which we include in the supporting information, including i) a comparison of memory allocation between {\tt PlasmoData.jl}, {\tt Networkx}, {\tt MetaGraphs.jl}, and {\tt Matlab}'s graph function for the image application; and ii) a brief discussion of how the graph structure and data impact classification results. In the above example, if gray-scale data is used instead of colored data (i.e., one weight instead of three weights are stored on each node), the classification accuracy decreases. 

\subsubsection{Classification using Graph Neural Networks}

An alternative method to TDA+SVM for graph classification are convolutional GNNs. {\tt PlasmoData.jl} facilitates the implementation of this method and enables systematic comparisons with other approaches. Since the data for nodes and edges is stored as a matrix within the {\tt PlasmoData.DataGraph} (or {\tt DataDiGraph}) structure, we can easily query this data to create the needed structure for GNNs. In Code Snippet \ref{code:GNNs}, we show how our data for the image classification can be easily transformed into a format used by a GNN. In this example, we use the packages {\tt GraphNeuralNetworks.jl} \cite{lucibello2021GNN} and {\tt Flux.jl} \cite{innes2018flux}. After the data is loaded, we build the {\tt PlasmoData.DataGraph} as before from the image data (Line \ref{line:mat_to_graph_gnn}). The {\tt PlasmoData.DataGraph} can then be passed to the {\tt GNNGraph} function from {\tt GraphNeuralNetworks.jl} (Lines \ref{line:gnn_graph_start} - \ref{line:gnn_graph_end}). This function takes as an argument a {\tt Graphs.SimpleGraph} (or {\tt Graphs.SimpleDiGraph}) to provide the adjacency lists, and a matrix of node and/or edge data. Since the {\tt PlasmoData.DataGraph} stores a {\tt Graphs.SimpleGraph}, this is readily provided to the {\tt GNNGraph} function. In addition, the node or edge data is provided by calling the API function {\tt get\_node\_data} (Line \ref{line:get_node_data}). The classes are also passed to {\tt Flux.onehotbatch} (Line \ref{line:onehotbatch}) to provide a one hot encoded array that can be used by the GNN.\\

\begin{figure}[!htp]
    \begin{minipage}[t]{\textwidth}
    \begin{scriptsize}
    \lstset{language=Julia, breaklines = true}
    \begin{lstlisting}[label = {code:GNNs}, caption = Interfacing {\tt PlasmoData.jl} with GNNs] 
using PlasmoData, JLD, Flux, GraphNeuralNetworks

function getdataset()
    # Load in data
    so2_data = load("so2_data.jld")["data"]
    so2_classes = load("so2_classes.jld")["classes"]

    GNN_graphs = []

    # Create GNNGraph from DataGraph
    for i in 1:288
        mat_graph = matrix_to_graph(data[i, :, :, :]) |\label{line:mat_to_graph_gnn}|
        gnn_graph = GraphNeuralNetworks.GNNGraph( |\label{line:gnn_graph_start}|
            mat_graph.g,
            ndata = get_node_data(mat_graph)|'| |\label{line:get_node_data}|
        ) |\label{line:gnn_graph_end}|
        push!(GNN_graphs, gnn_graph)
    end
            
    # One hot encode classes
    y = Flux.onehotbatch(so2_classes, 1:4) |\label{line:onehotbatch}|

    return GNN_graphs, y
end
    \end{lstlisting}
    \end{scriptsize}
    \end{minipage}
\end{figure}

With the data formatted as {\tt GNNGraph}s, we can build a full GNN through {\tt GraphNeuralNetworks.jl}. We ran our script on an AMD EPYC 7302 16-core processor with access to a NVIDIA Quadro RTX 6000 GPU (used for training the GNN). Using the same 288 images as before, we were able to classify the images with $89.2\%$ accuracy with a standard deviation of $5.1\%$ based on 5-fold CV. While the time for data processing was short ($3.8$ s), the GNN did take a long time to train and perform CV (947 s).\\

The ability to use {\tt PlasmoData.DataGraph}s for GNNs is a powerful capability. This framework for building GNNs from {\tt PlasmoData.DataGraph}s could be applied much more broadly than just to images (e.g., molecular property prediction). The user has significant flexibility in constructing their own graph data and using GNNs with that data, enabling application of {\tt PlasmoData.jl} to a variety of fields or use cases. For further details on constructing a {\tt GNNGraph} from data using {\tt PlasmoData.jl}, see the Supporting Information. 

\subsection{Multivariate Time Series Analysis}\label{sec:disease_data}

In this example, we illustrate how multivariate time series can be represented as graphs, and we highlight how the topology of this graph representation can provide interesting insights. Recently, de Souza and co-workers \cite{de2022euler} explored the use of TDA (and in particular the EC) as a potential tool for data-driven surveillance of epidemic outbreaks. Specifically, they were interested in finding indicators or fingerprints of when an outbreak is occurring (or likely to occur), and they based their efforts on locally reported disease data and on simulated data. In their work, they constructed correlation matrices as graphs, filtered these graphs, and looked at the resulting topology. They identified that there is a strong correlation between the EC curve and epidemic outbreaks, and they suggest that TDA could be used as a tool in disease surveillance. In this section, we will recreate some of their results using the {\tt DataGraph} abstraction and using {\tt PlasmoData.jl}, and we will suggest some other topological metrics that could be useful in analyzing datasets like these. In doing this, we emphasize that our purpose in this example is to highlight how data can be modeled with graphs and how {\tt PlasmoData.jl} facilitates much of this analysis. \\

\begin{figure}[!htp]
    \centering
    \includegraphics[width = .5\textwidth]{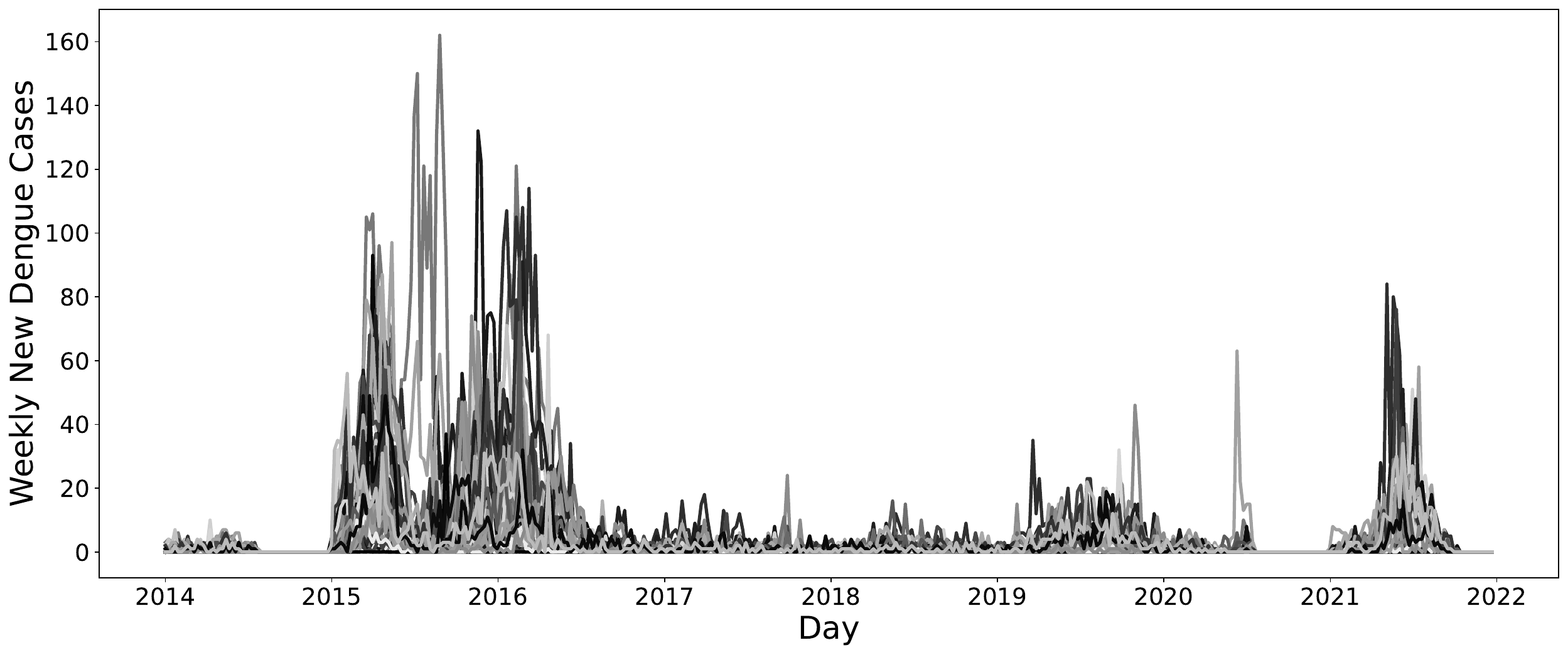}
    \captionof{figure}{New dengue cases by week for 93 districts in Recife, Brazil based on data from de Souza and co-workers \cite{de2022euler}}
    \label{fig:dengue_data}
\end{figure}

\begin{figure}[!htp]
    \centering
    \includegraphics[width = 0.8\textwidth]{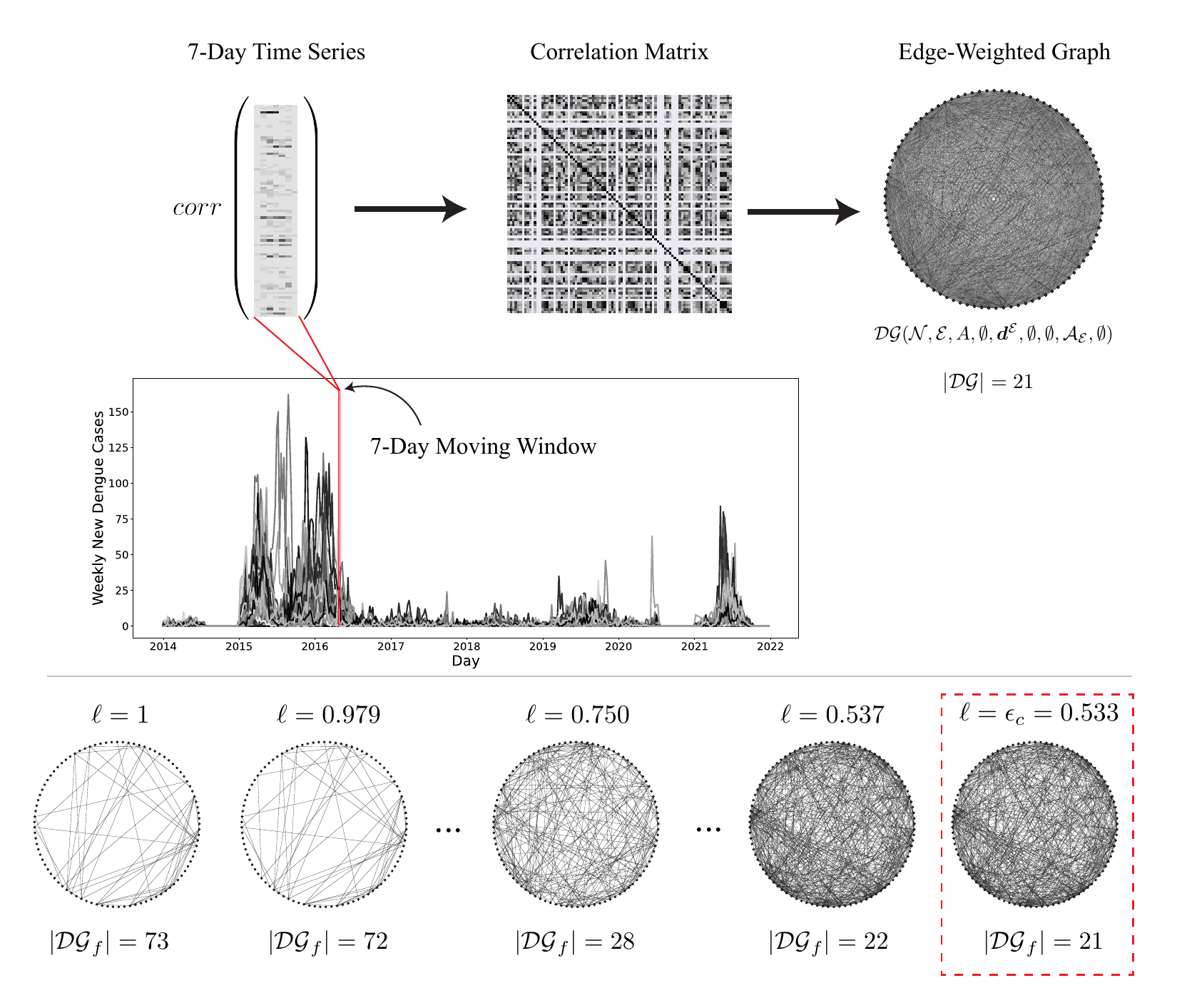}
    \captionof{figure}{General procedure used by de Souza and co-workers \cite{de2022euler}. For a 7-day moving time window, they formed a (Pearson) correlation matrix and then constructed an edge weighted graph. They then filtered out all edges and progressively added edges back into the graph by changing the filtration threshold, $\ell$, until the number of connected components was equal to that of the original graph}
    \label{fig:deSouza_process}
\end{figure}

First, we highlight the general data and methodology of de Souza and co-workers \cite{de2022euler}, but point the reader to their original paper to find an in depth explanation \cite{de2022euler}. For this analysis, we will use their data from Recife, Brazil of new daily dengue cases from 93 individual districts from 2014 - 2021 (Figure \ref{fig:dengue_data}). For each seven-day moving window of data, they computed the Pearson correlation for the 93 time series (one for each district) and then constructed the edge-weighted graph as outlined earlier (see for example Figure \ref{fig:sym_mat_to_graph}). They then computed a characteristic threshold value (what they call a critical percolation value) for this edge-weighted graph. This characteristic threshold value is the maximum possible threshold value that does not change the number of connected components from the number in the original graph when all edges with weights less than that threshold value are removed. A visualization of a this process can be seen in Figure \ref{fig:deSouza_process}, where $\ell$ is the threshold value used for filtration. 

Note that, in computing the EC of the filtered graph, de Souza and co-workers use methods different from that discussed in this present work. Where we have only discussed computing the EC for a graph (a 2D object), de Souza and co-workers \cite{de2022euler} treat the filtered graph as a CW-complex. CW-complexes are different than graphs, and involve the union of all {\it d}-cells in a graph (equivalently, the union of all $d$-node or -vertex cliques; see \cite{de2022euler, whitehead1949combinatorial}). A $d$-cell or $d$-vertex clique is a subset of nodes in a graph such that every node is connected by an edge to every other node in the subset (i.e., every two nodes in the subset are adjacent). A detailed discussion of CW-complexes is outside the scope of this paper; however, we mention this to highlight that the EC computed by de Souza and co-workers for CW-complexes is different than the EC we will show computed for a graph. Instead, their methods for computing the EC involve finding all $d$-cells up to some $d_{max}$ value. We will discuss later in this subsection some of the implications of this difference. \\

We can represent the Pearson correlation matrix at each time point within the {\tt DataGraph} abstraction, and we then can perform a similar analysis as de Souza and co-workers \cite{de2022euler} on the {\tt DataGraph} within {\tt PlasmoData.jl}. For the time series of data $\bX \in \mathbb{Z}_+^{2922 \times 93}$ and $\bX = [X_1, X_2, ..., X_{2922}]$, we first construct the Pearson correlation matrix $R_t = corr([X_{t - 6}, X_{t - 5}, ..., X_{t}])$ for each time $t$ between 7 and 2922. Since $R_t$ is a symmetric matrix, we can construct a {\tt DataGraph} following the method outlined in Section \ref{sec:sym_mat_to_graph}. The {\tt DataGraph} for $R_t$ is given in \eqref{eq:pc_to_graph}, where $r_{i, j}$ is the $i, j$ entry of $R_t$. 
\begin{equation}\label{eq:pc_to_graph}
    \begin{split}
    &\;\mathcal{DG}_t(\mathcal{N}, \mathcal{E}, A, \emptyset, \bd^\me, \emptyset, \emptyset, \ma_\me, \emptyset)\\
    \textrm{where} &\; \{ n_{i}: i \in \mathbb{N}_{93}\} \\
    &\; \mathcal{E} = \{(n_{i}, n_{j}): i \in \mathbb{N}_{92}, j \in \mathbb{N}_{93} \setminus \mathbb{N}_{i}\}\\
    &\; \mathcal{A}_\mathcal{E} = \{ a \} \\
    &\; d^\me_a((n_i, n_j)) = r_{i, j}, i \in \mathbb{N}_{92}, j \in \mathbb{N}_{93} \setminus \mathbb{N}_i
    \end{split}
\end{equation}
where $A$ is a symmetric matrix ($\mdg$ is undirected) matching the connectivity defined by $\me$.
\\

With the {\tt DataGraph} defined, we can perform the analysis shown in Figure \ref{fig:deSouza_process}. The graph edges are filtered using the logic set 
$$\ml_{t, \ell} = \begin{cases} a : [\ell, \max\left(\{ r: r \in R_t \}\right)] \end{cases}$$
\noindent for attribute $a$. We define the characteristic threshold value for time $t$ to be 
$$\epsilon_t = \sup \{ \epsilon : |f_\mn(\mdg_t, \ml_{t, \epsilon})| = |\mdg_t| \}. $$
\noindent The value of $\epsilon_t$ can be found by setting $\ell$ in $\ml_{t, \ell}$ to be the largest edge weights and then progressively choosing the next largest edge weights until $|f_\mn(\mdg_t, \ml_{t, \ell})| = |\mdg_t|$. This process is repeated for each time value. For each graph filtered by the characteristic threshold value, we can consider various topological metrics like the EC or the number of communities. \\

We performed the above analysis in {\tt PlasmoData.jl}, and we will show how the {\tt DataGraph} abstraction can enable topological analysis of the dengue data. {\tt PlasmoData.jl} readily facilitates the process of building the graph, filtering the edges, and performing TDA. Code Snippet \ref{code:denge} shows how these tasks can be implemented. First, the data for Recife, Brazil from de Souza and co-workers \cite{de2022euler} is loaded on line \ref{line:recife_data}. This data are daily reported new cases of dengue outbreaks in each of 93 city districts. We then define the function {\tt find\_smallest\_filtered\_graph} on Lines \ref{line:recife_function_start} - \ref{line:recife_function_end} that will add edges to the graph until the graph has only one connected componentent. The symmetric matrix for building the graph is formed by computing the Pearson correlation matrix on Line \ref{line:recife_correlation}. Because we use a limited time window (7 days), some values in the matrix were not defined (i.e., {\tt NaN} values). We removed all rows/columns of the symmetric matrix that were {\tt NaN} values (Lines \ref{line:recife_filter_nans_start} - \ref{line:recife_filter_nans_end}), and used the resulting symmetric matrix to construct the edge weighted graph, $\mathcal{DG}$ (line \ref{line:recife_build_graph}) (this results in a symmetric matrix with fewer than 93 nodes). We next define a sequence of values $\ell_1, \ell_2, ..., \ell_k$ which are equal to the edge weights of the graph ($\bd_a^\me$) in descending order (see Line \ref{line:recife_iters}; non-unique values are removed in the code). Once $\mathcal{DG}_t$ is defined, we then iterate through the sequnce of edges weights until $|f_\mn(\mdg_t, \ml_{t, \ell_i})| = 1$, where $\ell_i$ is the $i$th value in the sequence, and where $\ell_i = \epsilon_t$. Note that we use $|f_\mn(\mdg_t, \ml_{t, \ell_i})| = 1$ rather than $|f_\mn(\mdg_t, \ml_{t, \ell_i})| = |\mdg_t|$ because we have removed the rows and columns of the matrix which are not defined; this results in the edge-weighted {\tt PlasmoData.DataGraph} formed in line \ref{line:dengue_graph} having only one connected component.\\

\begin{figure}[!htp]
    \begin{minipage}[t]{\textwidth}
    \begin{scriptsize}
    \lstset{language=Julia, breaklines = true}
    \begin{lstlisting}[label = {code:denge}, caption = Code for converting time series data to an edge-weighted graph. The code follows the general procedure of de Souza and co-workers \cite{de2022euler} in filtering the graph] 
using PlasmoData, Graphs, DelimitedFiles    
    
# Read in data
data = readdlm("Recife_data.csv", ',', Int) |\label{line:recife_data}|

# Define function for performing filtration
function find_smallest_filtered_graph(graph) |\label{line:recife_function_start}|

    # Define values for iteration; this ensures that we only add one edge at a time
    iter_values = sort(get_edge_data(graph)[:], rev = true) |\label{line:recife_iters}|

    for i in iter_values
        # Filter out all edges below a given threshold
        # Only keeps edges where Base.isgreater(edge_weight, i) is true
        filtered_graph = filter_edges(graph, i; fn = Base.isgreater) |\label{line:dengue_graph}|

        # If the number of connected components is 1, return the TDA metrics
        if length(connected_components(filtered_graph)) == 1 |\label{line:TDA_start}|
            EC = get_EC(filtered_graph)
            num_max_cliques = length(maximal_cliques(filtered_graph))
            num_communities = length(clique_percolation(filtered_graph, k = 25))
            return EC, num_max_cliques, num_communities
        end |\label{line:TDA_end}|
    end
end |\label{line:recife_function_end}|

# Create array for storing solutions
ECs = zeros(size(data, 1) - 6)
num_max_cliques = zeros(size(data, 1) - 6)
num_communities = zeros(size(data, 1) - 6)

for i in 1:length(sols)
    # Form a correlation matrix based on 7 days of data
    cor_mat = cor(data[i:(i + 6), :], dims = 1) |\label{line:recife_correlation}|

    # Remove the matrix entries that are NaNs
    bit_vec = (!).(isnan.(cor_mat[:, 1])) |\label{line:recife_filter_nans_start}|
    sym_mat = cor_mat[bit_vec, bit_vec] |\label{line:recife_filter_nans_end}|

    # If there are not more than 2 nodes in the graph, skip this iteration
    node_count = size(sym_mat, 1)
    if node_count <= 2
        continue
    end

    sym_graph = symmetric_matrix_to_graph(sym_mat) |\label{line:recife_build_graph}|

    ECs[i], num_max_cliques[i], num_communities[i] = find_smallest_filtered_graph(sym_graph)
end
    \end{lstlisting}
    \end{scriptsize}
    \end{minipage}
\end{figure}

\begin{figure}[!htp]
    \centering
    \includegraphics[width = .9\textwidth]{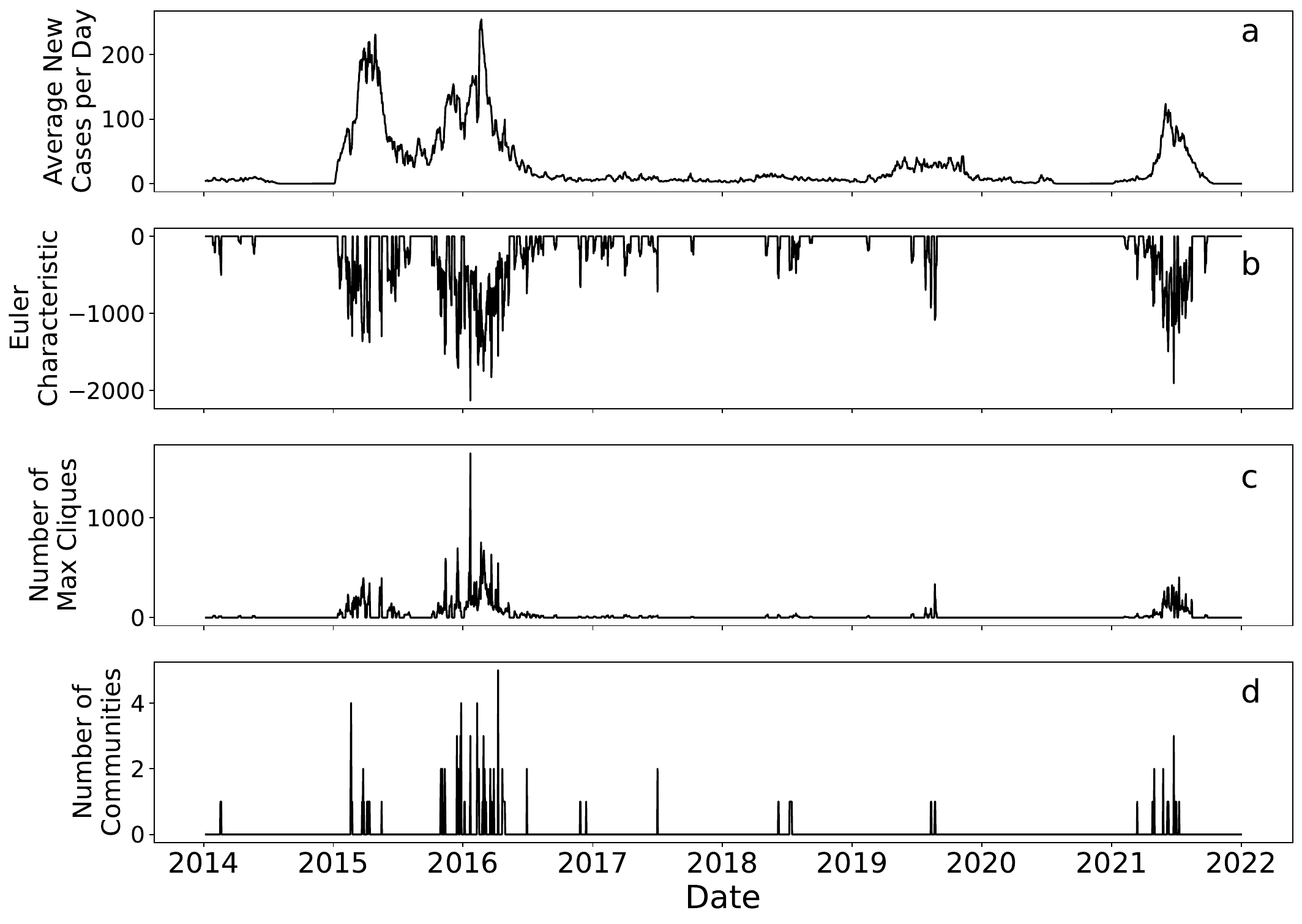}
    \captionof{figure}{Results for TDA of dengue data for 93 districts in Recife, Brazil \cite{de2022euler}. For each 7-day moving time window, an edge-weighted {\tt DataGraph} was formed. Results presented include the average number of new cases per day for the moving time window (a), the EC for the {\tt DataGraph} (b), the number of maximal cliques in the {\tt DataGraph} (c), and the number of communities in the {\tt DataGraph} using clique percolation ($k = 25$) \cite{palla2005uncovering}}
    \label{fig:dengue_EC}
\end{figure}

With the filtered {\tt DataGraph} defined, we can consider several topological metrics for analyzing our data. The goal of de Souza and co-workers \cite{de2022euler} was to identify ways of ``fingerprinting" when disease outbreaks were occurring. They chose the EC as their primary indicator, and they showed how the EC was able to help indicate when outbreaks were occurring, especially for larger $d_{max}$ values. This may be because the higher $d_{max}$ values indicate when there are several nodes (districts) are closely correlated. Consequently, we consider the EC of the graphs being formed, and then we also consider the size of cliques in the graph and the number of communities in the graph. In graph theory, a $k$-clique is a set of $k$ nodes which are connected to all other nodes in the clique. We hypothesize that a possible fingerprint for outbreaks may be when several nodes (districts) are closely correlated. Thus, we compute the EC for comparison with de Souza and co-workers \cite{de2022euler}, we  compute the total number of maximal cliques in the graph (i.e, the total number of $k_{max}$-cliques in the graph, where $k_{max}$ is the largest integer value for which there is a nonempty set of nodes forming a $k_{max}$-clique in the graph) and we compute the number of communities in the graph (found by clique percolation \cite{palla2005uncovering}, with $k$ = 25; see Lines \ref{line:TDA_start} - \ref{line:TDA_end}). The results of these computations, as well as the average daily new cases during each 7 day moving window, are shown in Figure \ref{fig:dengue_EC}.\\

The results shown in Figure \ref{fig:dengue_EC} are consistent with the results of de Souza and co-workers \cite{de2022euler}. As mentioned above, the EC we compute in the above code is for a graph, while de Souza and co-workers \cite{de2022euler} computed the EC for CW-complexes, where their computed EC was a function of $d_{max}$. They used various $d_{max}$ values up to $7$, and found that the noise of the EC decreased with increasing $d_{max}$. In this case, the EC in Figure \ref{fig:dengue_EC} does contain significant noise (visually similar to that seen for $d_{max} = 2$ and $d_{max} = 3$ of de Souza and co-workers' \cite{de2022euler} work). However, the trends are similar, and we note that it is computationally much easier to compute the EC for a graph (i.e., number of nodes minus number of edges) than to compute the EC for a CW-complex with larger $d_{max}$ (where each $d_{max}$-cell may have to be determined). In addition, the number of maximal cliques (Figure \ref{fig:dengue_EC}c) and the number of communities (Figure \ref{fig:dengue_EC}d) also correlate to the average number of new cases per day, and they appear slightly less noisy than the EC. A detailed analysis of these additional indicators as fingerprints for disease outbreaks is outside the scope of this work; instead, our purpose in showing these results is to highlight how the {\tt DataGraph} abstraction and how {\tt PlasmoData.jl} facilitates different TDA approaches.

\subsection{Connectivity Analysis}

In this subsection, we present a technology pathway analysis example to illustrate how the {\tt DataGraph} abstraction and {\tt PlasmoData.jl} can be used to navigate connectivity and provide insights into the technology pathway. The technology pathway we consider has five raw materials (petroleum naphtha, natural gas, corn stover, sugar beets, and sugarcane) that can be used in technologies to produce seven different polymers (low density polyethylene (LDPE), high density polyethylene (HDPE), polypropylene (PP), polyvinylchloride (PVC), polystyrene (PS), polyethylene terephthalate (PET), and Nylon66). There are 15 different intermediate chemicals produced and 27 different technologies that can be used to produce different chemicals. Each of these raw materials, products, intermediates, and technologies are represented by nodes.  The interconnectivity of materials and technologies is complex and difficult to navigate \cite{lopez2023graph}.  This type of connectivity also arises in complex, multi-product supply chains \cite{tominac2021economic}. 
\\

Within the above technology pathway, different types of nodes have specific data attached to them. For example, nodes can contain supply limitations, demands, costs, and carbon emissions, and these will be saved as node data. This problem also contains connections between raw materials and technologies, technologies and intermediates, intermediates and technologies, and technologies and products which can all be represented by directed edges, forming a directed graph, with 57 total edges. In addition, we also have flows that are represented as edge data. These flows are dependent on the connections in the graph and on the data stored in the nodes (e.g., it depends on the cost of technologies or the maximum available supply of the products).\\

The {\tt DataGraph} for this technology pathway is a heterogenous graph and contains multiple attributes for node and edge data. The {\tt DataGraph} contains the set of 54 nodes and 57 directed edges discussed above, which we notate here as $\mn$ and $\me$. There are four subsets of nodes, $\mn_{raw}, \mn_{prod}, \mn_{int}$, and $\mn_{tech}$ for the sets of raw materials, products, intermediates, and technologies, respectively, such that $\mathcal{N} = \mn_{raw} \cup \mn_{prod} \cup \mn_{int} \cup \mn_{tech}$. Each node stores data under different attributes indicating whether the node is a raw material ($a_{raw}$), product ($a_{prod}$), intermediate ($a_{int}$), or technology ($a_{tech}$) and data under different attributes for the raw material cost ($a_{cost}$), the CO\textsubscript{2} emissions per weight of material ($a_{CO_2}$), maximum raw material supply ($a_{sup}$), and maximum production demand ($a_{dem}$). Each edge has data for the optimal flow of material along the edge under attribute $a_{flow}$. 
\\

The {\tt DataGraph}, $\mdg$, can be represented mathematically as:
\begin{equation}\label{eq:digraph}
    \begin{split}
    &\;\mathcal{DG}(\mathcal{N}, \mathcal{E}, A, \bd^\mn, \bd^\me, \emptyset, \ma_\mn, \ma_\me, \emptyset)\\
    \textrm{where} &\; \mn = \mn_{raw} \cup \mn_{prod} \cup \mn_{int} \cup \mn_{tech} \\
    &\; \ma_\mn = \{ a_{raw}, a_{prod}, a_{tech}, a_{cost}, a_{CO_2}, a_{sup}, a_{dem} \}\\
    &\; \mathcal{A}_\mathcal{E} = \{ a_{flow} \} \\
    &\; d^\mn_{a_{raw}}(n) = 1, \quad n \in \mn_{raw}, \qquad d^\mn_{a_{raw}} = 0, \quad n \in \mn \setminus \mn_{raw}\\
    &\; d^\mn_{a_{prod}}(n) = 1, \quad n \in \mn_{prod}, \qquad d^\mn_{a_{prod}} = 0, \quad  n \in \mn \setminus \mn_{prod}\\
    &\; d^\mn_{a_{int}}(n) = 1, \quad n \in \mn_{int}, \qquad d^\mn_{a_{int}} = 0, \quad n \in \mn \setminus \mn_{int}\\
    &\; d^\mn_{a_{tech}}(n) = 1, \quad  n \in \mn_{tech}, \qquad d^\mn_{a_{tech}} = 0, \quad n \in \mn \setminus \mn_{tech}\\
    &\; d^\mn_{a_{cost}}(n) = \alpha_{cost}(n), \quad n \in \mn_{raw}\cup \mn_{tech}, \qquad d^\mn_{a_{cost}} = 0, \quad n \in \mn_{prod} \cup \mn_{int}\\
    &\; d^\mn_{a_{CO_2}}(n) = \alpha_{CO_2}(n), \quad n \in \mn_{raw} \cup \mn_{tech}, \qquad d^\mn_{a_{CO_2}}(n) = 0, \quad n \in \mn_{prod} \cup \mn_{int} \\
    &\; d^\mn_{a_{sup}}(n) = \alpha_{sup}(n), \quad n \in \mn_{raw}, \qquad d^\mn_{a_{sup}} = 0, \quad n \in \mn \setminus \mn_{raw}\\
    &\; d^\mn_{a_{dem}}(n) = \alpha_{dem}(n), \quad n \in \mn_{prod}, \qquad d^\mn_{a_{dem}} = 0, \quad n \in \mn \setminus \mn_{prod}\\
    &\; d^\me_{a_{flow}}(e) = \alpha_{flow}(e), \quad e \in \me
    \end{split}
\end{equation}
where $A$ is an asymmetric matrix ($\mdg$ is directed) matching the connectivity defined by $\me$. Here, $\alpha_{cost}(n)$ is the raw material cost for node $n$, $\alpha_{CO_2}(n)$ is the CO\textsubscript{2} emissions per weight of material for node $n$, $\alpha_{sup}(n)$ is the maximum raw material supply for node $n$, $\alpha_{dem}(n)$ is the maximum production demand for node $n$, and $\alpha_{flow}(e)$ is the optimal flow on edge $e$ (this value is computed separately by solving an optimization problem that uses the above data).\\

The code for creating $\mdg$ in {\tt PlasmoData.jl} is given in Code Snippet \ref{code:tech_path}. In this code, each type of node (Raw Material, Product, Intermediate, and Technology) is loaded as a separate CSV file, where the CSV contains the name of the node and data corresponding to that node. Each node is added using the {\tt add\_node!} function, and data is added using the {\tt add\_node\_data!} function. In addition, we add an attribute to each node (Lines \ref{line:raw_data}, \ref{line:prod_data}, \ref{line:int_data}, and \ref{line:tech_data}) where a value of $1$ indicates the type of node. The edges and edge data are then added to the graph using the {\tt add\_edge!} and {\tt add\_edge\_data!} functions (Lines \ref{line:edge_data_start} - \ref{line:edge_data_end}). \\

\begin{figure}[!htp]
    \begin{minipage}[t]{\textwidth}
    \begin{scriptsize}
    \lstset{language=Julia, breaklines = true}
    \begin{lstlisting}[label = {code:tech_path}, caption = Code for constructing a directed graph for a technology pathway problem] 
using PlasmoData, Graphs, DelimitedFiles, PlasmoDataPlots
    
# Read in data
raw_data  = readdlm("rawmaterial_data_array.csv", ',')
prod_data = readdlm("product_data_array.csv", ',')
int_data  = readdlm("intermediates_data_array.csv", ',')
tech_data = readdlm("technology_data_array.csv", ',')
edge_data = readdlm("edge_data.csv", ',')

# Define DataDiGraph
dg = DataDiGraph()

# Add nodes and node data
for i in 1:size(raw_data, 1)
    add_node!(dg, raw_data[i, 1])
    add_node_data!(dg, raw_data[i, 1], 1, "Raw Material") |\label{line:raw_data}|
    add_node_data!(dg, raw_data[i, 1], raw_data[i, 2], "Cost")
    add_node_data!(dg, raw_data[i, 1], raw_data[i, 3], "CO2 Cost")
    add_node_data!(dg, raw_data[i, 1], raw_data[i, 4], "Max Supply")
end

for i in 1:size(prod_data, 1)
    add_node!(dg, prod_data[i, 1])
    add_node_data!(dg, prod_data[i, 1], 1, "Product") |\label{line:prod_data}|
    add_node_data!(dg, prod_data[i, 1], prod_data[i, 2], "Demand Limit")
end

for i in 1:size(int_data, 1)
    add_node!(dg, int_data[i, 1]) 
    add_node_data!(dg, int_data[i, 1], 1, "Intermediate") |\label{line:int_data}|
end

for i in 1:size(tech_data, 1)
    add_node!(dg, tech_data[i, 1])
    add_node_data!(dg, tech_data[i, 1], 1, "Technology") |\label{line:tech_data}|
    add_node_data!(dg, tech_data[i, 1], tech_data[i, 2], "Cost")
    add_node_data!(dg, tech_data[i, 1], tech_data[i, 3], "CO2 Cost")
end

# Add edges and edge data
for i in 1:size(edge_data, 1) |\label{line:edge_data_start}|
    edge = (edge_data[i, 1], edge_data[i, 2])
    PlasmoData.add_edge!(dg, edge)
    PlasmoData.add_edge_data!(dg, edge, edge_data[i, 3], "Optimal Flow")
end |\label{line:edge_data_end}|

plot_graph(dg, dag_positions = true, nlabels = dg.nodes) |\label{line:plot_graph}|
    \end{lstlisting}
    \end{scriptsize}
    \end{minipage}
\end{figure}

\subsubsection{Visualization}

\begin{figure}[!htp]
    \centering
    \includegraphics[scale = .25]{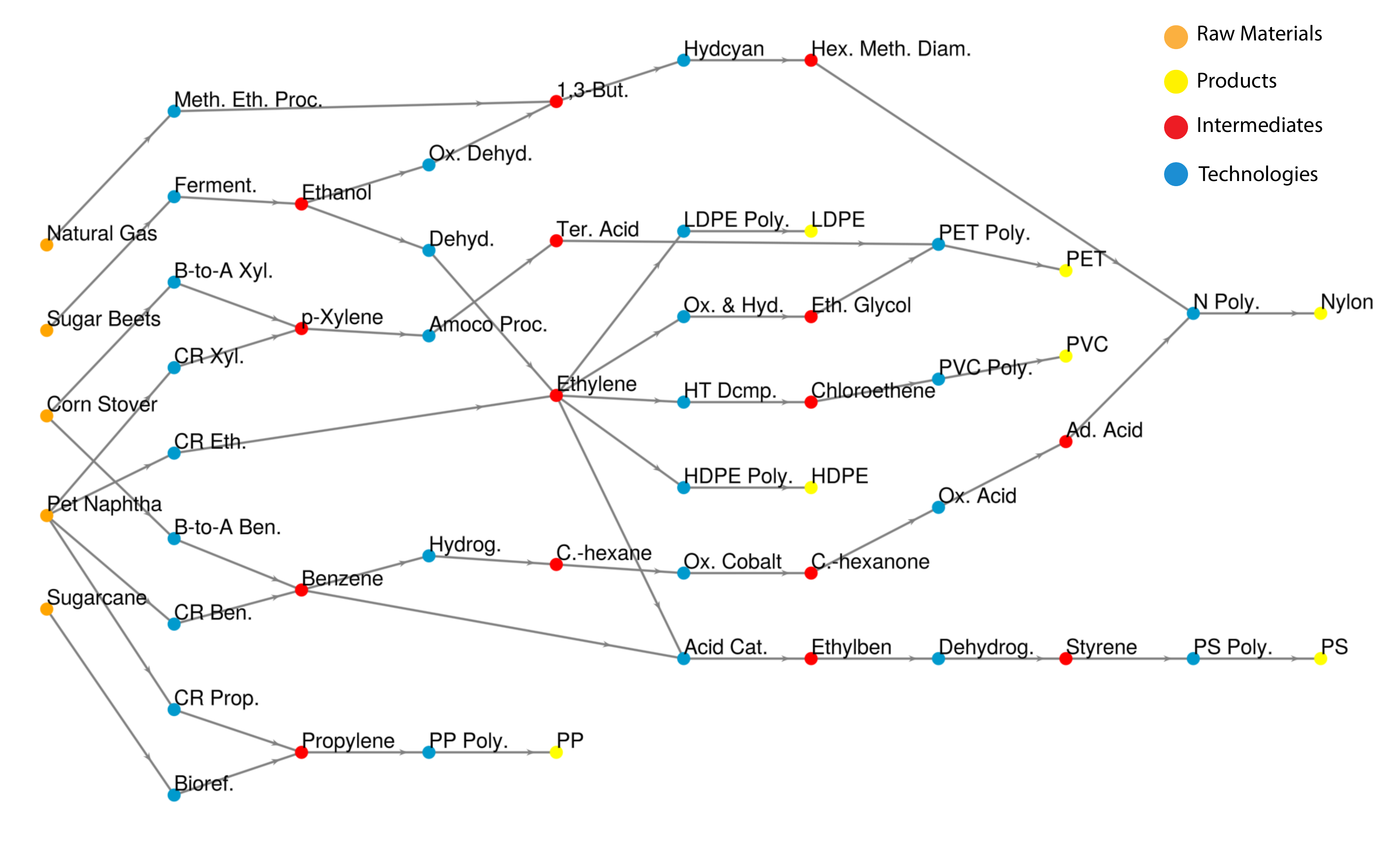}
    \captionof{figure}{Directed graph of a layered technology pathway example; created using {\tt PlasmoDataPlots.jl}}
    \label{fig:tech_path}
\end{figure}

One of the potential benefits of these graph representations is the ability to visualize the structure of the problem. In this example, $\mdg$ is a directed acyclic graph (DAG), which is a common structure in pathway and supply chain problems. Because it is a DAG, the graph can be presented in different layers, and different algorithms have been proposed for identifying these layers and the positions of nodes within the visualization \cite{sugiyama1981methods,zarate2018optimal}. The above technology problem is shown in Figure \ref{fig:tech_path} using the algorithm of Zarate and co-workers \cite{zarate2018optimal} to determine node positions (as implemented within the {\tt Julia} package {\tt LayeredLayouts.jl} \cite{LayeredLayouts.jl}). This visualization is automated through {\tt PlasmoDataPlots.jl}. In this case, Figure \ref{fig:tech_path} (without node coloring) can be easily constructed with just Line \ref{line:plot_graph} of Code Snippet \ref{code:tech_path} (abbreviation meanings are available in the supporting information).\\

\begin{figure}[!htp]
    \centering
    \includegraphics[scale = .25]{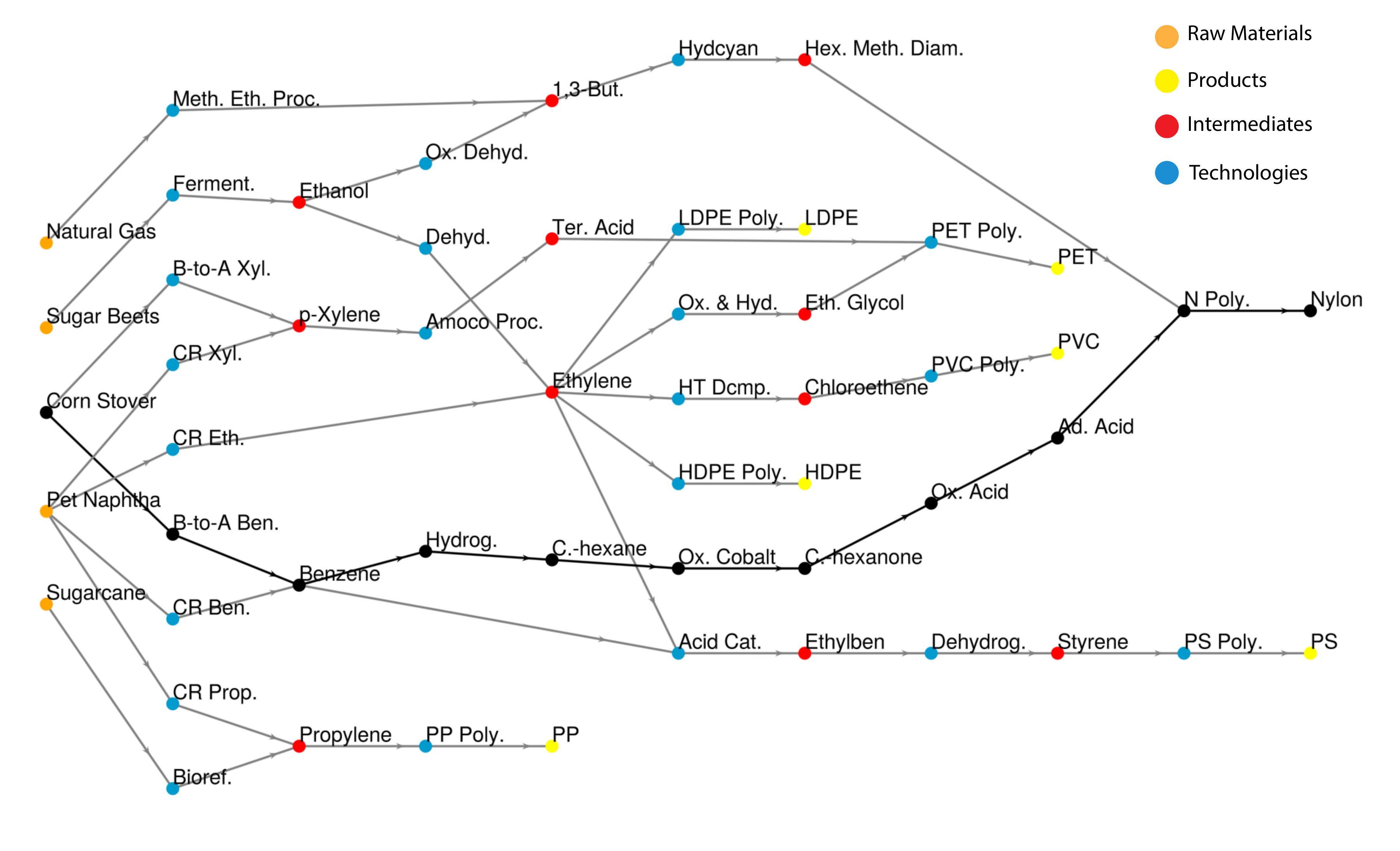}
    \captionof{figure}{Directed graph with a pathway between "Corn Stover" and "Nylon" highlighted, created using {\tt PlasmoDataPlots.jl}}
    \label{fig:tech_path_with_path}
\end{figure}

The ability to visualize these structures can be a significant resource in evaluating the data. For example, it can make clear what some of the paths are or what some of the most essential nodes are. For instance, from Figure \ref{fig:tech_path}, it is clear that the node "Ethylene" has more connections than any other node and is directly connected to several products. In addition, {\tt PlasmoDataPlots.jl} provides the ability to highlight paths within a graph. For example, by calling the function {\tt plot\_graph\_path} with the source ("Corn Stover") and destination ("Nylon") nodes of the path, Figure \ref{fig:tech_path_with_path} can be generated.  Visualizations for DAGs or for layered graphs are often not implemented directly in graph plotting tools, and other tools for visualization (e.g., yFiles \cite{yFiles}) can require subscriptions, so this ability for visualization is a powerful capability of {\tt PlasmoData.jl}.\\

\begin{figure}[!htp]
    \centering
    \includegraphics[scale = .38]{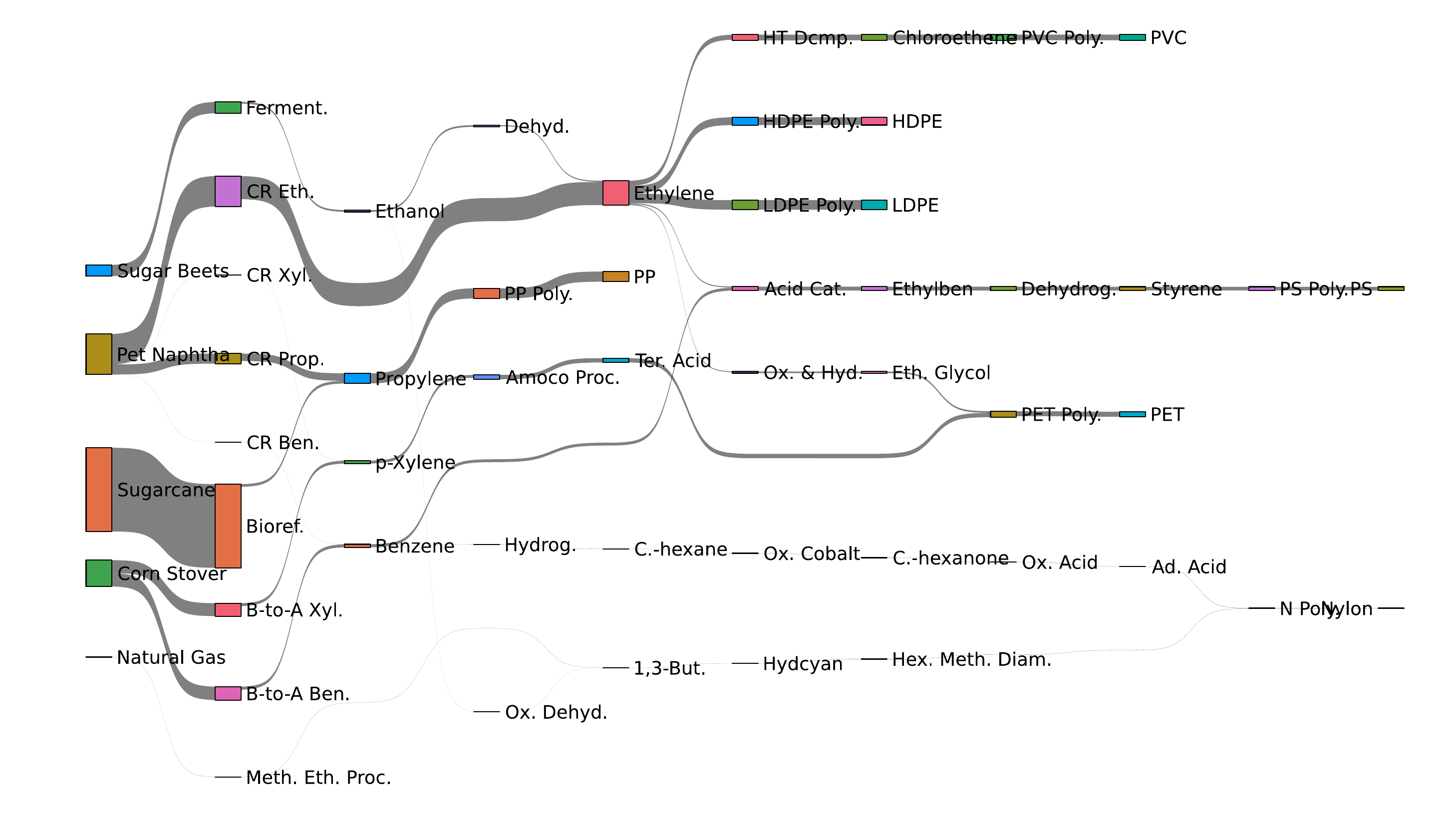}
    \captionof{figure}{Sankey diagram of a technology pathway example}
    \label{fig:sankey_plot}
\end{figure}

In addition to plotting the layered structure, incorporating the data of the graph could also be useful. One way of doing this would be to create a Sankey diagram. Often, Sankey diagrams are not thought of as graphs, but they are effectively just directed graphs with edge weights. Thus, to go from the directed graph defined in Code Snippet \ref{code:tech_path} to a Sankey diagram, we essentially need only five lines of code (see Code Snippet \ref{code:sankey_plot}). The resulting diagram can be seen in Figure \ref{fig:sankey_plot}. Thus, constructing data within a graph structure can provide unique and useful visualization capabilities. 

\begin{figure}[!htp]
    \begin{minipage}[t]{\textwidth}
    \begin{scriptsize}
    \lstset{language=Julia, breaklines = true}
    \begin{lstlisting}[label = {code:sankey_plot}, caption = Code for constructing a Sankey diagram from a {\tt PlasmoData.DataDiGraph}] 
using SankeyPlots

# Define src and dst for edges
src = [i for (i, j) in dg.edges]
dst = [j for (i, j) in dg.edges]
weights = get_edge_data(dg, "Optimal Flow")
        
# Plot sankey diagram
sankey(src, dst, weights; 
    size = (1400, 800), 
    node_labels = dg.nodes, 
    edge_color = :black, 
    label_size = 9
)
    \end{lstlisting}
    \end{scriptsize}
    \end{minipage}
\end{figure}

\subsubsection{Connectivity Measures}

An important challenge in working with graph models of data is how to navigate connections to get information out of the data. For example, in the above technology pathway, how nodes are connected can reveal information about each node. Different metrics can give insight into how ``important" or connected a node may be. Thus, we can consider metrics such as the number of upstream or downstream nodes or the number of products to which a node is connected. {\tt PlasmoData.jl} facilitates these kinds of metrics by providing functions for identifying paths between nodes and identifying sets of upstream or downstream nodes (in part through accessing functions in {\tt Graphs.jl}). In addition, one might also be interested in how manipulating the graph structure can impact these paths (e.g., If a node is removed, what happens to the structure?). We will show here how some of these analyses can be performed. \\

We consider how connectivity metrics can assist in understanding the data. Algorithms have been proposed in literature for identifying paths and connections between nodes in a graph \cite{bellman1958routing,dijkstra2022note, floyd1962algorithm, lee1961algorithm}. For example, we can consider to how many products each raw material can contribute, or we can consider how many raw materials contribute to each product. Analyses like these can give insight into the raw materials or products; for example this could help identify which raw materials or products are most susceptible to disruption (i.e., if a product only has one raw material connected to it, it may be more easily disrupted if anything happens to the supply of that material). For each raw material and product, we can also count the total number of upstream and downstream nodes to provide an idea of how "connected" each raw material and product is to the rest of the graph. All of the above metrics are easily performed in {\tt PlasmoData} as shown in Code Snippet \ref{code:connectivity_metrics}, using functions like {\tt has\_path} (Line \ref{line:has_path}), {\tt downstream\_nodes} (Line \ref{line:downstream_start}), and {\tt upstream\_nodes} (Line \ref{line:upstream_start}). The above metrics can also be added to the graph as new data by defining a new attribute for each metric (see Lines \ref{line:downstream_end}, \ref{line:upstream_end}, and \ref{line:has_path_attribute}).\\

\begin{figure}[!htp]
    \begin{minipage}[t]{\textwidth}
    \begin{scriptsize}
    \lstset{language=Julia, breaklines = true}
    \begin{lstlisting}[label = {code:connectivity_metrics}, caption = Code for determining the connectivity metrics of the technology graph given in Code Snippet \ref{code:tech_path}] 
# Compute metrics for raw materials
for r in raw_data[:, 1]
    count = 0
    num_downstream = length(downstream_nodes(dg, r)) |\label{line:downstream_start}|
    add_node_data!(dg, r, num_downstream, "Number Downstream") |\label{line:downstream_end}|
    for p in prod_data[:, 1]
        if PlasmoData.has_path(dg, r, p)
            count += 1
        end
    end
    add_node_data!(dg, r, count, "Connected Products")
end

# Compute metrics for products
for p in prod_data[:, 1]
    count = 0
    num_upstream = length(upstream_nodes(dg, p)) |\label{line:upstream_start}|
    add_node_data!(dg, p, num_upstream, "Number Upstream") |\label{line:upstream_end}|
    for r in raw_data[:, 1]
        if PlasmoData.has_path(dg, r, p) |\label{line:has_path}|
            count += 1
        end
    end
    add_node_data!(dg, p, count, "Connected Raw") |\label{line:has_path_attribute}|
end
    \end{lstlisting}
    \end{scriptsize}
    \end{minipage}
\end{figure}

In addition to computing measures on the original graph, we can also consider how these metrics change as the graph structure is altered such as through node- or edge-removal, node- or edge-filtration, or aggregation. For example, we can look at what happens if we remove different intermediates or technologies from the graph. Doing so can be thought of as considering the impact if an intermediate material or technology were no longer available or taken off-line unexpectedly. Using {\tt PlasmoData.jl}, we remove the intermediates Ethylene, Terephthalic Acid, and Cyclohexane separately from the graph and see what impact each has on the metrics (i.e., we call {\tt remove\_node!} and reevaluate Code Snippet \ref{code:connectivity_metrics}). The original metrics for each graph and the resulting changes can be seen in Tables \ref{tab:raw_material_metrics} and \ref{tab:product_metrics}. Of those three intermediates, the only intermediate that is completely essential is Ethylene (when ethylene is removed, LDPE, HDPE, and PVC have no connection to any raw materials). It is also clear from the metrics that removing Ethylene from the graph could impact the production of PS and PET since these are no longer connected to as many raw materials. This type of analysis can be valuable in analyzing the resilience of a system in the face of failure of certain elements. 

\begin{table}[!htp]
    \caption{Metrics on a technology graph with the number of connected products (Prod) to each raw material and the number of downstream nodes (Down) for each raw material. The nodes Ethylene, Ter Acid, and C.-hexane were each individually removed from the original graph and metrics recomputed}
    \setlength{\tabcolsep}{10pt}
    \centering
    \begin{tabular}{c|cc|cc|cc|cc}
    \hline\hline
      & \multicolumn{2}{c|}{Original Graph} & \multicolumn{2}{c|}{w/o Ethylene} &  \multicolumn{2}{c|}{w/o Ter. Acid} & \multicolumn{2}{c}{w/o C.-hexane}\\
    \hline
     Raw Material & Prod & Down & Prod & Down & Prod & Down & Prod  & Down\\
    \hline
    \hline
    Sugar Beets & 6 & 29 & 1 & 10 & 6 & 29 & 6 & 29 \\
    Sugarcane   & 1 & 5  & 1 & 5  & 1 & 5  & 1 & 5  \\
    Corn Stover & 3 & 23 & 3 & 23 & 2 & 20 & 2 & 16 \\
    Natural Gas & 1 & 7  & 1 & 7  & 1 & 7  & 1 & 7  \\
    Pet Naphtha & 7 & 39 & 4 & 28 & 7 & 38 & 6 & 32 \\
    \hline
    \end{tabular}
    \label{tab:raw_material_metrics}
\end{table}

\begin{table}[t]
    \caption{Metrics on the technology graph with the number of connected raw materials (Raw) to each product and the number of upstream nodes (Up) for each product. The nodes Ethylene, Ter Acid, and C.-hexane were each individually removed from the original graph and metrics recomputed}
    \setlength{\tabcolsep}{10pt}
    \centering
    \begin{tabular}{c|cc|cc|cc|cc}
    \hline\hline
      & \multicolumn{2}{c|}{Original Graph} & \multicolumn{2}{c|}{w/o Ethylene} &  \multicolumn{2}{c|}{w/o Ter. Acid} & \multicolumn{2}{c}{w/o C.-hexane}\\
    \hline
     Product & Raw & Up & Raw & Up & Raw & Up & Raw  & Up\\
    \hline
    \hline
    LDPE  & 2 & 9  & 0 & 2  & 2 & 9  & 2 & 9  \\
    HDPE  & 2 & 9  & 0 & 2  & 2 & 9  & 2 & 9  \\
    PP    & 2 & 7  & 2 & 7  & 2 & 7  & 2 & 7  \\
    PVC   & 2 & 11 & 0 & 4  & 2 & 11 & 2 & 11 \\
    PS    & 3 & 17 & 2 & 11 & 3 & 17 & 3 & 17 \\
    PET   & 4 & 17 & 2 & 11 & 2 & 11 & 3 & 17 \\
    Nylon & 4 & 22 & 4 & 22 & 4 & 22 & 2 & 15 \\
    \hline
    \end{tabular}
    \label{tab:product_metrics}
\end{table}

\begin{figure}[!htp]
    \centering
    \includegraphics[scale = .25]{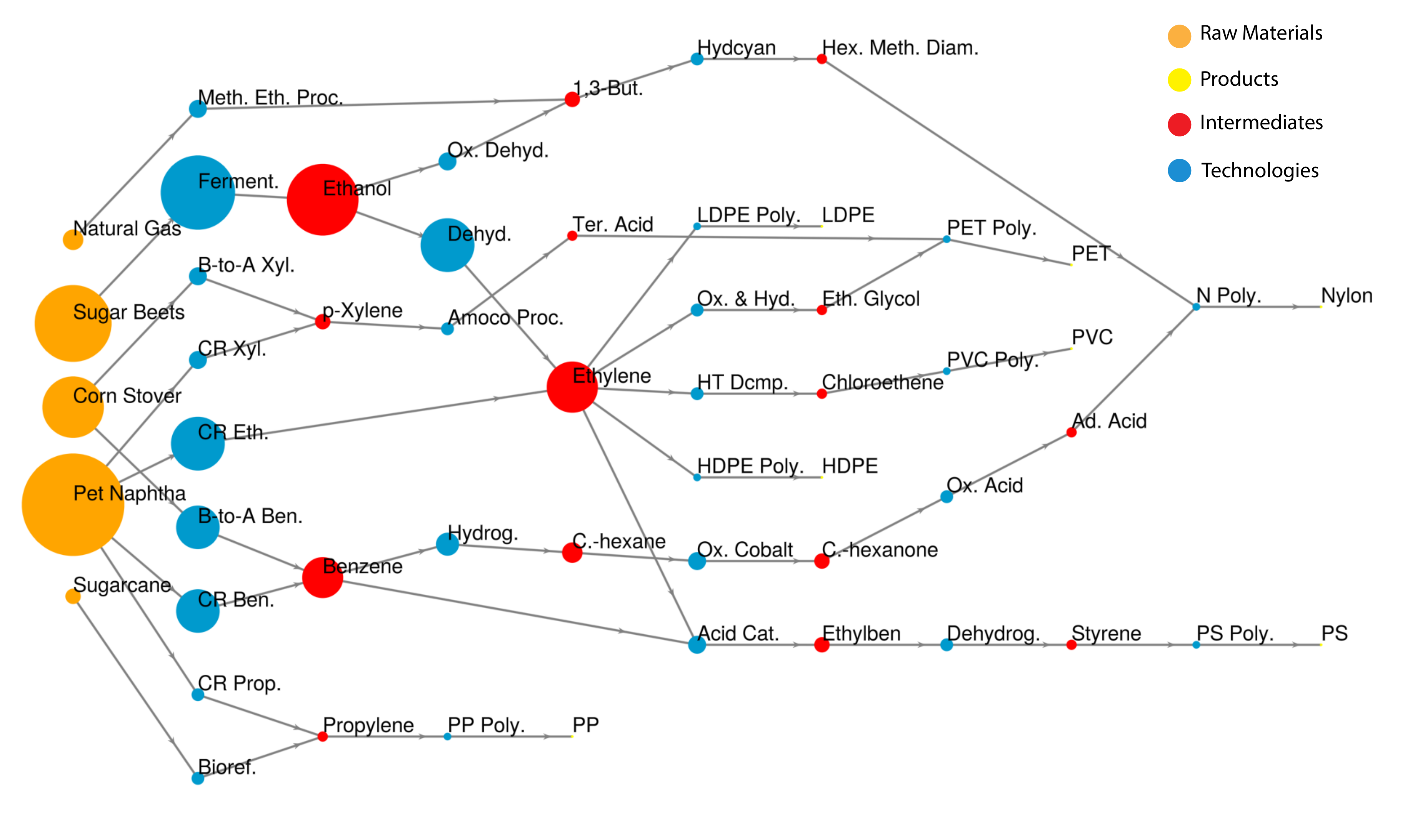}
    \captionof{figure}{Directed graph of a layered technology pathway example with nodes sized by the number of downstream nodes, created using {\tt PlasmoDataPlots.jl}}
    \label{fig:tech_path_downstream}
\end{figure}

The above tools can also be coupled with the visualization tools to provide an easier interface to understand these changes. For example, Figure \ref{fig:tech_path_downstream} shows the technology graph where nodes are sized by their number of downstream connections (i.e., sizing by node data). This shows visually which raw materials are connected to the most downstream nodes, but it also gives insight into which intermediate and technology nodes contain the most downstream connections. 

The {\tt DataGraph} abstraction and {\tt PlasmoData.jl} provide a framework for examining the connections within data structures to gain insight into the data or system. In the above example, we have highlighted just a couple ways that this could be done, but several other methods (enabled within {\tt PlasmoData.jl}) could likewise be used. For example, the structure of the graph could be filtered based on node weights (e.g., cost or carbon emissions) or edge weights (e.g., flows) which would alter the structure and connections of the graph. Further, other structure manipulations (such as adding edges) could be important in studying these graphs. For example, you could consider where edges could be placed to increase the number of connections between raw materials and products (hypothetically increasing the "resilience" of the structure). These are just a couple examples of other ways {\tt PlasmoData.jl} could be used to elucidate more information about a graph. 


\section{Conclusions and Future Work}\label{sec:conclusion}

We presented a graph-theoretic  abstraction for modeling data; we call this abstraction{\tt DataGraph} and provide a software implementation in the {\tt Julia} package that we call {\tt PlasmoData.jl}. We show that common data structures (e.g., matrices, tensors, images) can be represented under the {\tt DataGraph} abstraction, and there are a variety of mathematical tools available for analyzing the data modeled as a graph (e.g., filtration, aggregation, topological analysis). We presented case studies to show this abstraction can be used for a variety of tasks including determining features for machine learning, analyzing time series, and performing path analysis. In doing this, we have highlighted how representing data can be a modeling' task and having the correct abstraction for the data can strongly influence the value of the data.\\

There are a couple of areas that we would like to focus on in future work. First, there are more applications we would like to explore. Using graphs to represent molecules has garnered significant interest and has been applied in many studies and we would like to streamline molecular representations within {\tt PlasmoData.jl}, such as enabling construction of the graph from a SMILES (Simplified Molecular Input Line Entry System) string, and explore how {\tt PlasmoData.jl} can enhance this analysis. We are also interested in other applications of the pathway analysis performed above. Additionally, we would like to expand the capabilities of {\tt PlasmoData.jl}. There are other TDA metrics and path functions that could be implemented, and there are topological descriptors that could be approximated with graphs. For example, in the work by Jiang and co-workers \cite{jiang2023scalable}, they use a marching-square algorithm to approximate the Minkowski functionals of a filtered image. A marching-square algorithm could be implemented for an image represented by a graph by iterating through the nodes and analyzing the connections to the adjacent nodes. This could lead to more in-depth image analysis. Also, we would like to expand the integration of {\tt PlasmoData.jl} with GNN packages; this could include improving the interface with GNN packages and enabling different graph manipulation functions with GNN algorithms, such as using aggregation for pooling operations or message-aggregator operators. \\

In addition, while graphs can be powerful tools for modeling data, they do have some limitations. Nodes and edges do not have any notion of placement in space. Oftentimes, the systems being represented by graphs have a fixed location (e.g., supply chains may have a fixed production location) which can significantly influence the system. Furthermore, graphs only capture connectivity between nodes; in many instances, it may be more accurate to include a link between more than two nodes. In these cases, hypergraphs (or other representations) may be more appropriate. Finally, graphs are restricted to edges between no more than two nodes; in future work, we would like to explore generalizations of hypergraphs and simplicial complexes (which may capture more complex and higher-order relations) and how to capture more complex attributes for these objects, such as spatial locations. 


\section*{Acknowledgments}

This  material  is  based  upon  work  supported  by  the  U.S.  Department  of  Energy, Office  of  Science,  Office  of  Advanced  Scientific  Computing  Research  (ASCR)  under  contract  DE-SC0023361 and in part by the National Science Foundation under award CBET-2315963. 


\bibliography{./DataGraphs}

\end{document}



\title{{\bf Supporting Information}\\ {\tt PlasmoData.jl} -- A Software Framework for Representing, Modeling, and Analyzing Data as Graphs}

\author{David L. Cole${}^{\dag}$and Victor M. Zavala${}^{\dag\ddag}$\thanks{Corresponding Author: victor.zavala@wisc.edu}
}

\date{\small
  ${}^\dag$Department of Chemical and Biological Engineering, \\[0in]
  University of Wisconsin-Madison, Madison, WI 53706 United States of America\\[.05in]
  ${}^\ddag$Mathematics and Computer Science Division, \\
  Argonne National Laboratory,  Lemont, IL 60439 United States of America\\[-2in]
}


\maketitle
\tableofcontents

\section{Data Structure and Memory Allocation}

\subsection{Memory Allocation Comparisons}
For our software implementation, how we chose to store node, edge, and graph data was an important consideration that deserves some explanation. {\tt PlasmoData.jl} stores this data differently than some alternative tools. All data is stored within a matrix for the node data, a matrix for the edge data, and a vector for the graph data. In {\tt Julia}, these data structures can be defined to take a variety of data types (e.g., a matrix can be defined as {\tt Matrix\{Any\}} to take any data structure, or it can be defined as {\tt Matrix\{Float64\}} to restrict the matrix entires to be floats), and this matches the {\tt DataGraph} abstraction. However, many tools (including {\tt Networkx}\cite{hagberg2008networkx} and {\tt MetaGraphs.jl}\cite{Graphs2021}) store data as dictionaries, where each node or edge has a dictionary of attribute names and corresponding data for that attribute. This can be convenient when storing few attributes or for storing attributes that are unique to each node or edge (i.e., each node or edge does not contain the same attributes). However, when each node or edge contains the same attribute (such as the case with the {\tt DataGraph} abstraction) and when there are large amounts of these attributes, storing data as a dictionary is not efficient, because each node must store a list of all dictionary keys. For example, if you store an weight on every node of a graph with the weight name "weight", every node of the graph will have a dictionary containing the key "weight" and the corresponding weight value. If every node of a graph contains the same weight name, that name is being stored in the graph $|\mathcal{N}|$ times, where $\mathcal{N}$ is the set of all nodes in the graph. Thus, we instead choose to store the data in {\tt PlasmoData.jl} as a matrix, meaning each weight name is stored only once, and each node or edge is required to have some value for that weight name. This is also similar to what is done in {\tt Matlab}'s {\tt graph} and {\tt digraph} functions.
\\

In addition, for some data analysis, it is more efficient (and in some cases, it could be necessary) to have a vector or matrix of weights. If the data is stored as a dictionary, then a vector or matrix has to be formed by iterating through all nodes or edges. For example, GNN software may be designed to take node or edge data as matrices (this is the case with {\tt GraphNeuralNetworks.jl}). Thus, {\tt PlasmoData.jl} does not have to iteratively construct vectors or matrices of data when these data types are required.
\\

To validate this data storage decision, we compared the memory allocation for four graphs formed in {\tt PlasmoData.jl}, {\tt NetworkX}, {\tt MetaGraphs}, and {\tt Matlab}'s {\tt graph} \cite{matlab_graphs} function. All graphs are formed from a $100 \times 100$ matrix. Each graph has a different number of node weights saved on its nodes ($1$, $10$, $25$, and $100$ weights). The scripts used to replicate these results are available at \url{https://github.com/zavalab/JuliaBox/tree/master/PlasmoData_examples}. Every node in the graph was given a node name of a tuple of the form $(i, j)$ corresponding to the $i, j$th matrix entry. For the {\tt Matlab} script, this name was given as a string (rather than as a tuple) due to apparent naming restrictions for nodes. Each weight was named by the string "weighti" where "i" was an integer corresponding to the weight number. \\

Comparing memory size can be difficult since there were three different programming languages used ({\tt Julia}, {\tt Python}, and {\tt Matlab}), so we used the following methods for getting memory. In {\tt Julia}, we used the function {\tt Base.summarysize} which returns the memory size in bytes of all unique objects reachable from the argument. In {\tt Python}, there is no base function of which we are aware that gives the memory allocaiton of all unique reachable objects. Consequently, we only give a lower bound on the memory allocated in {\tt NetworkX} by iterating through the nodes, edges, and node attribute dictionary and calling {\tt \_\_sizeof\_\_()} to get the memory of each component of these objects. For {\tt Matlab}, we used a function to iterate through all properties of the graph and sum the size of each property. The results of these methods can be seen in Table \ref{tab:mem_alloc}. \\

\begin{table}[t]
    \caption{Memory Allocaiton comparisons for the graph representation of a $100 \times 100$ matrix containing $1$, $10$, $25$, and $100$ weights on each node. Comparisons were made between a {\tt PlasmoData.jl} {\tt DataGraph}, a {\tt MetaGraphs.jl} {\tt MetaGraph}, a {\tt NetworkX} {\tt graph}, and a {\tt Matlab} {\tt graph}.}
    \setlength{\tabcolsep}{12pt}
    \centering
    \begin{tabular}{c|c|c|c|c}
    \hline\hline
    & \multicolumn{4}{c}{\shortstack{Memory Allocations \\ (million bytes)}}\\
    \hline
    \# Node Weights    & 1 & 10 & 25 & 100\\
    \hline \hline
    {\tt PlasmoData.jl} & 4.14 & 4.86 & 6.06 & 12.07\\
    {\tt MetaGraphs.jl} & 6.19 & 15.07 & 16.27 & 54.91\\
    {\tt NetworkX} & $\ge$ 4.14 & $\ge$ 5.42 & $\ge$ 13.58 & $\ge$ 48.78 \\
    {\tt Matlab graph} & 10.71 & 11.43 & 12.64 & 18.65\\
    \hline
    \end{tabular}
    \label{tab:mem_alloc}
\end{table}

When constructing a matrix as a graph (and when saving node names within this graph), the {\tt DataGraph} structure of {\tt PlasmoData.jl} can be significantly less than alternative data storage problems. When there are many weights saved on each node, the memory allocation can be 3-5 times less. The graph defined in {\tt Matlab} is comparable in size for the larger problems since it is also using a table format for defining the graph. However, we chose not to use {\tt Matlab} as a basis for our package since we wanted an open source package that can be easily extended. \\

It is important to consider that there are an innumerable number of graphs that could be constructed within graph-based modeling tools, and we want to emphasize that {\tt PlasmoData.jl} is not intended to be the ideal tool for all applications. Rather, {\tt PlasmoData.jl} has been developed for representing and modeling data as graphs. Other considerations could be important for a user. For example, in many instances, Each node in a graph may NOT have the same weight/attribute types, and in this case, a dictionary may be a better storage method. Lastly, we recognize that making comparisons between programming languages can be difficult because of implementation differences between the programming languages; further, the choice of programming language to use can be dependent on many factors. Consequently, this analysis is not intended to claim that {\tt PlasmoData.jl} is superior to other modeling tools, but instead, we are showing that {\tt PlasmoData.jl} can be efficient and comparable to some other modeling tools for certain applications. 

\subsection{Memory Usage within {\tt PlasmoData.jl}}

For reference, we also provide a general breakdown of the memory storage for a single {\tt DataGraph} example. For the $100 \times 100$ matrix representation with $100$ node weights mentioned above, Table \ref{tab:datagraph_alloc} contains the memory allocation to each attribute of the {\tt DataGraph}. The largest portion of the memory allocation is for storing the data on the node. The next largest contributors are the edge map storage, the {\tt Graphs.SimpleGraph}, and the edges storage, respectively. This also shows how storing node names and node or edge data can significantly increase memory allocation. The graph structure itself can be represented only by a {\tt Graphs.SimpleGraph} which accounts for just under $10\%$ of the total required memory for the {\tt DataGraph}. The code for generating this breakdown is available at \url{https://github.com/zavalab/JuliaBox/tree/master/PlasmoData_examples}. \\

\begin{table}[t]
    \caption{Memory Allocation within a {\tt DataGraph} for the graph representation of a $100 \times 100$ matrix with $100$ weights on each node}
    \setlength{\tabcolsep}{12pt}
    \centering
    \begin{tabular}{c|c|c}
    \hline\hline
    Property & \shortstack{Thousands \\of Bytes} & \% Total \\
    \hline \hline
    {\tt nodes} & 240. & 1.99\\
    {\tt node\_map} & 439. & 3.64\\
    {\tt node\_data} & 8,007. & 66.36\\
    {\tt edges} & 630. & 5.22\\
    {\tt edge\_map} & 1,639 & 13.58\\
    {\tt edge\_data} & 0.6 & $<$0.01\\
    {\tt graph\_data} & 0.6 & $<$0.01\\
    {\tt g} ({\tt Graphs.SimpleGraph}) & 1,110. & 9.20\\
    \hline
    \end{tabular}
    \label{tab:datagraph_alloc}
\end{table}

\section{Image Analysis -- Considerations}

\subsection{Considerations for Image Classification}

In performing the image classification outlined in the primary manuscript for the liquid crystal problem of Jiang and co-workers \cite{jiang2023scalable}, there are a number of decisions that can influence the accuracy of the methods. We will briefly discuss three of these decisions.\\

Three considerations in this problem are how much data to use (i.e., gray scale data vs. all three channels of data), whether to include diagonal connections in the graph mesh representation (e.g., matrix entry $i, j$ is connected also to $i + 1, j + 1$), and whether we should scale the Euler Characteristic (EC). Because the EC can get very large (especialy as the dimension of the image grows), there is an option within {\tt PlasmoData.jl}'s {\tt run\_EC\_on\_nodes} function to divide the matrix by the number of components (total nodes plus total edges) in the original, unfiltered graph. This ensures that the values returned by the function are never greater than one. We tested each of these decisions to see what impact it could have on the final solution. We did this using the scripts at \url{https://github.com/zavalab/JuliaBox/tree/master/PlasmoData_examples}, with one script for the 3-channel data and another for the gray-scale data. We changed the variables {\tt diagonal} and {\tt scale} at the beginning of each script to alter whether diagonal edges were placed and to change the scaling of the EC, respectively. The results can be seen in Table \ref{tab:image_class}.\\

\begin{table}[t]
    \caption{Results for image classification of for liquid crystals from the data of Jiang and co-workers\cite{jiang2023scalable}}
    \setlength{\tabcolsep}{12pt}
    \centering
    \begin{tabular}{c|c|c|c|c}
    \hline\hline
    Data Used & With Diagonals / With Scaling & Total Time (s) & Accuracy & \shortstack{Standard\\ Deviation} \\
    \hline \hline
    \multirow{4}{*}{\shortstack{3-Channel \\ Data}} & True / True & 37.3 & 91.0 \% & 4.2 \% \\
    & True / False & 36.5 & 94.8 \%  & 2.1 \% \\
    & False / True & 25.1 & 86.8 \% & 2.7 \% \\
    & False / False & 24.5 & 95.1 \% & 1.9 \% \\
    \hline
    \multirow{4}{*}{\shortstack{Gray-Scale \\ Data}} & True / True & 50.9 & 82.3 \% & 2.7 \% \\
    & True / False * & 67.0 & 72.6 \% & 3.9 \% \\
    & False / True & 34.2 & 80.5 \% & 3.0 \% \\
    & False / False * & 53.2 & 76.4 \% & 5.6 \% \\
    \hline
    \multicolumn{5}{l}{* - denotes that training the SVM reached max number of iterations}\\
    \hline
    \end{tabular}
    \label{tab:image_class}
\end{table}

Several interesting points can be taken from the results shown in Table \ref{tab:image_class}. First, using the 3-channel data rather than the gray-scale data resulted in better classification rates. We also note that this could be for different reasons than SVM size. For the gray-scale data, we defined the filtration threshold range as being from $0$-$1$ in increments of $0.001$. In contrast, for the 3-channel data, the filtration threshold range was from $0$-$1$ in increments of $0.005$, so the length of the vector passed to the SVM was actually slightly smaller for the 3-channel data. In addition, we note that not including the diagonals resulted in a shorter overall time compared with including the diagonals. Lastly, the scaling seemed to help the classification rates for the gray-scale images but not for the 3-channel images. Thus, depending on the application, some small changes could improve computation time or classification accuracy. 

\subsection{Graph Neural Network Interface}

Here we make a brief note about constructing and training a graph neural network (GNN) from data defined in {\tt PlasmoData.jl}. In the primary manuscript, we give an example of training a GNN for a graph with node data only (i.e., no edge data). The script for performing this is available at \url{https://github.com/zavalab/JuliaBox/tree/master/PlasmoData_examples}. However, problems with edge data can be slightly more complex. This is because the edge data in the {\tt PlasmoData.jl} graph have an arbitrary order; the edges are added in whatever order the user defines them. However, for {\tt GraphNeuralNetworks.jl} \cite{lucibello2021GNN}, edge data is to have a specific order that matches the adjacency lists of the {\tt Graphs.SimpleGraph} or {\tt Graphs.SimpleDiGraph} passed to the {\tt GNNGraph} function. For {\tt Graphs.SimpleGraph}s, the edge order is stored implicitly through an adjacency list for each node (i.e., explicit node-node pairs for edges are not stored; instead, lists are stored for every neighbor containing the index of that nodes immediate neighbors). Similarly, for {\tt Graphs.SimpleDiGraph}s, two lists are stored for each node: a list of nodes that are connected by an outgoing edge from the node of interest, and a list of nodes that are connected by an incoming edge to the node of interest (i.e., immediate upstream and downstream neighbors). 

The order in which {\tt GraphNeuralNetworks.jl} (and presumably other GNN software) expects the edge data is based on these adjacency lists, and we have added functions in {\tt PlasmoData.jl} to comply with this expected order. For {\tt Graphs.SimpleGraph}s, the expected edge order is dependent on the node order, where all edges connected to the first node are first, then all edges connected to the second node (with the exception of any edge between node 2 and node 1), then all edges connected to the third node (with the exception of edges between nodes 3 and 2 or 3 and 1), etc. For {\tt Graphs.SimpleDiGraph}s, the expected order is the same, but only the outgoing node list is considered (i.e., edges originating at node 1 are considered first, then edges originating at node 2, etc.). As the user defined edge order may not match this expected order, we have defined the functions {\tt get\_ordered\_edge\_data} and {\tt order\_edges!} in {\tt PlasmoData.jl}. The first function returns the edge data on a graph in the order defined above (as expected by {\tt GraphNeuralNetworks.jl}). The second function changes in place the edge order for the {\tt DataGraph} or {\tt DataDiGraph} to match the expected order. In other words, after calling {\tt order\_edges!}, {\tt get\_ordered\_edge\_data} and {\tt get\_edge\_data} return the same matrix. For users passing edge data to a GNN, we recommend using functions in {\tt PlasmoData.jl} to ensure that the edge data is in the proper order.

\section{Connectivity Analysis}

The node abbreviations and names used for the technology pathway example are given in Tables \ref{tab:raw_mat}, \ref{tab:products}, \ref{tab:intermediates}, and \ref{tab:tech}.

\begin{table}
    \caption{Node abbreviations for raw materials in the technology pathway example}
    \begin{center}
        \begin{filecontents*}{raw_materials.csv}
Abbreviation,Name
Sugar Beets,Sugar Beets
Sugarcane,Sugarcane
Corn Stover,Corn Stover
Natural Gas,Natural Gas
Pet Naphtha,Petroleum Naphtha
\end{filecontents*}
\csvautotabular[respect all]{raw_materials.csv}
    \end{center}
    \label{tab:raw_mat}
\end{table}

\begin{table}
    \caption{Node abbreviations for products in the technology pathway example}
    \begin{center}
        \begin{filecontents*}{products.csv}
Abbreviation,Name
LDPE,LDPE
HDPE,HDPE
PP,PP
PVC,PVC
PS,PS
PET,PET
Nylon,Nylon66
\end{filecontents*}
\csvautotabular[respect all]{products.csv}
    \end{center}
    \label{tab:products}
\end{table}

\begin{table}
    \caption{Node abbreviations for intermediates in the technology pathway example}
    \begin{center}
        \begin{filecontents*}{intermediates.csv}
Abbreviation,Name
Ethanol,Ethanol
Ethylene,Ethylene
Eth. Glycol,Ethylene Glycol
Chloroethene,Chloroethene
Propylene,Propylene
Benzene,Benzene
p-Xylene,p-Xylene
Ethylben,Ethylbenzene
Styrene,Styrene
"1,3-But.","1,3-Butadiene"
C.-hexane,Cyclohexane
C.-hexanone,Cyclohexanone
Ad. Acid,Adipic Acid
Hex. Meth. Diam.,Hexamethylenediamine
Ter. Acid,Terephthalic Acid
\end{filecontents*}
\csvautotabular[respect all]{intermediates.csv}
    \end{center}
    \label{tab:intermediates}
\end{table}

\begin{table}
    \caption{Node abbreviations for technologies in the technology pathway example}
    \begin{center}
        \begin{filecontents*}{technologies.csv}
Abbreviation,Name
Dehyd.,Dehydration
Ox. & Hyd.,Oxidation & Hydration
HT Dcmp.,Halogenation Thermal Decomposition
Acid Cat.,Acid Catalysis
Dehydrog.,Dehydrogenation
Ox. Dehyd.,Oxdiative Dehydration 
Bioref.,Biorefinery
Ferment.,Fermentation
CR Eth.,Catalytic Reforming for Ethylene
CR Prop.,Catalytic Reforming for Propylene
CR Xyl.,Catalytic Reforming for Xylene
CR Ben.,Catalytic Reforming for Benzene
Meth. Eth. Proc.,MethanolEthylene Process
Hydrog.,Hydrogenation
Ox. Cobalt,Oxidation Cobalt
Ox. Acid,Oxidation Acid
Hydcyan,Hydrocyanation and Hydrogenation
Amoco Proc.,Amoco Process
B-to-A Xyl.,Biomass-to-Aromatic Process for Xylene
B-to-A Ben.,Biomass-to-Aromatic Process for Benzene
LDPE Poly.,LDPE Polymerization
HDPE Poly.,HDPE Polymerization
PP Poly.,PP Polymerization
PVC Poly.,PVC Polymerization
PS Poly.,PS Polymerization
PET Poly.,PET Polymerization
N Poly.,Nylon66 Polymerization
\end{filecontents*}
\csvautotabular[respect all]{technologies.csv}
    \end{center}
    \label{tab:tech}
\end{table}

\bibliography{supporting_info}